\documentclass[journal]{IEEEtran}
\usepackage{cite}
\usepackage{amsmath,amssymb,amsfonts}
\usepackage{algorithmic}
\usepackage{graphicx}
\usepackage{subcaption}
\usepackage{textcomp}
\usepackage{array}
\usepackage{enumitem}
\usepackage{authblk}
\usepackage{url}
\usepackage{hyperref}

\title{Influence of Network Topology and Vaccination Strategies on HPV Dynamics: A Simulation Study Using the SeCoNet Growth Model}

\author[1,2]{Weiyi Wang}
\author[1,3]{Mahendra Piraveenan}

\affil[1]{Modelling and Simulation Research Group, School of Computer Science, Faculty of Engineering, University of Sydney}
\affil[2]{Sydney Medical School, Faculty of Medicine and Health, University of Sydney}
\affil[3]{Charles Perkins Centre, University of Sydney, and Johns Hopkins University}

\begin{document}

\maketitle

\begin{abstract}
This study examines how contact network topology influences the effectiveness of vaccination programs in the context of human papillomavirus (HPV) transmission. Using the SeCoNet sexual contact network growth model, we evaluate age-based, ring-based, and several centrality-based vaccination strategies across the overall, male, and female cohorts, focusing on peak incidence, timing of peak prevalence, and cumulative incidence. The simulations show that degree-, betweenness-, and percolation centrality-based strategies are generally the most effective, while ring vaccination achieves the greatest reduction in cumulative incidence among females. Network topology also plays a critical role: higher average degree reduces vaccination effectiveness, whereas higher power-law exponent, longer average shortest path length, and stronger clustering improve vaccination outcomes. The results highlight the importance of incorporating network structure into the design of HPV vaccination programs.
\end{abstract}

\section{Introduction}
\label{intro}
Human papillomavirus (HPV) is a widespread virus transmitted via sexual contact primarily, affecting individuals of all genders \cite{AustralianInstituteOfHealthAndWelfare2018HumanPapillomavirus}. It's estimated that approximately 90\% of the global population will contract HPV infection at least once in their lifetime \cite{AustralianDepartmentOfHealthAndAgedCare2023HumanPapillomavirus}. While most HPV infections are asymptomatic and can resolve spontaneously, persistent infection may lead to the development of carcinomas
\cite{Burchell2006Chapter6, Castellsague2009HPVVaccination, WorldHealthOrganization2022HumanPapillomavirus}. Over 200 distinct HPV types have been identified so far, of which 12 HPV genotypes are classified as oncogenic or the high-risk strains, including types 16, 18, 31, 33, 35, 39, 45, 51, 52, 56, 58 and 59 \cite{WorldHealthOrganization2022HumanPapillomavirus}. The oncogenic potential of HPV varies by type, with HPV-16 exhibiting the highest likelihood of persistence and the strongest transmissibility \cite{Burchell2006Chapter6, WorldHealthOrganization2022HumanPapillomavirus, Lowy2008HumanPapillomavirus}. In addition, HPV-16 is also the most prevalent type worldwide, followed by HPV-18 \cite{WorldHealthOrganization2022HumanPapillomavirus, Lowy2008HumanPapillomavirus}. The high-risk strains can cause cancers of the anogenital areas, head and neck, whereas the low-risk ones, such as HPV-6 and -11, contribute to warts on the genitals and surrounding skin \cite{AustralianInstituteOfHealthAndWelfare2018HumanPapillomavirus, WorldHealthOrganization2022HumanPapillomavirus, Lowy2008HumanPapillomavirus}.

Among HPV-related malignancies, the most common one is cervical cancer \cite{AustralianDepartmentOfHealthAndAgedCare2023HumanPapillomavirus}. Back in the 1980s, Schwarz et al. \cite{Schwarz1985StructureTranscription} first demonstrated the link between HPV infection and cervical cancer by identifying the presence and transcription of HPV-16 and -18 in cervical carcinoma biopsies. As research on this topic advanced, it became clear that certain HPV strains were the etiological agents of cervical cancer, which are now referred to as the high-risk strains \cite{Villa1997HumanPapillomaviruses}. Compared to low-risk strains, high-risk HPV infections persist longer, and chronic infections can lead to cervical intraepithelial neoplasia
(CIN), which may progress to invasive cervical cancer (ICC) if untreated. More than 95\% of cervical cancer cases are attributed to persistent infections with oncogenic HPV types, and about 70\% of these are related to ongoing HPV-16 or -18 infection \cite{AustralianDepartmentOfHealthAndAgedCare2023HumanPapillomavirus, WorldHealthOrganization2022HumanPapillomavirus, Carter2011HPVInfection}. In addition to oncogenic HPV infection, other factors contributing to the risk of cervical cancer include the presence of other sexually transmitted infections (STIs), early onset of sexual activity, multiple sexual partners, immunodeficiency, frequent pregnancies, and unhealthy lifestyles, all of which may affect susceptibility to infection and the progression to malignancy \cite{Sung2021GlobalCancer, WorldHealthOrganization2022CervicalCancer}.

Natural infection does not typically induce a strong immunological response, and it usually takes 8 to 12 months after HPV infection for seroconversion to occur \cite{WorldHealthOrganization2022HumanPapillomavirus}. After viral clearance, lower than two thirds of women develop antibodies against HPV, and the proportion is even smaller among men \cite{AustralianDepartmentOfHealthAndAgedCare2023HumanPapillomavirus}. Furthermore, empirical studies suggest that women can be repeatedly infected with the same type or concurrently or subsequently infected with other types; however, the currently available data do not provide evidence that the antibodies induced by natural HPV infection can confer protection against reinfection \cite{WorldHealthOrganization2022HumanPapillomavirus}.

Vaccination has been introduced as the primary preventive intervention against HPV infection, as well as associated cervical premalignant lesions and cancer \cite{AustralianDepartmentOfHealthAndAgedCare2023HumanPapillomavirus, WorldHealthOrganization2022HumanPapillomavirus}. The first HPV vaccine was licensed in 2006, and to date, six prophylactic vaccines have been approved. All of these vaccines contain virus-like particles (VLPs) targeting the two most prevalent HPV types, 16 and 18 \cite{WorldHealthOrganization2022HumanPapillomavirus, Lowy2008HumanPapillomavirus}. Compared to natural infection, HPV vaccination is more immunogenic and able to trigger a higher polyclonal antibody response, and the response avidity does not significantly increase after boosting \cite{WorldHealthOrganization2022HumanPapillomavirus}. All prophylactic HPV vaccines have performed high efficacy in the HPV-naive population \cite{AustralianDepartmentOfHealthAndAgedCare2023HumanPapillomavirus, WorldHealthOrganization2022HumanPapillomavirus}. HPV vaccination primarily targets girls aged 9-14 years, ideally before their sexual debut, whereas following the onset of sexual activity, adolescents and young females are at the highest risk of acquiring HPV infection, with the risk decreasing as age increases \cite{Castellsague2009HPVVaccination, Lowy2008HumanPapillomavirus}. The secondary target population includes females aged 15 and older, boys and men who have sex with men (MSM), provided that high vaccination coverage is achieved in the primary target group. According to statistics published by the World Health Organisation (WHO), till 2022, 125 countries have added HPV vaccination to their national immunisation program for girls, and 47 of them also extend the vaccine to boys \cite{WorldHealthOrganization2022HumanPapillomavirus}. Among developed countries, Australia has implemented one of the most comprehensive HPV immunisation programs \cite{Drolet2017ImpactHuman}. The Australian National Immunisation Program (NIP) initially included adolescent girls aged 12-13 years and offered catch-up vaccination for females up to 26 years of age starting in 2007. In 2013, the program was expanded to include boys as part of a gender-neutral vaccination strategy  \cite{AustralianInstituteOfHealthAndWelfare2018HumanPapillomavirus, Drolet2017ImpactHuman}. The current immunisation program also extends to individuals with severely immunocompromising conditions and MSM \cite{AustralianDepartmentOfHealthAndAgedCare2023HumanPapillomavirus}. 

Modelling approaches have been utilised to simulate HPV transmission and evaluate the effectiveness of prevention strategies. Most of these models are developed within the classical infrastructures, in which populations are assumed to be homogeneously mixed, and there is no demonstration of relationships among individuals. However, in the context of STIs including HPV, infectivity is not homogeneous, and sexual contact provides the prerequisite route for infection dissemination \cite{Munoz-Quiles2024QuantifyingDelay}. Cost-effectiveness studies have been conducted to evaluate the HPV vaccination program. In all settings, adolescent girls, prior to exposure to HPV, are prioritised to be vaccinated to achieve optimal efficacy, as guided by WHO \cite{WorldHealthOrganization2022HumanPapillomavirus}. Once the high vaccination coverage in girls is ensured, vaccination of young women and boys may also be considered \cite{Kim2008HealthEconomic, Kim2009CostEffectiveness}. Though vaccinating boys is exemplified to require considerable cost and has trivial incremental value to  herd immunity, it provides partial protection for MSM, a key population that exists as a reservoir for HPV infections and cannot benefit from the herd effect generated by vaccinating girls only \cite{Munoz-Quiles2021EliminationInfections, Villanueva2022MathematicalModel}. 

Although there are various types of sexual contacts, this study focuses specifically on heterosexual contacts through vaginal intercourse. The majority of cervical cancer cases occur when HPV is transmitted from a male to a female during vaginal intercourse \cite{Herbert2008ReducingPatient}. While women are predominantly affected by cervical cancer and other complications resulting from HPV infection, research has shown that the risk of transmission through women-to-women contact is negligible \cite{Burchell2006Chapter6}. Therefore, while recognising the diversity of sexual orientations and behaviours in society, we assume that heterosexual interactions through vaginal intercourse are the primary means of HPV transmission and its associated complications \cite{Herbert2008ReducingPatient, Baldwin2003HumanPapillomavirus}. 

Hence, we employ SeCoNet \cite{Wang2024SeCoNetHeterosexual}, a heterosexual contact network growth model, to simulate real-world relationship building processes. The SeCoNet sexual contact network growth model produces networks exhibiting scale-free properties, which are ubiquitous in sexual contact networks \cite{DeBlasio2007PreferentialAttachment, Bell2017NetworkGrowth}. The resultant contact networks have been validated for their utility in analysing HPV transmission. Originally developed to reflect Australian demographics, the model can be calibrated to fit different settings. 

Whereas the SeCoNet growth model was originally introduced to simulate the formation and dynamics of heterosexual contact network, the present study substantially extends the framework to examine how network topology interacts with targeted vaccination interventions. In this paper, we present: (i) a comparative assessment of seven vaccination strategies based on network topological properties, including age-based, vaccination ring and five centrality-based schemes (degree, betweenness, closeness, percolation and eigenvector centrality); (ii) a gender-disaggregated evaluation revealing sex-specific disease dynamics to capture distinct cohort behaviours; and (iii) a comprehensive sensitivity analysis evaluating strategy performance across systematic variations in topological metrics (average degree, scale-free exponent, shortest path length, and clustering coefficient) as well as epidemiological conditions.

\section{Methodology}
\subsection{The SeCoNet Sexual Contact Network Growth Model}

\begin{table}[htbp]
\caption{Age Distribution Used in SeCoNet Growth Model}
\begin{center}
\begin{tabular}{c|c|c|c|c|c}
\hline
\hline
 Age Group & \% & Age Group & \% & Age Group & \% \\
\hline
15 to 19 & 22.1 & 20 to 24 & 55.5 & 25 to 29 & 14.1 \\
30 to 34 & 4.4 & 35 to 39 & 1.8 & 40 to 44 & 1.8 \\
45 to 49 & 0.1 & 50 to 54 & 0.1 & 55 to 59 &  0.1 \\
\hline
\hline
\end{tabular}\label{age}
\end{center}
\end{table}

The SeCoNet sexual contact network growth model \cite{Wang2024SeCoNetHeterosexual} was originally proposed to create scale-free networks with constraints specific to sexual contact networks. As introduced in the previous section, this study focuses solely on heterosexual contacts, hence the generated networks are bipartite. In reality, each individual has a unique set of characteristics, such as age, appearance, education, income, social status, which can make some people more sexually attractive. Meanwhile, as people gain more dating experience, their flirting skills get polished with practice, which potentially increases their success in forming sexual connections, reflecting a ``rich-gets-richer'' dynamic \cite{Bell2017NetworkGrowth, Perc2008StochasticResonance, Jusup2022SocialPhysics}. This results in a heterogeneous distribution of sexual partners, exhibiting scale-free properties \cite{DeBlasio2007PreferentialAttachment}. The SeCoNet sexual contact network growth model was initially developed based on Australian demographics; here, we calibrate the model to reflect the student demographics at the University of Sydney at the end of 2021 \cite{TheUniversityOfSydney2022SnapshotStudent}.

Our investigation was intentionally designed as an in-depth case study of HPV dynamics within a dense, highly mobile young-adult community rather than a national projection for Australia as a whole. We specifically take the approach of calibrating the model to the student demographics at a Group of Eight university in Australia because a university campus presents a uniquely challenging scenario for epidemic magnitude control \cite{Mossong2008SocialContacts, Weeden2020SmallWorldNetwork}. Unlike the broader Australian population, the university environment features a highly concentrated age structure (predominantly young adults) associated with denser interaction networks. Crucially, epidemiological data demonstrate that sexually transmitted HPV incidence and point prevalence peak sharply among young adults aged 20–25 years \cite{DeSanjose2018NaturalHistory}, aligning directly with the core tertiary student demographic. Furthermore, instead of continuous population turnover throughout the year, a university setting experiences a regular, pulsed influx of new, susceptible individuals at the start of academic terms. This regular influx of susceptibles within a dense, age-homogeneous network creates an intensive transmission environment, providing an ideal stress test for evaluating the limits and efficacy of targeted, network-based vaccination strategies.

\subsubsection{Initialisation}
The generated contact networks consist of $N$ individuals and the relationships among them. In order to capture the individual heterogeneity in sexual contact dynamics, each node represents an individual, whereas links denote relationships in the network. Each node $i$ is assigned several attributes at initialisation: (1) age $g_i$; (2) gender $b_i$ (where 1 denotes females and -1 denotes males); (3) the estimated average relationship duration $\delta_i$. The age of each individual is assumed to range from 15 to 59 years. The age and gender distributions reflect the cohort structure of students at the University of Sydney in 2021, with a sex ratio of approximately 69.5 males to 100 females \cite{TheUniversityOfSydney2022SnapshotStudent}. The age structure is summarised in Table \ref{age}, assuming that the age groups 35-44 and 45+ years follow a homogeneous distribution. Given that the simulated population consists of tertiary students, most are under 26, aligning with the age criteria for HPV vaccination in Australia \cite{AustralianInstituteOfHealthAndWelfare2018HumanPapillomavirus}.  

Similar to the Barabási-Albert model \cite{Barabasi1999EmergenceScaling, Nepomuceno2020ComputationalChaos} and the Bianconi-Barabási model \cite{Bianconi2001CompetitionMultiscaling}, at the initial time $t_0 = 0$, there are $m_0$ bipartite links connecting $m_0$ pairs of nodes in the network, each representing a monogamous heterosexual relationship. A node $i$ is selected randomly, and the choice of a sexual partner (neighbour) is made preferentially, based on the fitness of the nodes. The fitness of node $j$ is defined as:
\begin{equation}
    \phi_j = \frac{|b_i - b_j|}{max\{\langle \eta \rangle, |g_i - g_j|\} * max\{1, |{lsp}_i - {lsp}_j|\} }
\end{equation}
where ${lsp}_i = T / \delta_i$ denotes the estimated Lifetime Sexual Partners (LSP) for node $i$. Their pairing depends on the age, gender, and LSP of nodes' $i$ and $j$; thus the fitness for node $j$ of the same gender as node $i$ is zero. The probability of node $i$ choosing node $j$ is defined as:
\begin{equation}
    p_j = \frac{\phi_j}{\sum_{h\in N}\phi_h}
\end{equation}

At initialisation, individuals under the age of 18 are assumed to be virgins, with no sexual partners. To parameterise age preferences in partnership formation, the baseline mean preferred age difference between partners was initially set to $\langle\eta\rangle = 3.5\text{ years}$ based on an empirical study \cite{Conroy-Beam2019WhyAge}. However, considering that tertiary student populations exhibit higher age homogeneity and tighter age assortativity than general population samples, we conducted a dedicated sensitivity analysis setting the preferred age gap to a tighter value of $\langle\eta\rangle = 1.5\text{ years}$ in Appendix \ref{aa} to compare general population mixing assumptions against highly age-clustered campus dynamics.

During the network growth, there are three mechanisms implemented in the model:
\begin{itemize}
    \item New node introduction;
    \item Link removals;
    \item Secondary link formation
\end{itemize}

\subsubsection{Mechanism I - New Node Introduction}
At each time step, $n$ nodes are introduced to the network, and each of these new nodes is connected to $m$ preexisting nodes, continuing until there are no new nodes available. The selection of $m$ nodes to which a new node $i$ will connect is made preferentially, depending on both the degree and fitness of the existing nodes. Here, the probability of node $i$ choosing preexisting node $j$ is defined as:
\begin{equation} \label{probability}
    q_j = \frac{(k_j + \epsilon)\phi_j}{\sum_{h\in N}(k_h + \epsilon)\phi_h}
\end{equation}
where $k$ represents the current degree of node $j$, and the presence of $\epsilon$ ensures that nodes with no existing links (for example, if all previous relationships have terminated) can still be preferentially selected according to their fitness. Therefore, $\epsilon$ must be a positive real number, and its value can be adjusted to calibrate the growth model. Links formed through this mechanism are referred to as primary links to distinguish them from links created by mechanism III hereafter.

\subsubsection{Mechanism II - Link Removals}
Classical growth models primarily focus on establishing connections, but in this context, not all relationships are assumed to last indefinitely. Therefore, in addition to relationship formation, a new mechanism is proposed for relationship discontinuation.

When each link is created, it is assigned an expected relationship duration $\Delta_{ij}$, which is based on the average expected relationship duration of the two partners (nodes). Specifically, the expected duration of the relationship (link) is determined by the minimum of the expected durations assigned to both individuals, such that:
\begin{equation} \label{relationship duration}
    \Delta_{ij} \sim Exp(min\{{\delta}_i, {\delta}_j\})
\end{equation}
Once the age of the link reaches its expected duration, the link will be removed, signifying the termination of the relationship. Consequently, at each time step $t$, the rate at which links are removed, $\theta$, is given by $1 / \langle \Delta \rangle$, where $\langle \Delta \rangle$ represents the average expected duration of relationships.

\subsubsection{Mechanism III - Secondary Link Formation}
In the context of sexual contact network formation, two individuals who are already part of the contact network may eventually form a relationship with each other, and this is not restricted to the time when either individual first joins the network. To account for this, the third mechanism is introduced to generate secondary links between existing nodes. It is postulated that the creation of these secondary links occurs at a rate that matches the deletion of primary links, meaning that once the first mechanism (primary link formation) is completed, the total number of links in the network becomes stable. 

This mechanism also uses the fitness-based preferential attachment in Mechanism I to create secondary links among nodes that are already in the network. The probability of node $i$ choosing node $j$ also follows \eqref{probability}.

The growth model can be divided into two distinct phases. In Phase 1, all three mechanisms operate simultaneously, meaning individuals continue to join the contact network, and relationships are formed according to the model's rules. During this period, at each time step $t$, the number of secondary links $m_i$ is set as:
\begin{equation} \label{mi1}
    m_i = (nmt + m_0)\theta
\end{equation}
While in Phase 2, only the last two mechanisms are active, whereby it is assumed that all individuals in the simulated population have already joined the network, and new relationships are formed only among people (nodes) which have had at least one relationship previously. Here, the number of secondary links $m_i$ is set as:
\begin{equation} \label{mi2}
    m_i = M\theta
\end{equation}
The key distinction of this model is its ability to go beyond simply generating scale-free networks, as many existing growth models do. Instead, this model specifically simulates the process of relationship formation and maintenance within a heterosexual contact network, closely reflecting real-world dynamics. This aspect of the model is significant, as it is tailored to the unique characteristics of sexual contact networks.

\begin{figure}[htbp]
\centering
\includegraphics[width=\linewidth]{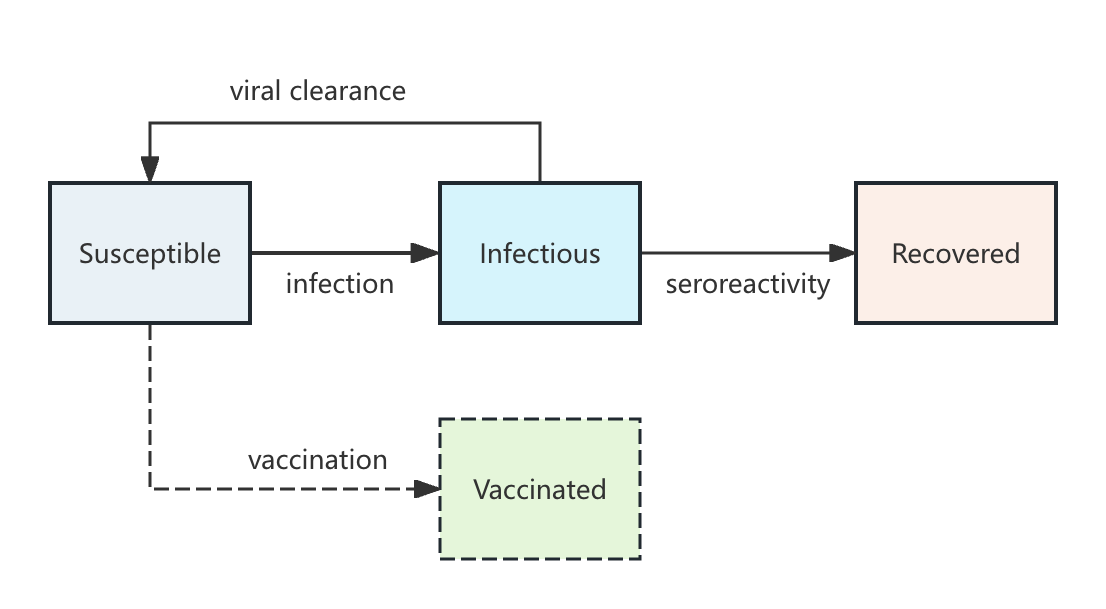}
\caption{Compartments of HPV Transmission Dynamics - The model was initially implemented using an SIRS epidemic framework; subsequently, various vaccination strategies were evaluated to investigate the impact of network topology on containing HPV transmission.}\label{Figure_1}
\end{figure}

\subsection{HPV Transmission Dynamics}
\label{ii-b}
The HPV disease transmission dynamics were initially modelled using an SIRS (Susceptible-Infected-Recovered-Susceptible) epidemic framework without vaccination. The population is divided into three compartments: (1) Susceptible, (2) Infected, and (3) Recovered, in terms of health status \cite{Helbing2015SavingHuman, Wang2016StatisticalPhysics, Wang2017VaccinationEpidemics, Gosak2018NetworkScience}. Individuals and their sexual contacts are directly represented in the network as nodes and links, with the underlying topology reflecting the heterogeneity in sexual contacts and the transmission dynamics of the infection. The structure of the compartmental model can be seen in Fig. \ref{Figure_1}.

The calibrated model is designed to simulate high-risk HPV infection within the target population. The infection is seeded at initialisation, following the gender-specific prevalence in Australia before the inclusion of HPV vaccination in the NIP \cite{Tabrizi2014HPVGenotype, Bruni2023GlobalRegional}. During the simulation, it is assumed that the transmission rate per sexual act $\beta$ is the same for both female-to-male and male-to-female transmission, whereas the recovery period $\alpha_i$ varies among individuals according to an exponential distribution with a mean of 11 months \cite{Olsen2010HumanPapillomavirus}.

To clarify the mathematical distinction between genders in the model, gender ($b_i\in \{1, -1\}$) governs two key structural and epidemiological aspects of the parametrisation beyond serving as a bipartite node attribute:
\begin{enumerate}
    \item Demographic Ratio and Network Degree Scaling: The population is initialised with a female-to-male ratio of 59:41, reflecting student cohort diversity at the University of Sydney \cite{TheUniversityOfSydney2022SnapshotStudent}. In a bipartite heterosexual network, the total partnerships across genders must balance, in other words, the total degree of female nodes must equal the total degree of male nodes ($\sum_{i\in female} k_i = \sum_{j\in male} k_j$). Hence, the demographic asymmetry mathematically forces males to maintain a higher mean degree:
    \begin{equation}
        \langle k_{male}\rangle = \frac{N_{female}}{N_{male}}\langle k_{female}\rangle = \frac{59}{41}\langle k_{female} \rangle \approx 1.44\langle k_{female}\rangle
    \end{equation}
    This elevates individual male degree centrality and exposure rates across the network, as illustrated in Fig. \ref{11}.

    \item Gender-Specific Natural Immunity ($\rho$): Natural infection typically does not lead to immunity against future infections, and the immunological response is generally stronger in females than in males. Following viral clearance, the probability of developing immunity $\rho$ is implemented to be gender-specific, set to $p_{female} = 42.7\%$ for females and $p_{male} = 18.8\%$ for males based on empirical seroreactivity data \cite{AustralianDepartmentOfHealthAndAgedCare2023HumanPapillomavirus}. Consequently, males re-enter the susceptible state after viral clearance at a higher rate, sustaining higher reinfection dynamics.  
\end{enumerate}

More epidemic-related variables and parameters can be seen in Table \ref{variables}. The simulations run for $T=1000$ days, roughly corresponding to the duration of time that tertiary students spend at the university.

To evaluate the sensitivity and robustness of the model against key parameter assumptions, extensive scenario explorations were conducted beyond the baseline parametrisation. Specifically, we explored tighter age assortativity ($\langle\eta\rangle = 1.5 \text{ years}$), alternative infection transmissibility ($\beta = 0.3$), extended average relationship durations ($\langle\delta\rangle = 300, 500,\text{ and } 700 \text{ days}$), higher vaccine coverage levels ($c = 20\%, 40\%, \text{ and }80\%$), and scaled population sizes ($N = 5000$). The complete methodology and corresponding simulation results for these sensitivity analyses are presented in Appendices \ref{aa} through \ref{ae}.

\begin{table*}[htbp]
\caption{List of Variables and Parameters for the Calibrated SeCoNet Growth Model}
\begin{center}
\begin{tabular}{|m{6cm} m{1cm} m{6cm} m{3cm}|}
\hline
Variable / Parameter & Symbol & Value & Source\\
\hline
\textbf{Variable} & $ $ & &   \\
Timestep & $t$ &  &   \\
Number of contacts & $M$ &  & \\
Age  & $g_i$ & 15-59 &  calibrated from \cite{TheUniversityOfSydney2022SnapshotStudent} \\
Gender & $b_i$ & 1: female, -1: male & \\
Average Relationship duration for the node (in days) & ${\delta}_i$ & Gamma distribution with a mean of  $\langle\delta\rangle = 100$ & \\
Lifetime Sexual Partners & ${lsp}_i$ & $T / {\delta}_i$ (rounded) & \\
Node Degree & $k_i$ & & \\

Expected Relationship duration & $\Delta_{i,j}$ &  & \eqref{relationship duration}\\
High risk HPV transmissibility per coital act & $\beta$ & 0.13 & calibrated from \cite{Olsen2010HumanPapillomavirus} \\
High-risk HPV clearance period (in days) & $\alpha_i$ & Exponential distribution with a mean of 11 months & calibrated from \cite{Olsen2010HumanPapillomavirus} \\
HPV immunity acquiring probability & $\rho$ & female: 42.7\%, male: 18.8\% & calibrated from \cite{AustralianDepartmentOfHealthAndAgedCare2023HumanPapillomavirus}\\

& $ $ & &   \\

\textbf{Parameter} & $ $ & &   \\
Total timesteps & $T$ & 1000 & \\
Population  size & $N$ & 3000 & \\

Initial number of links  & $m_0$ & 10 & \\
Number of joining node per time step & $n$ & Phase 1: 100 \newline Phase 2: 0 & \\
Links added per joining node & $m$ &  & \cite{Bell2017NetworkGrowth, Law2020PlacementMatters}\\
Removed links per node time step & $m_r$ & & \\
Link removal rate per node per time step & $\theta$ & $1 / \langle\Delta\rangle$ & \\
Secondary links added per time step & $m_i$ &  & Phase 1: \eqref{mi1} \newline Phase 2: \eqref{mi2}\\

Average age difference & $\langle \eta \rangle$ & 3.5 & Conroy-Beam and Buss \cite{Conroy-Beam2019WhyAge} \\
Frequency of intercourse & $f$ & $1/2$ during the first two weeks \newline $1/7$ after the first two weeks & Althaus et al. \cite{Althaus2012TransmissionChlamydia} \\

Scale-free calibration parameter & $\epsilon$ & 0.5 & This can be calibrated.  \\

Average degree & $\langle k \rangle$ & & \cite{Barabasi1999EmergenceScaling, Law2020PlacementMatters}\\
Power law exponent & $\gamma$ & & \cite{Bell2017NetworkGrowth, Law2020PlacementMatters}\\
Average shortest path length & $L$ & & Piraveenan et al.\cite{Piraveenan2013PercolationCentrality}\\
Clustering coefficient & $C$ & &Law et al.\cite{Law2020PlacementMatters}\\
Vaccination coverage & $c$ & 10\% & \\

\hline
\end{tabular}\label{variables}
\end{center}
\end{table*}

\subsection{Vaccination Strategies}
The introduction of HPV vaccination has achieved significant results in reducing the risk of cervical cancer in Australia \cite{Drolet2017ImpactHuman}. However, since the immunisation program is school based, vaccination coverage is considerably lower among women who have left high school, compared to those under 18, as indicated by statistics in 2007, prior to the inclusion of males in the NIP \cite{Barbaro2015MeasuringHPV}. The recommended vaccination schedule for both females and males under 26 years old without immunocompromising conditions has recently been revised in Australia from three doses to one dose \cite{AustralianInstituteOfHealthAndWelfare2018HumanPapillomavirus}. Clinical data suggest a single dose can provide similar protection as the three-dose regimen \cite{WorldHealthOrganization2022HumanPapillomavirus}.

In the simulated scenarios, a single-dose, gender-neutral vaccination with full efficacy is administered. The adoption of a single-dose schedule reflects Australia's updated National Immunisation Program (NIP) guidelines and recent empirical evaluations demonstrating that a single dose offers comparable immunogenicity and protective efficacy to multi-dose regimens in individuals under 26 \cite{AustralianInstituteOfHealthAndWelfare2018HumanPapillomavirus, Steffens2026ImmunisationProgram}.

Four vaccination sessions are scheduled at time steps $t = 6, 13, 20, 27$ (corresponding to weekly roll-out sessions during an academic month). This phased delivery model mimics real-world university health centre vaccination campaigns and catch-up programs, where vaccines are distributed over multiple rolling sessions to accommodate student availability and logistical constraints \cite{Kessler2023StrategiesImprove}. Under the baseline scenario, a constrained vaccine allocation equal to 10\% of the population under 26 is evenly distributed across the four sessions to evaluate strategy performance under resource-limited catch-up conditions. To assess the impact of higher coverage and potential herd immunity effects, alternative coverage levels (20\%, 40\%, and 80\%) were systematically evaluated in Appendix \ref{ad}.

To systematically evaluate targeted interventions, our simulation framework couples graph-theoretic targeting algorithms with epidemiological dynamics. Targeted vaccination strategies rely on identifying key structural nodes whose removal maximally disrupts viral dissemination across the contact network. The efficacy of these network-based schemes is fundamentally constrained by both epidemiological parameters, such as transmission probability per coital act $\beta$, recovery rates $\alpha_i$, and vaccine coverage level $c$, and underlying topological parameters, including link density $\langle k \rangle$, power law exponent $\gamma$, clustering coefficient $C$, and average path length $L$. Prior studies on complex network immunisation confirm that network topology dictates herd immunity thresholds and determines whether localised targeting outperforms uniform random vaccination under resource constraints \cite{Piraveenan2013EffectVaccination, Thedchanamoorthy2014InfluenceVaccination}. By evaluating these targeting heuristics across varying topological parameters and coverage allocations, our simulation framework isolates the structural mechanisms that maximise epidemic flattening and peak delay.

The primary vaccination heuristics evaluated in this study include:

\subsubsection{Age-based Vaccination Strategy}

All vaccination strategies follow the current vaccination eligibility criteria in Australia \cite{AustralianInstituteOfHealthAndWelfare2018HumanPapillomavirus, AustralianDepartmentOfHealthAndAgedCare2023HumanPapillomavirus}, thus susceptible individuals (nodes) under 26 years of age, who have joined the network, are eligible for vaccination. During each vaccination session, eligible individuals will be randomly selected to receive the vaccine.

\subsubsection{Ring Vaccination Strategy}

At each vaccination session, infected nodes are ranked by their degree (the number of connections they have). Susceptible neighbours of infected nodes with high degrees are prioritised for vaccination.

\subsubsection{Degree Centrality-Based Vaccination Strategy}

In each vaccination session, nodes are ordered in terms of their degree, also known as degree centrality, and susceptible individuals with higher degrees are given priority for vaccination.

\subsubsection{Betweenness Centrality-Based Vaccination Strategy} \label{betweenness}

Betweenness centrality \cite{Piraveenan2013PercolationCentrality} quantifies the proportion of shortest paths that pass through a given node, averaged across all pairs of nodes in a network. This measure identifies nodes that function as bridges or intermediaries, connecting different components of the network. Nodes with high betweenness centrality tend to have a significant impact on information flow within the network:
\begin{equation}\label{bc}
    {BC}_i = \frac{1}{(N-1)(N-2)}\sum_{s \neq i \neq r}\frac{\sigma_{s,r}(i)}{\sigma_{s,r}}
\end{equation}
where $\sigma_{s,r}$ is the number of shortest paths between the source node $s$ and the target node $r$, while $\sigma_{s,r}(i)$ is the number of shortest paths between the source node $s$ and the target node $r$ that pass through node $i$.

In this scenario, nodes are ranked by their betweenness centrality during each vaccination session. Susceptible nodes with higher betweenness centrality are more likely to be vaccinated.

\subsubsection{Closeness Centrality-Based Vaccination Strategy}

Closeness centrality \cite{Piraveenan2013PercolationCentrality} measures how close a node is to the rest of the nodes in the network, in terms of the average shortest path length. It essentially assesses the average geodesic distance (the shortest path length) between a given node and every other node in the network, with higher closeness centrality indicating a node is more central and can reach others more quickly:
\begin{equation}
    {CC}_i = \frac{1}{\sum_{i \neq j}L_g(i, j)}
\end{equation}
where $L_g(i, j)$ is the shortest path (geodesic) distance between nodes $i$ and $j$.

Here, nodes are ranked according to their closeness centrality in each vaccination session. Susceptible nodes with higher closeness centrality have a higher likelihood of being vaccinated.

\subsubsection{Percolation Centrality-Based Vaccination Strategy}

Percolation centrality is proposed by Piraveenan et al. \cite{Piraveenan2013PercolationCentrality} to measure the importance of nodes in facilitating percolation (or spread) through the network. The percolation state of node $i$ is denoted as $\chi_i$. Here, for an infected node $i$, $\chi_i$ is set to 1; otherwise, $\chi_i$ is set to 0. Percolation centrality is defined as the proportion of percolation paths that pass through a given node. These paths are the shortest paths between a pair of nodes, with the source node being percolated (infected):
\begin{equation}
    {PC}_i = \frac{1}{N-2}\sum_{s \neq i \neq r} \frac{\sigma_{s,r}(i)}{\sigma_{s,r}}\frac{\chi_s}{[\sum_j \chi_j] - \chi_i}
\end{equation}
where $\sigma_{s,r}$ and $\sigma_{s,r}(i)$ are defined the same as \eqref{bc} in Section \ref{betweenness}. Percolation centrality quantifies how crucial a node is in the spread of infection in the network.

Nodes are ranked by their percolation centrality at each vaccination session. Susceptible nodes with higher percolation centrality are more likely to be vaccinated.

\subsubsection{Eigenvector Centrality-Based Vaccination Strategy}

Eigenvector centrality is a measure of a node's centrality score, which is proportional to the sum of the centrality scores of the neighbours. Given matrix $X$ as the centrality scores of nodes and $A$ as the adjacency matrix of the network, then $x$ can be defined as:
\begin{equation}
    x \propto Ax
\end{equation}

During each vaccination session, nodes are ranked according to their eigenvector centrality. Susceptible individuals with higher eigenvector centrality are prioritised for vaccination. 

\subsection{Computational Implementation}
All simulations were developed in Python and executed on high-performance computing nodes (NCI Gadi supercomputer, 48 CPUs, 192 GB RAM). Simulation wall times scale primarily with network size $N$ and link density $\langle k \rangle$; complete job array runs require approximately 17 hours for $N=3000$ and 34 hours for $N=5000$, with centrality-based heuristics (e.g., betweenness centrality) introducing an additional computational overhead of 1–2 hours. Detailed code architecture, dependencies, and reproduction guidelines, as well as the source code and summarised results for all simulation scenarios and strategy implementations, are publicly available on GitHub \cite{Wang2026SeCoNet}.

\section{Results and Discussion}

In this section, the results of our simulation experiments are presented. We will present them in the following manner. The four specific questions that these simulation experiments aim to answer are as follows:

\begin{enumerate}[label=(\roman*)]
    \item How can the effectiveness of vaccination strategies be measured, and what metrics can be used to illustrate this?
    \item How does the network topology of the contact network influence the effectiveness of various vaccination strategies?
    \item How do various vaccination strategies compare with each other?
    \item How does the effectiveness of vaccination strategies vary between females and males?
\end{enumerate}
   
To answer the first question, we measure the following epidemiological metrics, in the context of varying topologies of contact networks and varying vaccination strategies.

\begin{enumerate}[label=(\alph*)]
    \item Maximum HPV incidence (the maximum number of people infected on a given day, during the course of the simulations)
    \item Peak day of HPV prevalence (the day on which the HPV prevalence was highest - the first day of simulation was considered day one)
    \item Maximum HPV cumulative incidence (the cumulative incidence on the last day of the simulation)
    \item Maximum female HPV incidence (the maximum number of females infected on a given day, during the course of the simulations)
    \item Peak day of female HPV prevalence (the day on which the HPV prevalence was highest among females - the first day of simulation was considered day one)
    \item Maximum female HPV cumulative incidence (the female cumulative incidence on the last day of the simulation)
    \item Maximum male HPV incidence  
    \item Peak day of male HPV prevalence 
    \item Maximum male HPV cumulative incidence
\end{enumerate}

The influence of contact network topology and vaccination strategies on each of these epidemiological metrics is illustrated in a figure in this section. Therefore, Fig. \ref{general incidence}-\ref{male cumulative} correspond, in order, to each of these epidemiological metrics. In each of these figures, the comparative effect of a number of vaccination strategies is illustrated. The vaccination strategies compared are as follows: 

\begin{enumerate}[label =(\arabic*)]
    \item No vaccination (Null model)
    \item Age-based vaccination
    \item Ring vaccination
    \item Degree centrality-based vaccination
    \item Betweenness centrality-based vaccination
    \item Closeness centrality-based vaccination
    \item Percolation centrality-based vaccination
    \item Eigenvector centrality-based vaccination
\end{enumerate}

In the last five cases, the corresponding centrality measure of nodes (people) in the corresponding contact network is used to prioritise vaccination. Nodes with higher centrality are more likely to be vaccinated.

Each figure also considers four topology metrics of the corresponding contact networks:
\begin{enumerate}[label = (\Roman*)]
    \item Average degree $\langle k \rangle$
    \item power law exponent $\gamma$
    \item Average shortest path length $L$
    \item Network clustering coefficient $C$
\end{enumerate}

In each figure, there are four subfigures which illustrate the variation of each of these four topological metrics in the contact network. Therefore, in each subfigure, the y-axis denotes the epidemiological metric under consideration, while the x-axis denotes the topological metric under consideration. To account for stochastic variation in network generation and disease transmission, each reported data point represents the average of 10 independent simulation runs.

\subsection{Simulation Results for the Overall Cohort}

\begin{figure}[htbp]
\centering
\includegraphics[width=\linewidth]{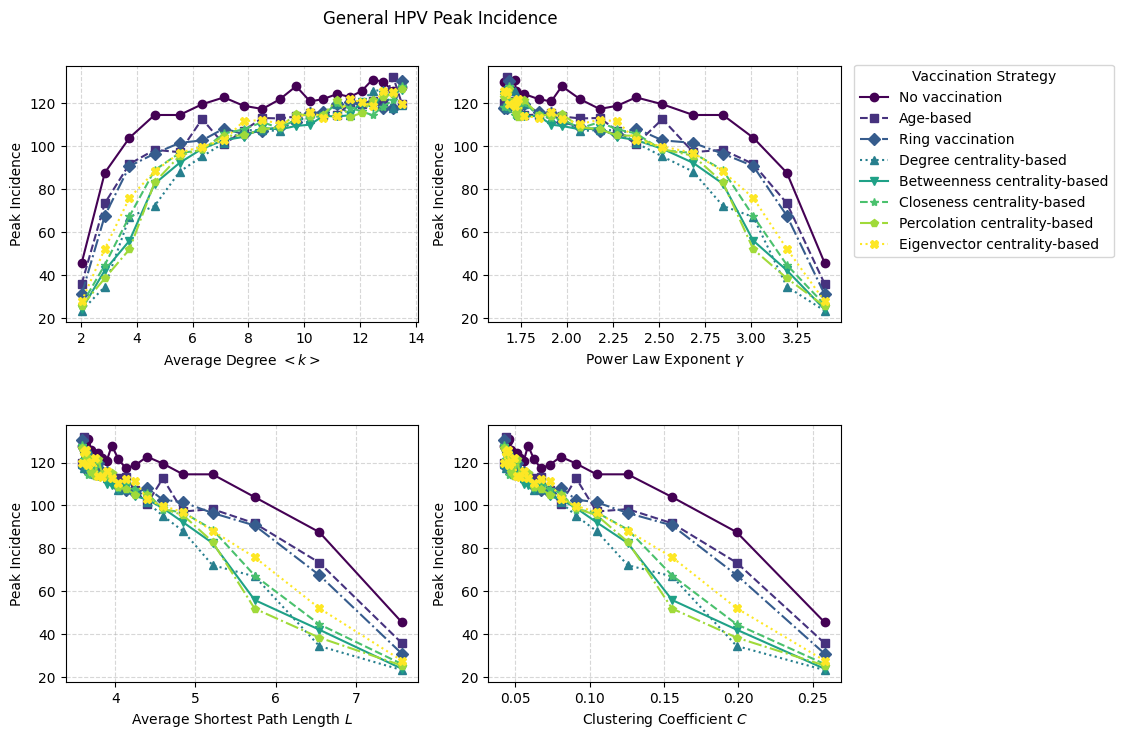}
\caption{Maximum Daily General HPV Incidence across Network Structures and Vaccination Strategies - The degree-, betweenness- and percolation centrality-based strategies are effective in reducing new incidence among individuals. However, as the network becomes more connected, the differences in effectiveness between the vaccination strategies become less notable.}\label{general incidence}
\end{figure}
 
Let us consider Fig. \ref{general incidence}. This figure shows the influence of vaccination strategies and topological metrics of the contact network on the maximum HPV incidence. We may surmise that vaccination strategies which reduce this maximum HPV incidence are comparatively more effective. Understandably, administering no vaccine results in the highest peak incidence always, regardless of the topological properties of the contact network. On the other hand, we may observe that the degree-, betweenness- and percolation centrality-based strategies are most effective, as they reduce peak incidence values. In general, it should be noted that there is not a significant difference between the various vaccination strategies in terms of effectiveness.

In terms of the influence of topological properties, some interesting observations can be made. We can observe, for example, that when the average degree $\langle k \rangle$ increases, peak incidence values increase, that is, vaccination becomes less effective, regardless of the vaccination  strategy used. In other words, increased link density makes any vaccination strategy less effective. In terms of the power law exponent $\gamma$, higher values of this seem to decrease peak incidence, that is, higher values of power law exponents (more heterogeneous degree distributions) seem to help vaccination. In terms of average shortest path length $L$, when this increases, peak incidence becomes less, that is, all vaccination strategies become more effective. In terms of clustering coefficient $C$, also, higher clustering coefficients seem to favour lower peak incidence values, that is, they make all vaccination strategies more effective. It should be noted that these last two metrics are used to quantify the `small-worldness' of a network. A network which has relatively high clustering yet relatively low average path lengths is said to be small-world \cite{Watts1998CollectiveDynamics}.  Here the increases in average shortest path length and clustering coefficient both boost vaccination efficiency; hence, we may surmise that small-worldness itself plays no role in influencing vaccination efficiency. Similarly, it should be noted that scale-freeness itself (that is, scale-free fitness) plays no role in vaccination efficiency, though the scale-free exponent does. Therefore, the extent to which a network is scale-free does not influence vaccination efficiency, though the value of the scale-free coefficient which corresponds to the underlying degree distribution does.

\begin{figure}[htbp]
\centering
\includegraphics[width=\linewidth]{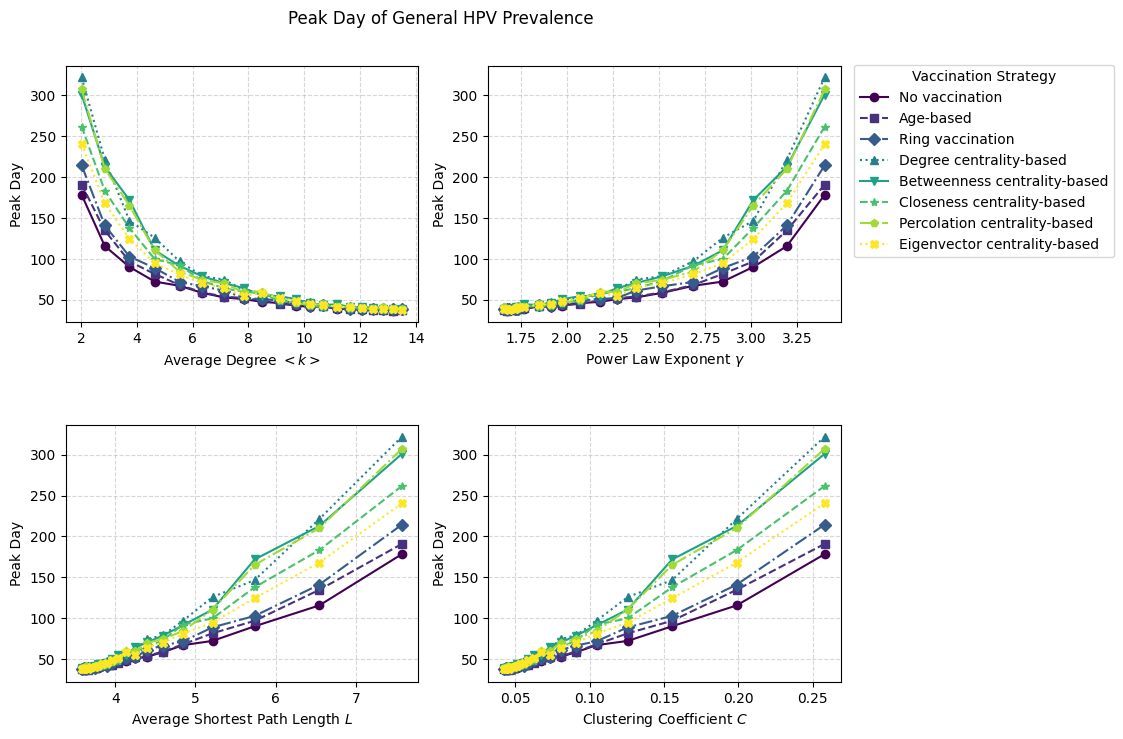}
\caption{Effect of Network Properties and Vaccination Strategies on Peak Day of General HPV Prevalence - The degree-, betweenness- and percolation centrality-based vaccination strategies are the most effective in delaying the peak of the general prevalence. However, as the network gets denser, the differences in effectiveness between the vaccination strategies become less pronounced.}\label{general prevalence}
\end{figure}
 
Now let us consider Fig. \ref{general prevalence}. This figure shows the influence of vaccination strategies and topological metrics of the contact network on the temporal progression of the infection, by looking at the day on which the maximum prevalence occurs. Therefore, we may surmise that vaccination strategies that increase the time to peak HPV prevalence - that is, delay the spread of infection as much as possible - are comparatively more effective. Again, administering no vaccine results in the lowest peak day - that is, the infection understandably peaks quickly in the absence of vaccination. The degree-, betweenness- and percolation centrality-based vaccination seem to delay the peak prevalence the most, regardless of the topological properties of the contact network. In terms of the influence of topological properties, we can observe that when average degree $\langle k \rangle$ increases, peak prevalence occurs faster - that is, vaccination becomes less efficient. When the power law exponent $\gamma$ increases, peak prevalence occurs later - that is, the vaccination becomes more efficient. When the average shortest path length $L$ increases, the peak prevalence occurs later - the vaccination in general becomes more efficient. When the average clustering coefficient $C$ increases, the peak prevalence is delayed - that is, the vaccination becomes more efficient. These results in terms of topological properties qualitatively match the results from Fig. \ref{general incidence}, despite a different epidemiological metric being used to measure vaccination efficiency. Therefore, these results further confirm our observations about the influence of topological properties that we made earlier.

\begin{figure}[htbp]
\centering
\includegraphics[width=\linewidth]{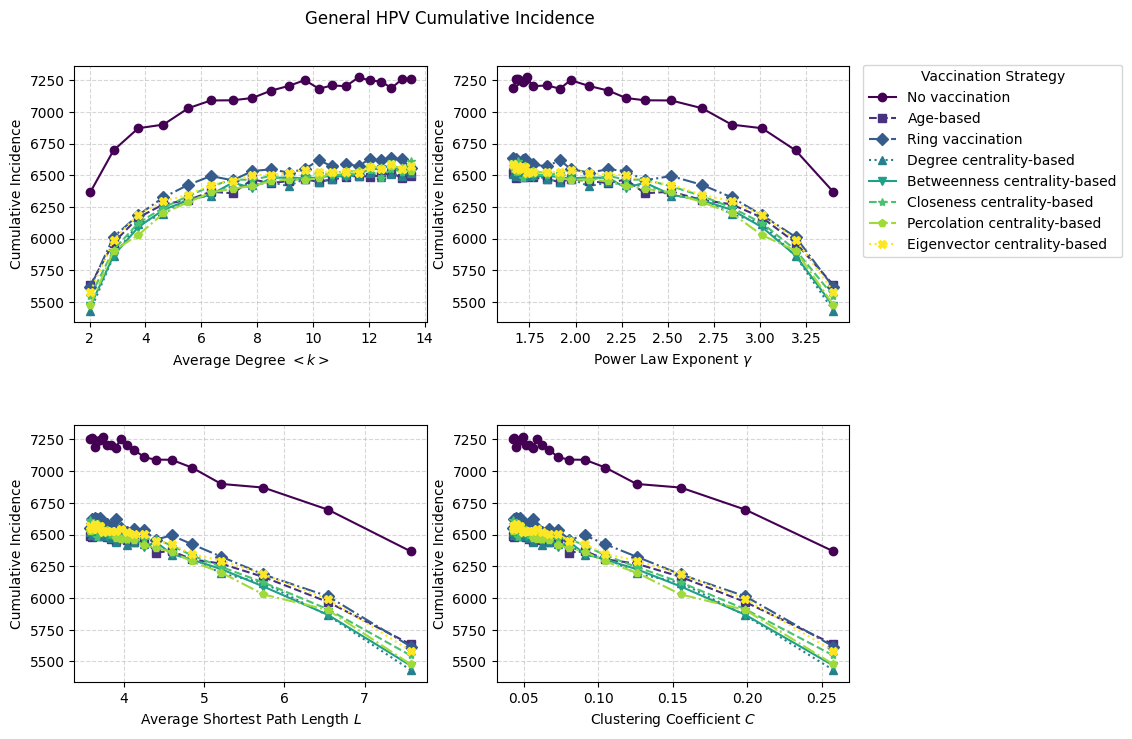}
\caption{Effect of Network Structure and Vaccination Strategy on General HPV Cumulative Incidence - The degree-, betweenness- and percolation centrality-based vaccination strategies work effectively in containing the cumulative incidence in the population.}\label{general cumulative}
\end{figure}
 
Now we consider a different epidemiological metric, the maximum cumulative incidence, that is, the cumulative incidence at the end of the simulation. This indicates the total number of individuals affected by the disease during the simulation period. Here, therefore, a lower value will indicate better vaccination efficiency. The results corresponding to this metric are shown in Fig. \ref{general cumulative}. The absence of vaccination consistently produces the maximum cumulative incidence. Among the strategies tested, the degree-, betweenness- and percolation centrality-based vaccination are most effective in reducing cumulative incidence. In terms of the impact of topological properties, similar to the previous two figures, with increasing average degree $\langle k \rangle$, cumulative incidence increases, which means that vaccination becomes less efficient. With increasing power law exponent $\gamma$, cumulative incidence decreases, which means that vaccination becomes more efficient. With increasing average shortest path length $L$, cumulative incidence reduces, which means that vaccination becomes more efficient. With increasing clustering coefficient $C$, cumulative incidence drops, which means that vaccination becomes more efficient.

\subsection{Simulation Results for the Female Cohort}

\begin{figure}[htbp]
\centering
\includegraphics[width=\linewidth]{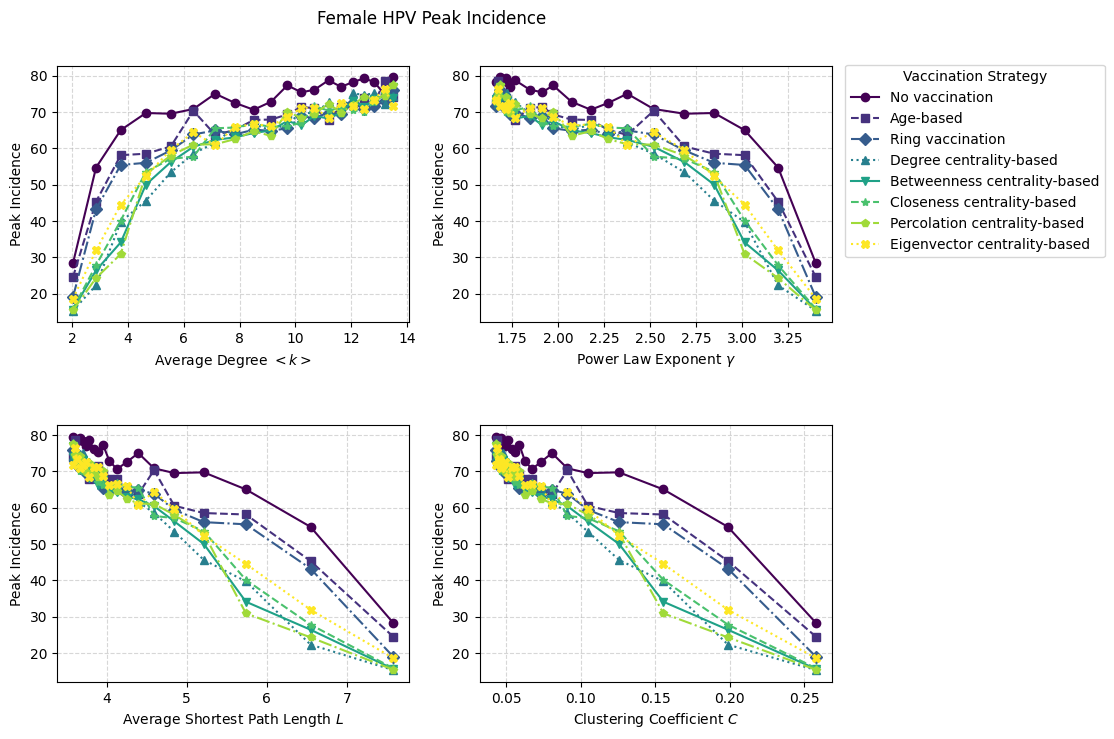}
\caption{Maximum Daily Female HPV Incidence across Network Structures and Vaccination Strategies - The percolation centrality-based vaccination strategy plays a crucial role in reducing new incidence among females overall.}\label{female incidence}
\end{figure}

\begin{figure}[!htbp]
\centering
\includegraphics[width=\linewidth]{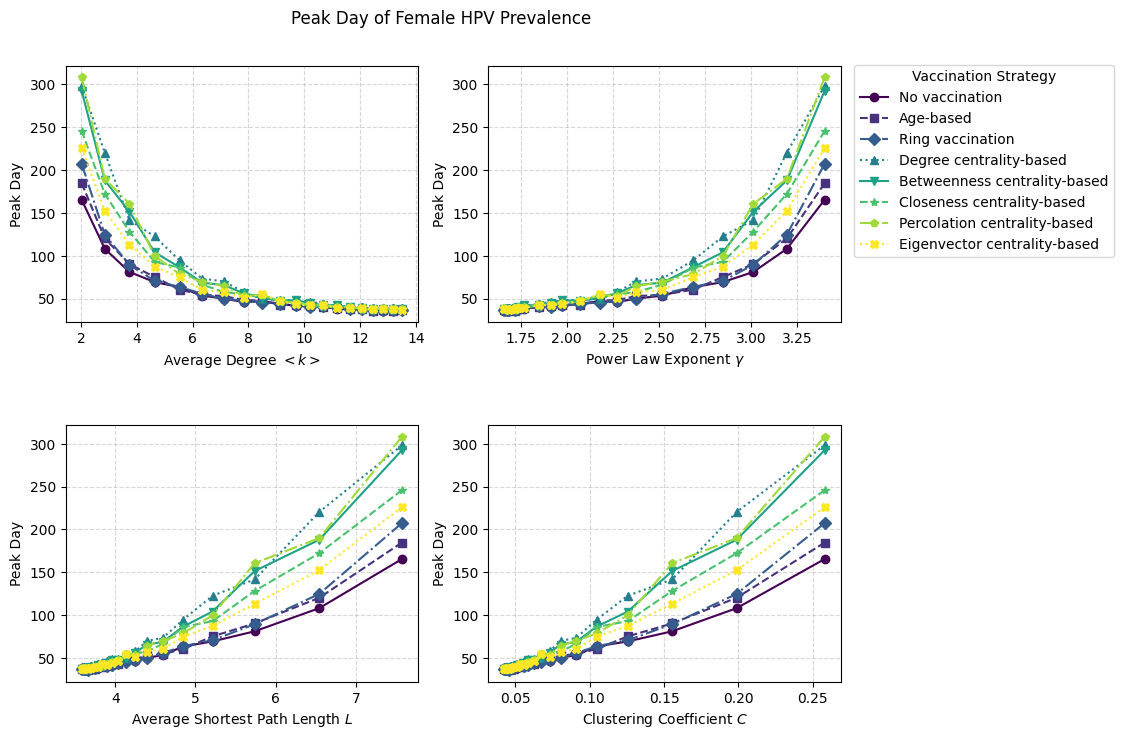}
\caption{Effect of Network Properties and Vaccination Strategies on Peak Day of Female HPV Prevalence - The degree centrality-based vaccination strategy is successful in delaying the peak of female infectious proportion.}\label{female prevalence}
\end{figure}

Fig. \ref{female incidence} is similar to Fig. \ref{general incidence}, in that, it presents the effect of vaccination strategies and network topological metrics on the maximum daily incidence, but focuses on females only. However, we observe that there is no qualitative difference between the overall cohort and the female only cohort in terms of which vaccination strategies are the most useful, with degree-, betweenness- and percolation centrality-based vaccinations being the most useful. Similarly, Fig. \ref{female prevalence} focuses on the female cohort in terms of HPV prevalence and observes that the vaccination strategy which shows the highest effectiveness is the degree centrality-based vaccination, similar to the overall cohort. 

\begin{figure}[htbp]
\centering
\includegraphics[width=\linewidth]{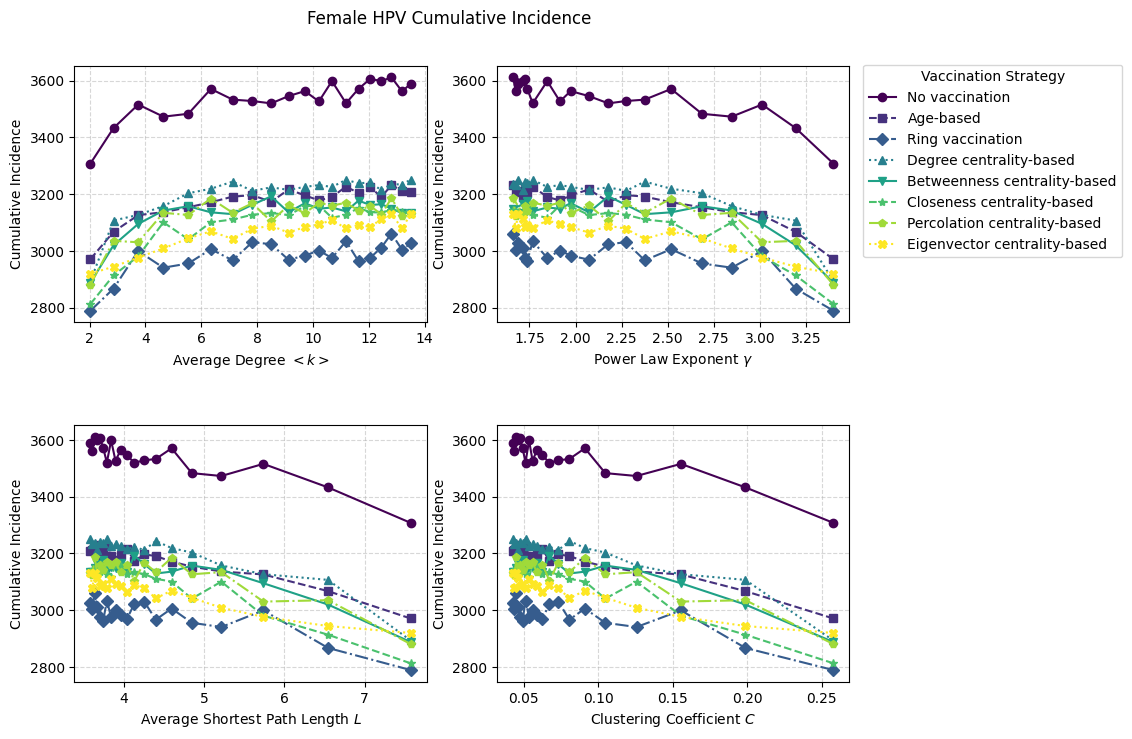}
\caption{Effect of Network Structure and Vaccination Strategy on Female HPV Cumulative Incidence - Ring vaccination strategy works the best in reducing the cumulative incidence within female cohort.}\label{female cumulative}
\end{figure}

Fig. \ref{female cumulative} examines the cumulative incidence among women at the conclusion of the simulation experiment. 
Here we can observe that when no vaccination is administered, the highest level of female HPV cumulative incidence occurs. In contrast to Fig. \ref{general cumulative}, the ring vaccination strategy, rather than the percolation centrality-based strategy, appears to be the most effective in reducing female cumulative incidence. This suggests that removing local transmission chains involving infected individuals may have benefits in reducing infection spread. 
The reason for this could best be understood by contrasting the degree distributions of males and females, which is shown in Fig. \ref{11}. We could observe here that males have a more heterogeneous degree distribution compared to females, indicating that among males, there are a very small number of males who have a high number of relationships, while the remaining males have relatively few relationships. By contrast, most females seem to fall between these two extremes having a relatively lower number of relationships. This explains why centrality-based vaccination strategies, which can efficiently target hubs, are relatively less effective in the female cohort than in the overall cohort, which includes both males and females. The real contrast here is not between the overall cohort and the females, but between males and females. Nonetheless, the influence of network topological structure is consistent with earlier findings that increasing average degree worsens cumulative incidence, while higher power law exponent, longer average shortest path length, and stronger clustering reduce it.

\subsection{Simulation Results for the Male Cohort}

\begin{figure}[htbp]
\centering
\includegraphics[width=\linewidth]{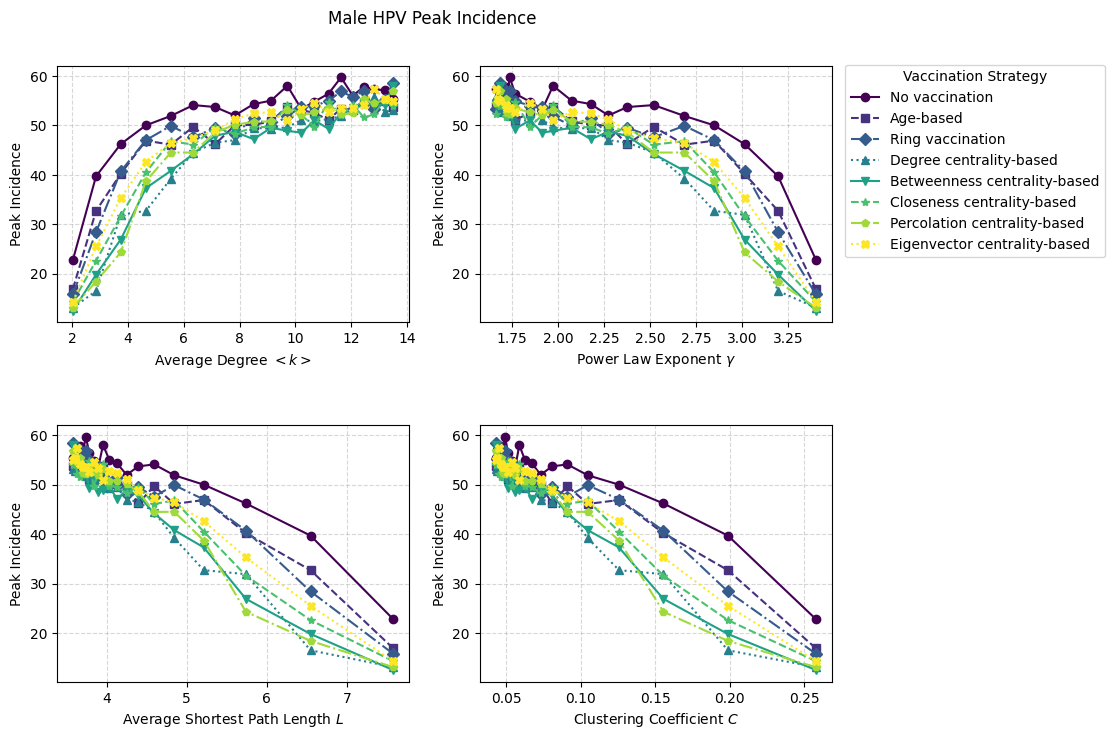}
\caption{Maximum Daily Male HPV Incidence across Network Structures and Vaccination Strategies - The degree-, betweenness- and percolation centrality-based vaccination strategies are effective in controlling the new incidence in the male group.}\label{male incidence}
\end{figure}

\begin{figure}[htbp]
\centering
\includegraphics[width=\linewidth]{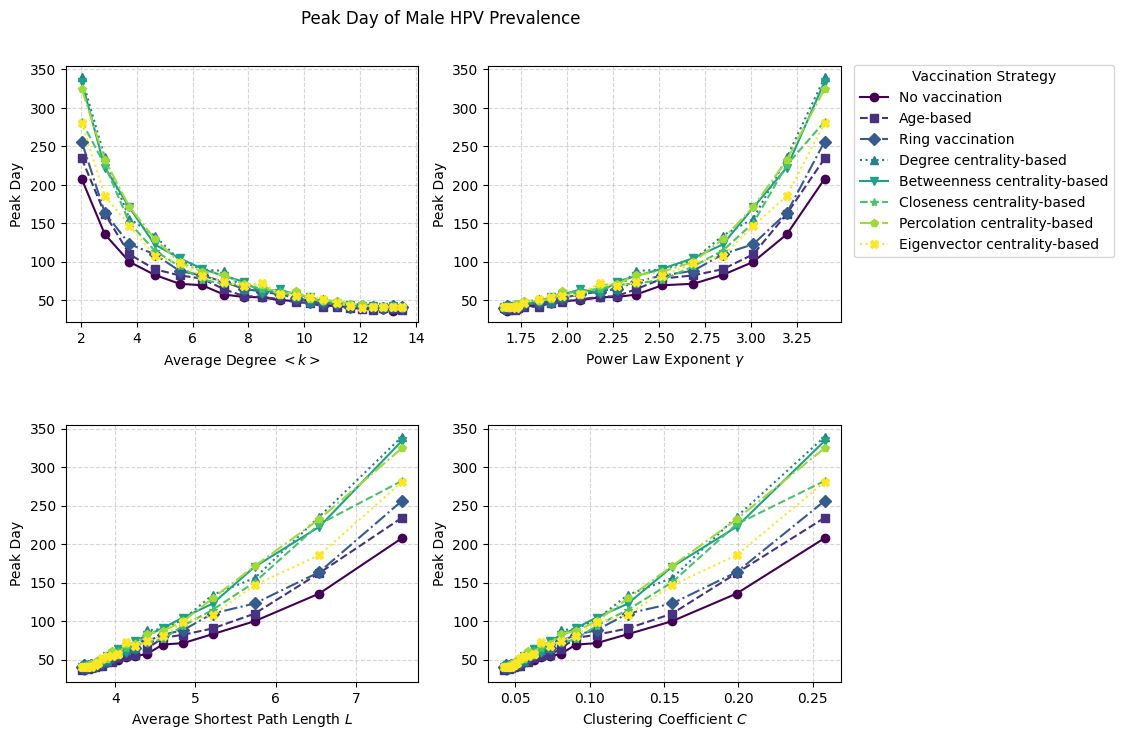}
\caption{Effect of Network Properties and Vaccination Strategies on Peak Day of Male HPV Prevalence - The degree-, betweenness- and percolation centrality-based vaccination strategies are critical in postponing the peak of the male infectious proportion.}\label{male prevalence}
\end{figure}

\begin{figure}[htbp]
\centering
\includegraphics[width=\linewidth]{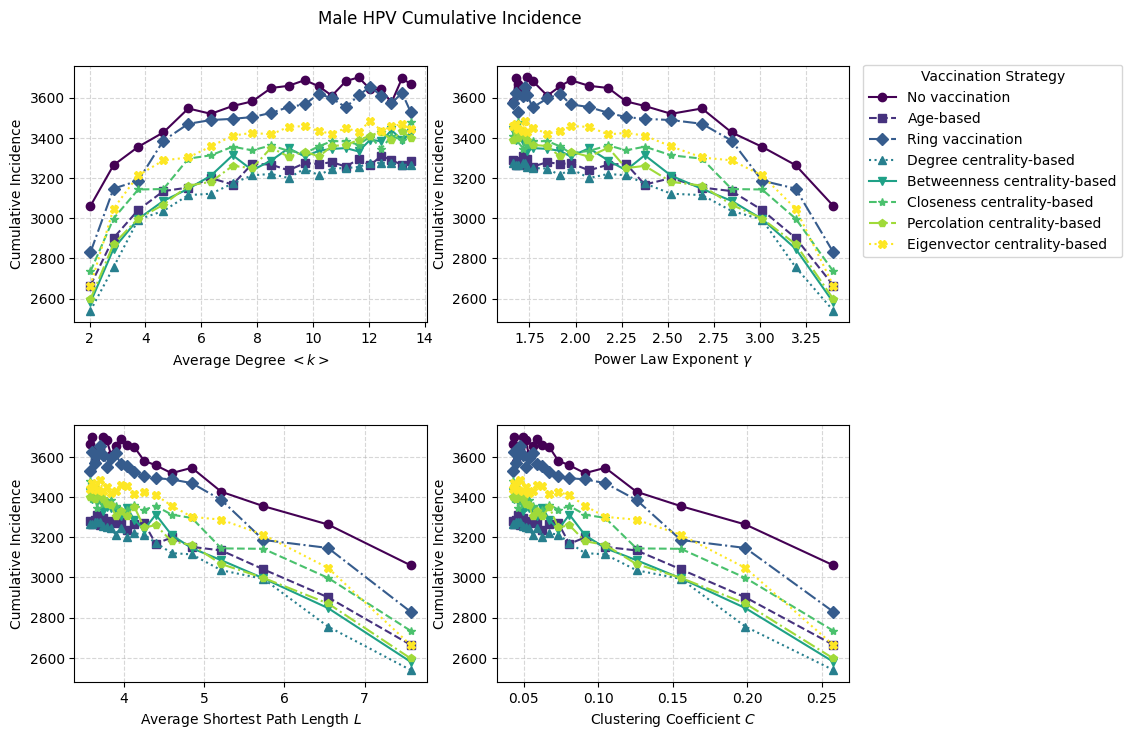}
\caption{Effect of Network Structure and Vaccination Strategy on Male HPV Cumulative Incidence - The degree centrality-based vaccination strategy is the most effective in reducing the cumulative incidence within male cohort.}\label{male cumulative}
\end{figure}

\begin{figure*}[htbp]
\centering
\includegraphics[width=\textwidth]{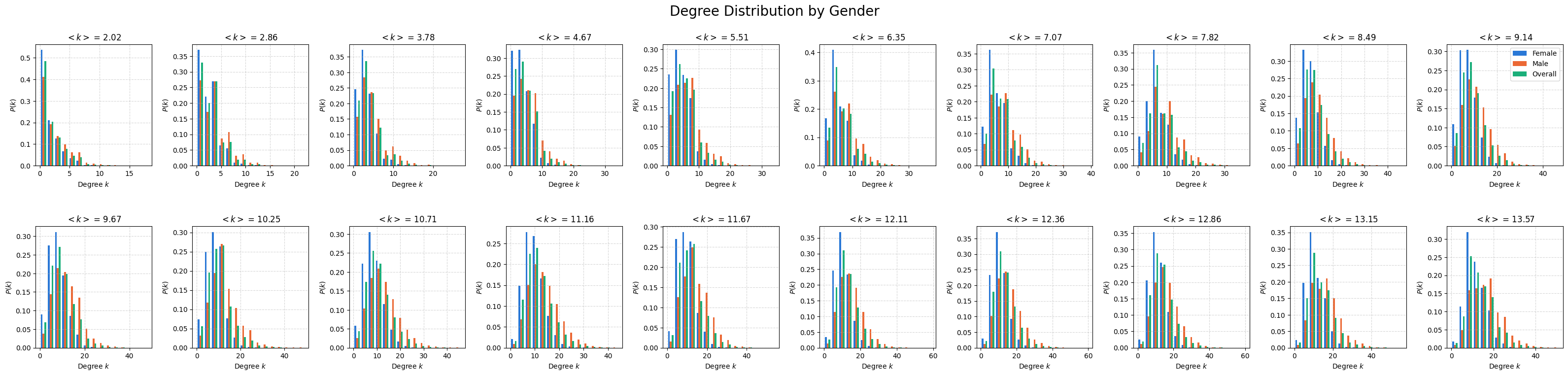}
\caption{The Degree Distribution of the cohort for different simulation runs which corresponds to different average degrees $\langle k \rangle$ - The figure contrasts the degree distribution of the overall cohort $N = 3000$ with that of females and males where the ratio of female-to-male is $59:41$. }\label{11}
\end{figure*}

For completeness, we also show the maximum daily incidence, peak day of prevalence and cumulative incidence for the male cohort only in Fig. \ref{male incidence}-\ref{male cumulative}. We observe no qualitative difference between the overall cohort and the male cohort. Again, this can be explained by the fact that most of the dominant hubs in the sexual contact network seem to be males as shown in Fig. \ref{11}. As such, degree-, betweenness- and percolation centrality-based vaccination strategies which target the dominant hubs work well among males as they do in the overall cohort.

Across all simulations, vaccination substantially reduces HPV incidence, delays the timing of peak prevalence, and lowers cumulative incidence burden, compared to the null model. Among the strategies implemented, degree-, betweenness- and percolation centrality-based strategies emerge as the most effective, although their relative advantages vary by cohort and epidemiological metric. The topology of the sexual contact network has played a decisive role in shaping outcomes. Greater degree heterogeneity, longer average shortest path length, and stronger clustering coefficient can enhance the effectiveness of vaccination strategies, whereas higher average degree consistently undermines their efficiency. While the overall and male cohorts benefit most from degree-, betweenness- and percolation centrality-based strategies, the female cohort exhibits a distinct pattern, with ring vaccination proving the most effective strategy in reducing cumulative incidence. Together, these findings highlight both the universal importance of targeting highly connected individuals and the potential for gender-specific optimisation of vaccination strategies.

An interesting pattern that emerges from the simulation results is that ring vaccination appears to be particularly effective in reducing cumulative incidence among females. We have observed by analysing the degree distributions that there are certain males who act as hubs and have more connections than even the most highly connected females. This results in centrality-based vaccination being more effective among males than it is among females. It is well established in the literature that ring vaccination can be highly effective in slowing infection spread when it can be successfully implemented \cite{Vassallo2020RingVaccination, Scianna2025ComparingEffectiveness}. Therefore, in scenarios where centrality-based vaccination is less effective, ring vaccination becomes more effective, and this is what seems to happen within the female cohort. As such, we may conclude that the relative effectiveness of ring vaccination among females is actually the result of centrality-based vaccination becoming less effective among females due to the degree distribution of the female cohort being less heterogeneous. 

Another interesting observation from the simulation results is that a higher power law exponent appears to aid the vaccination program, by reducing incidence and cumulative incidence, and delaying the timing of the epidemic peak. A higher power law exponent reflects a more heterogeneous network structure, where prominent hubs play a greater role. Intuitively, one might expect that such prominent hubs would accelerate the spread of infection. However, the simulations suggest the opposite: greater heterogeneity seems to strengthen the effectiveness of vaccination programs. This can be explained by the fact that infections are unlikely to originate in the most prominent hubs. When infection is seeded randomly, higher heterogeneity delays the time it takes for the pathogen to reach these super-hubs. As a result, vaccination efforts have a greater window of opportunity to suppress transmission before the infection reaches the hubs that could otherwise drive rapid spread. Thus it seems that higher heterogeneity in scale-free contact networks, indicated by higher scale-free exponents, aids vaccination efforts more.

\section{Conclusion}

In this paper, we investigated how contact network topology influences the effectiveness of vaccination programs in the context of HPV infection spread. We examined several topological parameters, including the average degree, the power law exponent, the average shortest path length, and the network clustering coefficient. For vaccination strategies, we considered age-based vaccination, ring vaccination, degree centrality-based vaccination, betweenness centrality-based vaccination, closeness centrality-based vaccination, percolation centrality-based vaccination, and eigenvector centrality-based vaccination. As a benchmark, we also analysed the null model with no vaccination program. Our analysis was conducted separately for the overall cohort, the male cohort, and the female cohort. To evaluate the effectiveness of the different vaccination programs, we focused on three outcome measures: peak incidence, timing of peak prevalence, and cumulative incidence of HPV infection dynamics, in relation to the network topological metrics studied.

Our simulation experiments highlight consistent patterns in how both vaccination strategies and network topology shape epidemic outcomes. Among vaccination strategies, the degree-, betweenness- and percolation centrality-based approaches were generally the most effective across the overall, male, and female cohorts, with respect to peak incidence, timing of peak prevalence, and cumulative incidence. The notable exception was in the female cohort, where ring vaccination achieved the greatest reduction in cumulative incidence, suggesting that targeted containment of early transmission chains can be especially beneficial. By contrast, closeness centrality-based, eigenvector centrality-based, and age-based strategies consistently performed less well. In terms of network metrics, higher average degree was associated with higher peak incidence, earlier peak prevalence, and greater cumulative incidence, reflecting reduced vaccination effectiveness in denser networks. Conversely, higher power law exponent, longer average shortest path length, and stronger clustering coefficient were each associated with lower peak incidence, delayed peak prevalence, and reduced cumulative incidence, indicating that heterogeneity, longer path lengths, and local clustering enhance the ability of vaccination programs to suppress spread. These tendencies were robust across all cohorts, underscoring the importance of network structure in determining the success of vaccination interventions. These findings suggest that future vaccination programs should explicitly consider network topology in their design.

This study has several limitations that warrant acknowledgement. First, the analysis relies on the SeCoNet contact network simulation model to generate the underlying networks. Although this model captures key features of sexual contact formation within a community, real-world contact networks may exhibit different topological characteristics, potentially affecting the relative effectiveness of vaccination strategies. Whether such differences would substantially alter the qualitative patterns reported here remains to be established through validation using empirical sexual contact network data. Second, the simulations were conducted on networks of relatively limited size, and extending the analysis to substantially larger networks may yield quantitatively different outcomes. Third, our study focused on four widely used topological metrics—average degree, power-law exponent, average shortest path length, and clustering coefficient—while other measures, such as assortativity and modularity, may provide additional insights into how network structure influences vaccination effectiveness. Finally, the model was specifically parameterised using the student demographics of a Group of Eight university in Australia, which limits the direct generalisability of the quantitative findings to the broader Australian population or to settings with different sociodemographic characteristics. In particular, the concentrated age structure (predominantly young adults), periodic influx of susceptible individuals at the beginning of academic terms, and dense interaction networks characteristic of a university campus create an intensive transmission environment in which vaccination strategies can be stress-tested. Consequently, the simulated transmission dynamics may not fully reflect general population mixing patterns or environments characterised by lower partner turnover. Future research incorporating empirical contact network data, larger and more diverse populations, and additional topological measures would help address these limitations and establish the extent to which the findings generalise across different settings. Nevertheless, the present study provides valuable insights into how sexual contact network topology influences the effectiveness of HPV vaccination strategies.

\appendices
\section{Simulation Results for a Narrowed Preferred Age Difference}\label{aa}

To evaluate how age-homogeneous mixing, a structural feature characteristic of tertiary student populations, influences disease transmission and intervention performance, we conducted sensitivity experiments setting the preferred partner age difference to a tighter lower-bound boundary condition of $\langle\eta\rangle = 1.5\text{ years}$. This is evaluated in contrast to the general population baseline parametrisation of $\langle\eta\rangle = 3.5\text{ years}$. All other network and epidemic parameters ($N = 3000$, $\beta = 0.13$, $\langle\delta\rangle = 100\text{ days}$, and vaccine coverage $c = 10\%$) were held constant. Fig. \ref{12} - \ref{20} present the maximum daily incidence, peak day of prevalence, and cumulative incidence across overall, female, and male cohorts under tighter age assortativity.

\subsection{Simulation Results for the Overall Cohort}

\begin{figure}[htbp]
\centering 
\includegraphics[width=\linewidth]{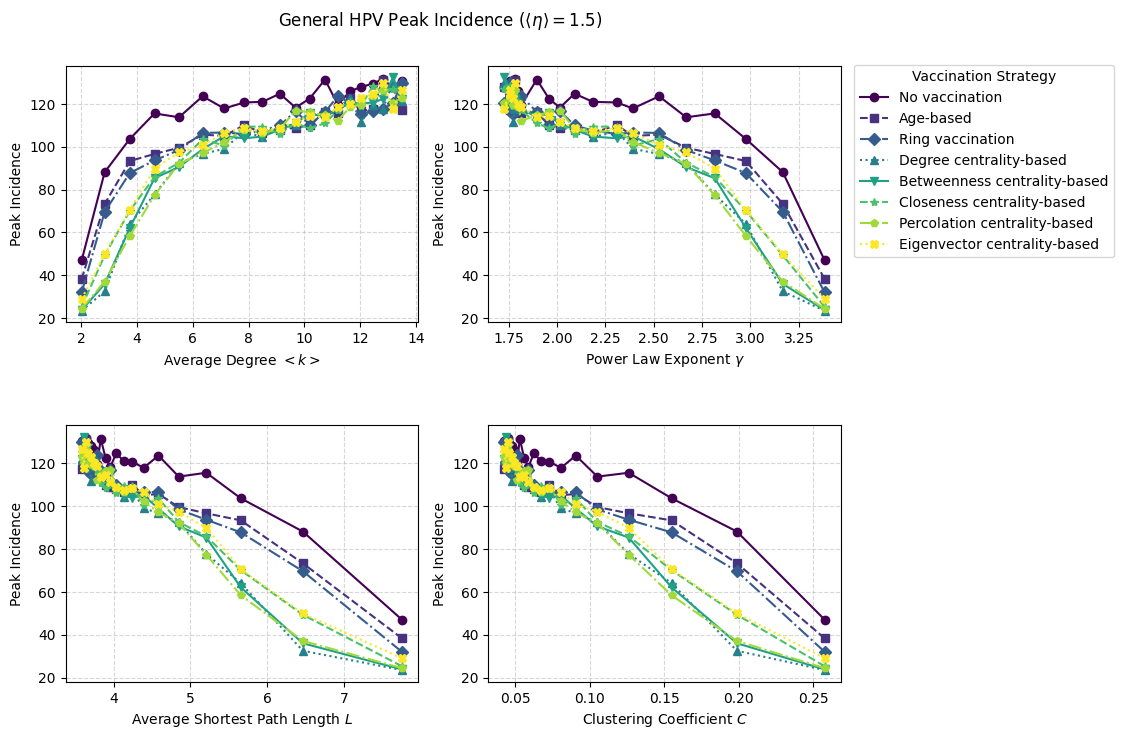}
\caption{Maximum Daily General HPV Incidence across Network Structures and Vaccination Strategies ($N = 3000$, $\beta = 0.13$, $\langle\delta\rangle = 100$, $c =10\%$, $\langle \eta \rangle = 1.5$) - Narrowing the preferred age difference to 1.5 years maintains baseline incidence trends. Degree- and percolation centrality-based strategies remain optimal in suppressing peak daily incidence, though higher network connectivity narrows strategy performance gaps.}\label{12}
\end{figure}

Restricting partnership formation to tighter age bands ($\langle\eta\rangle = 1.5\text{ years}$) concentrates infection within closely matched age cohorts. Maximum daily general incidence reaches slightly higher peak levels across moderate link densities compared to baseline, as dense age-clustered subgroups accelerate local viral turnover, as shown in Fig. \ref{12}.

\begin{figure}[htbp]
\centering
\includegraphics[width=\linewidth]{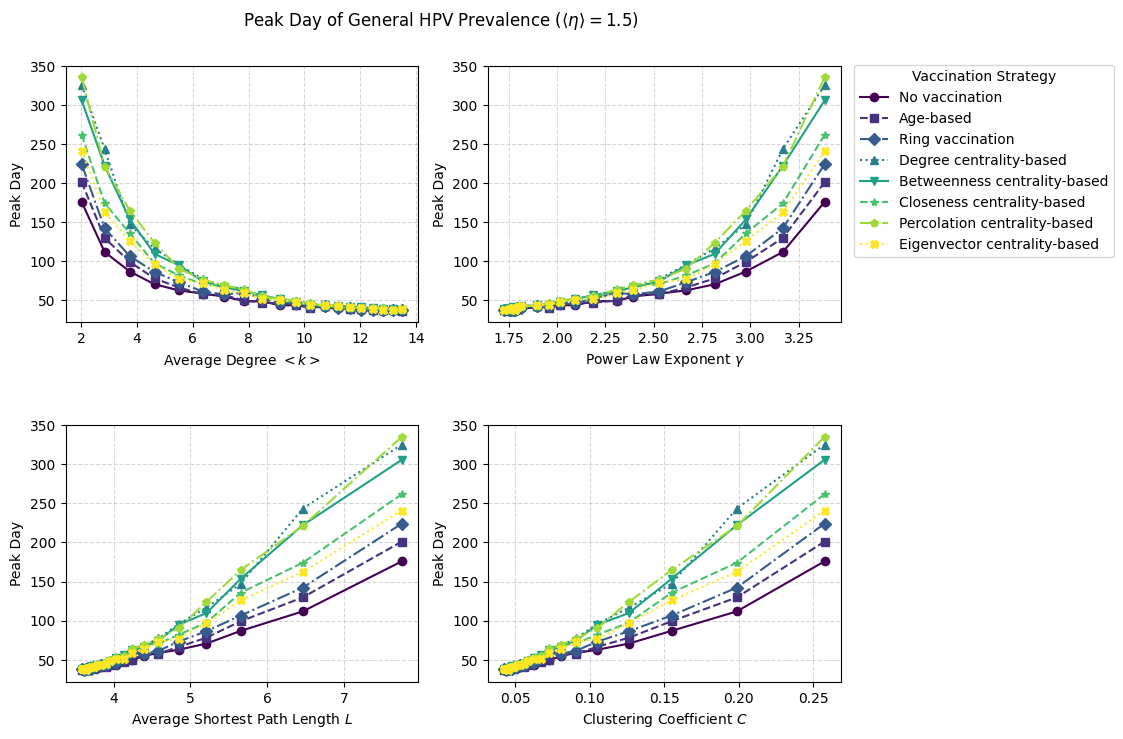}
\caption{Effect of Network Properties and Vaccination Strategies on Peak Day of General HPV Prevalence ($N = 3000$, $\beta = 0.13$, $\langle\delta\rangle = 100$, $c =10\%$, $\langle \eta \rangle = 1.5$) - Tighter age assortativity slightly extends the timeline to peak general prevalence. The degree centrality-based strategy consistently achieves the greatest temporal delay across all network topologies.}\label{13}
\end{figure}

However, centrality-based vaccination schemes, specifically degree-, betweenness-, and percolation centrality maintain their superior performance in suppressing peak daily case counts. The timeline to peak general prevalence shows minor temporal compression, but degree centrality-based targeting consistently delivers the longest delay in epidemic progression in Fig. \ref{13}.

\begin{figure}[htbp]
\centering
\includegraphics[width=\linewidth]{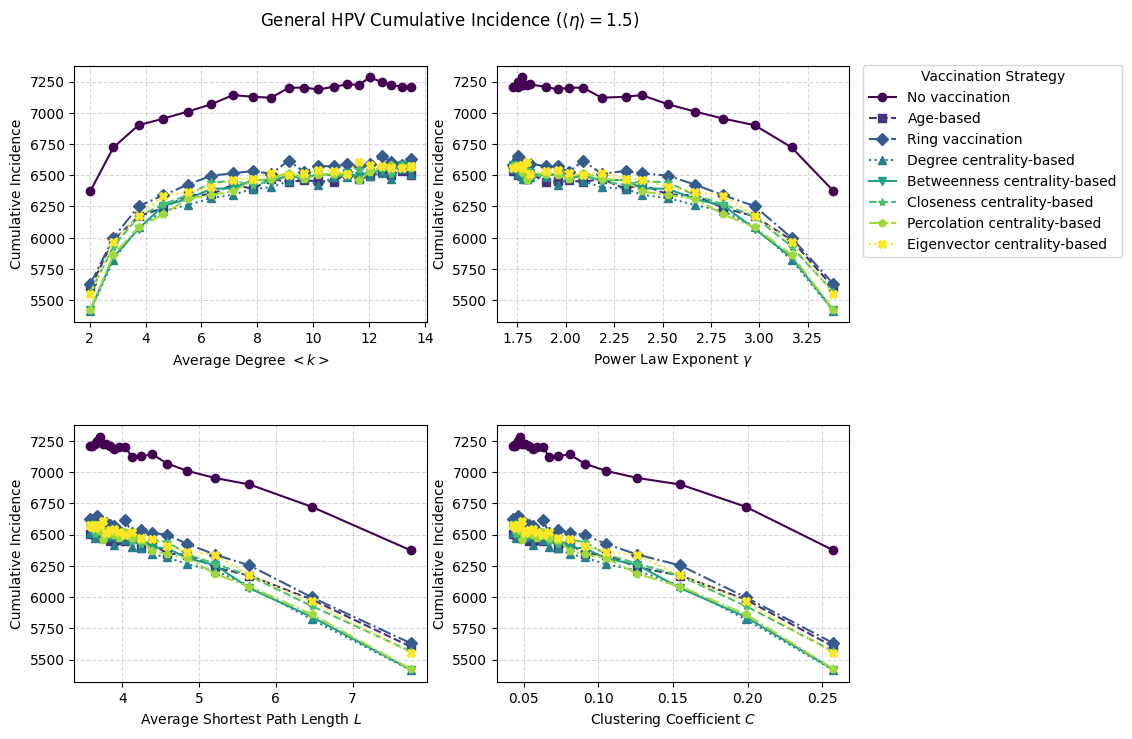}
\caption{Effect of Network Structure and Vaccination Strategy on General HPV Cumulative Incidence ($N = 3000$, $\beta = 0.13$, $\langle\delta\rangle = 100$, $c =10\%$, $\langle \eta \rangle = 1.5$) - Degree-, betweenness- and percolation centrality-based vaccination strategies effectively contain population-wide cumulative infection burden, exhibiting robust performance under stricter age-matching constraints.}\label{14}
\end{figure}

Overall cumulative general incidence remains within baseline ranges ($\sim 5,000–7,000\text{ cases}$), with degree-, betweenness-, and percolation-based strategies forming the primary high-efficacy cluster in Fig. \ref{14}.

\subsection{Simulation Results for the Female Cohort}
\begin{figure}[htbp]
\centering
\includegraphics[width=\linewidth]{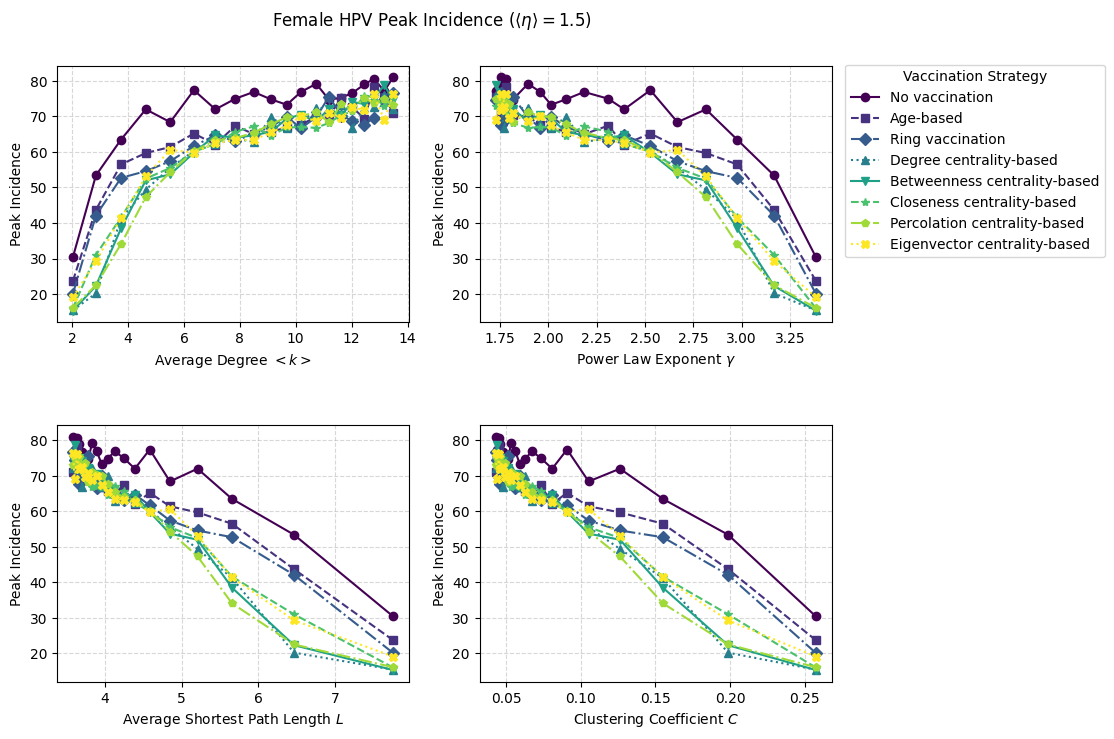}
\caption{Maximum Daily Female HPV Incidence across Network Structures and Vaccination Strategies ($N = 3000$, $\beta = 0.13$, $\langle\delta\rangle = 100$, $c =10\%$, $\langle \eta \rangle = 1.5$) - The percolation centrality-based strategy plays a crucial role in reducing peak daily female incidence, matching baseline observations.}\label{15}
\end{figure}

In the female cohort, tighter age assortativity enhances the efficacy of percolation centrality-based vaccination in controlling peak daily female incidence, particularly at lower network densities ($\langle k \rangle < 4$) in Fig. \ref{15}.

\begin{figure}[htbp]
\centering
\includegraphics[width=\linewidth]{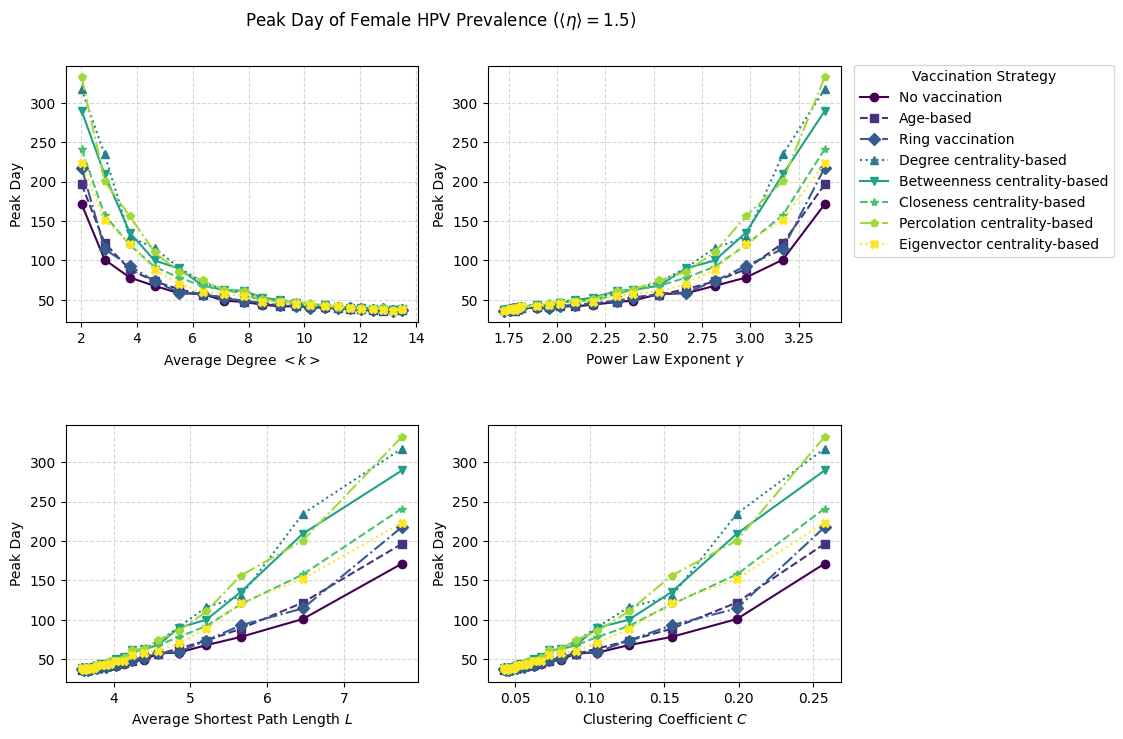}
\caption{Effect of Network Properties and Vaccination Strategies on Peak Day of Female HPV Prevalence ($N = 3000$, $\beta = 0.13$, $\langle\delta\rangle = 100$, $c =10\%$, $\langle \eta \rangle = 1.5$) - Degree centrality-based targeting demonstrates superior efficacy in postponing peak female infection prevalence under narrow age-preference constraints.}\label{16}
\end{figure}

Temporal trends indicate that while age-homogeneous mixing slightly advances the peak day of female prevalence, degree centrality-based targeting provides the most reliable buffer prior to peak prevalence as shown in Fig. \ref{16}. 

\begin{figure}[htbp]
\centering
\includegraphics[width=\linewidth]{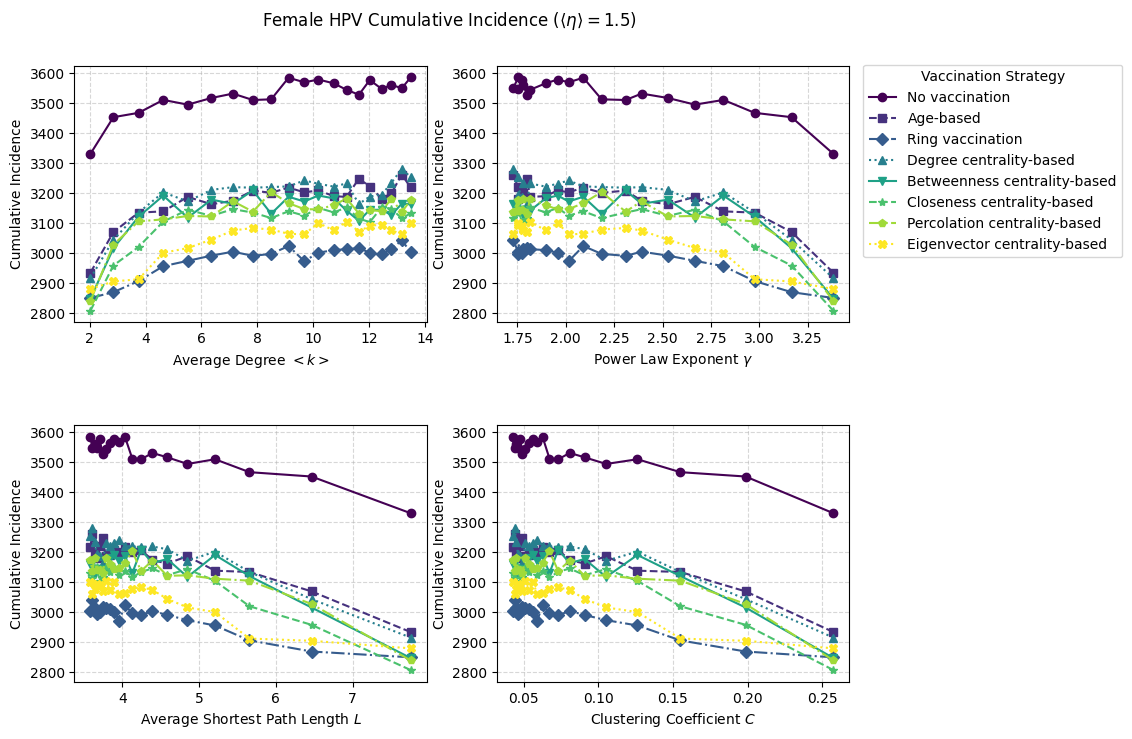}
\caption{Effect of Network Structure and Vaccination Strategy on Female HPV Cumulative Incidence ($N = 3000$, $\beta = 0.13$, $\langle\delta\rangle = 100$, $c =10\%$, $\langle \eta \rangle = 1.5$) - The ring vaccination strategy consistently achieves the greatest reduction in female cumulative infection burden, reinforcing its gender-specific advantage across varying age-assortativity settings.}\label{17}
\end{figure}

Crucially, ring vaccination consistently achieves the lowest cumulative incidence within the female cohort under tighter age mixing, confirming that its gender-specific structural advantage in containing localised female transmission chains is robust across both general-population and age-clustered student mixing regimes in Fig. \ref{17}.

\subsection{Simulation Results for the Male Cohort}
\begin{figure}[htbp]
\centering
\includegraphics[width=\linewidth]{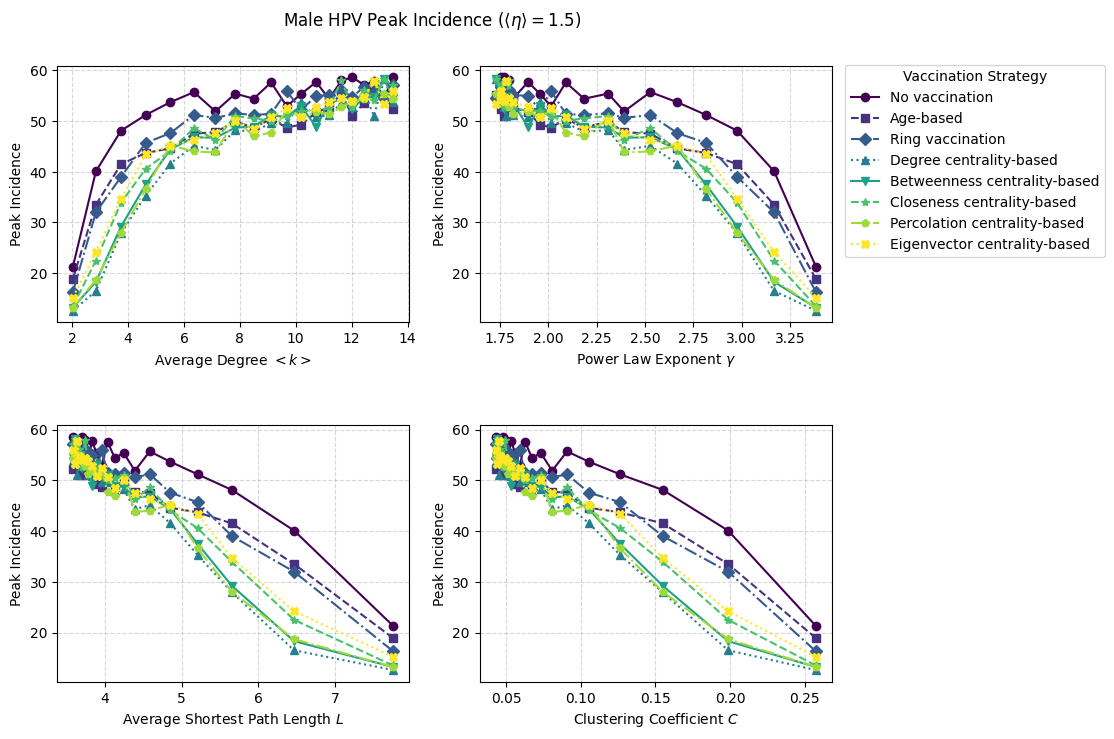}
\caption{Maximum Daily Male HPV Incidence across Network Structures and Vaccination Strategies ($N = 3000$, $\beta = 0.13$, $\langle\delta\rangle = 100$, $c =10\%$, $\langle \eta \rangle = 1.5$) - Degree-, betweenness- and percolation centrality-based strategies remain highly effective in controlling peak male incidence.}\label{18}
\end{figure}

For the male cohort, degree-, betweenness-, and percolation centrality-based strategies effectively control peak daily male incidence under tighter age preferences ($\langle\eta\rangle = 1.5\text{ years}$) in Fig. \ref{18}.

\begin{figure}[htbp]
\centering
\includegraphics[width=\linewidth]{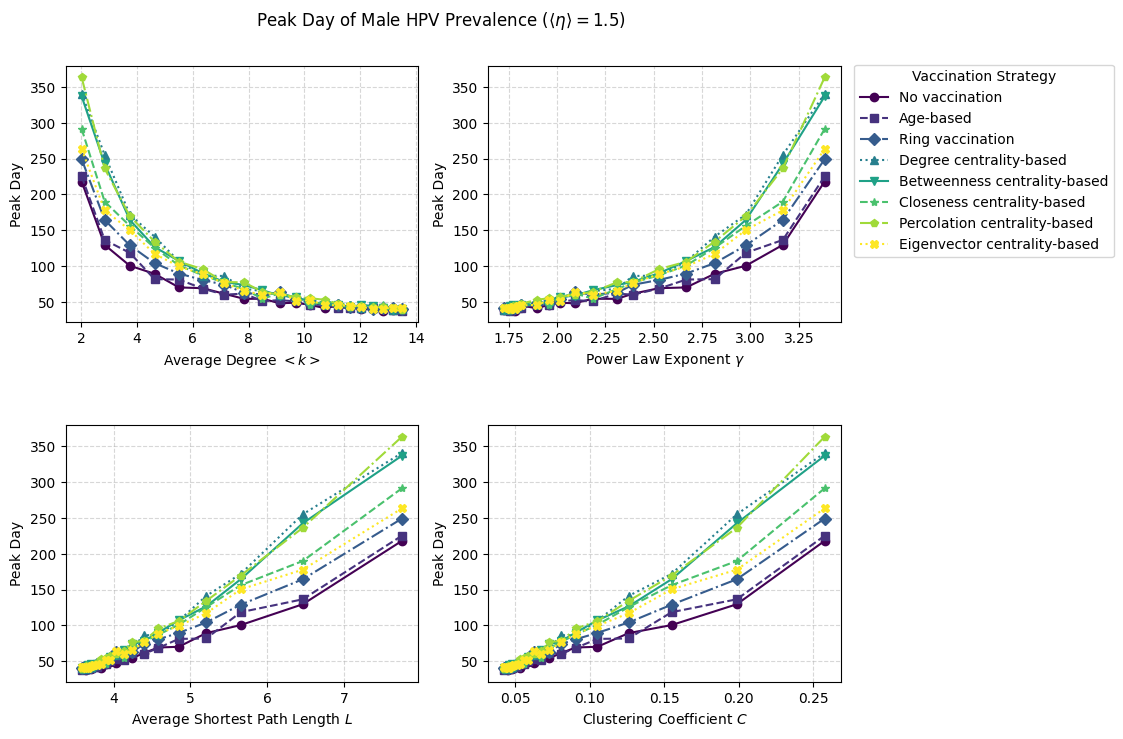}
\caption{Effect of Network Properties and Vaccination Strategies on Peak Day of Male HPV Prevalence ($N = 3000$, $\beta = 0.13$, $\langle\delta\rangle = 100$, $c =10\%$, $\langle \eta \rangle = 1.5$) - Postponement of peak male prevalence responds symmetrically to topological metrics, with centrality targeted schemes providing optimal delays.}\label{19}
\end{figure}

Temporal dynamics of peak male prevalence mirror baseline observations, demonstrating that centrality-targeted schemes maintain scale- and topology-invariant delays in male epidemic progression in Fig. \ref{19}. 

\begin{figure}[htbp]
\centering
\includegraphics[width=\linewidth]{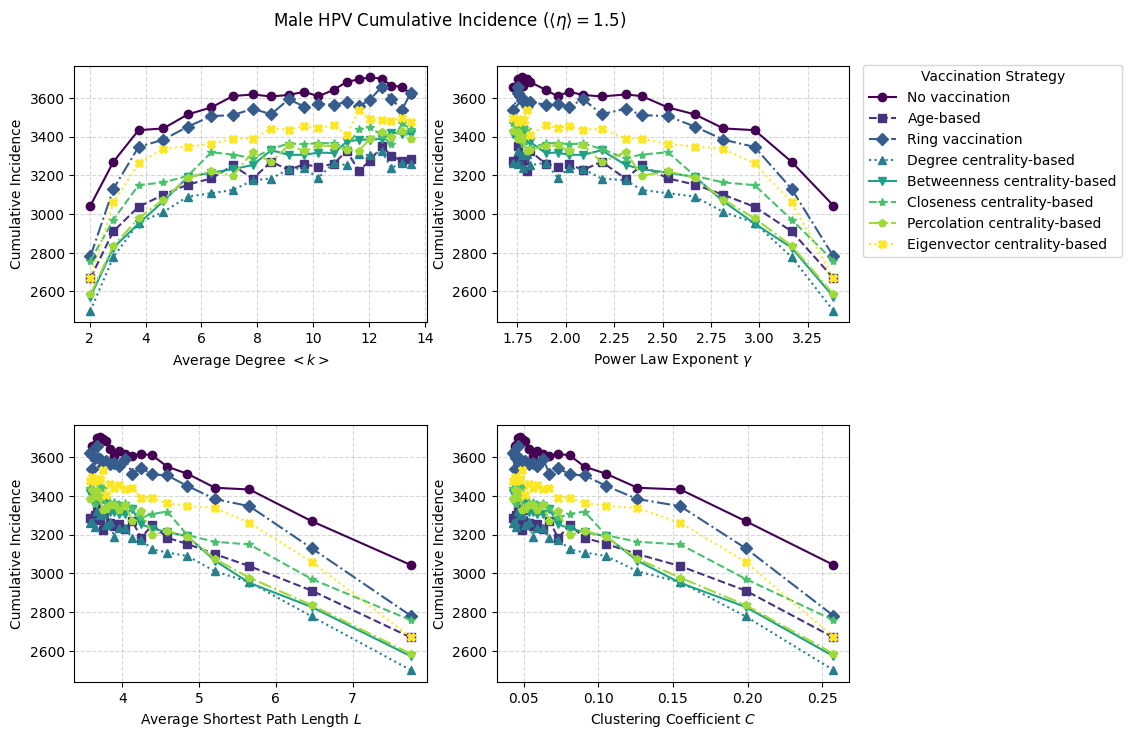}
\caption{Effect of Network Structure and Vaccination Strategy on Male HPV Cumulative Incidence ($N = 3000$, $\beta = 0.13$, $\langle\delta\rangle = 100$, $c =10\%$, $\langle \eta \rangle = 1.5$) - Degree centrality-based targeting remains the most effective approach for mitigating cumulative disease burden in the male cohort under stricter age-matching dynamics.}\label{20}
\end{figure}

Finally, degree centrality-based targeting remains the single most effective intervention for mitigating cumulative male disease burden, demonstrating that male-targeted centrality schemes are resilient to variations in age-assortative partnership formation in Fig. \ref{20}.

\section{Simulation Results with an Elevated Transmissibility}\label{ab}

To evaluate the sensitivity and robustness of our findings against variations in infection transmission, we conducted additional simulation experiments with an elevated high-risk HPV transmissibility per sexual act ($\beta = 0.3$), compared to the baseline setting of $\beta = 0.13$ \cite{Olsen2010HumanPapillomavirus}. All other structural and epidemiological parameters were held constant at their baseline values ($N = 3000$, $\langle\delta\rangle = 100 \text{ days}$, vaccine coverage $c = 10\%$, and preferred age difference $\langle\eta\rangle = 3.5 \text{ years}$). Fig. \ref{21} - \ref{29} present the corresponding maximum daily incidence, peak day of prevalence, and cumulative incidence across overall, female, and male cohorts.

\subsection{Simulation Results for the Overall Cohort}

\begin{figure}[htbp]
\centering
\includegraphics[width=\linewidth]{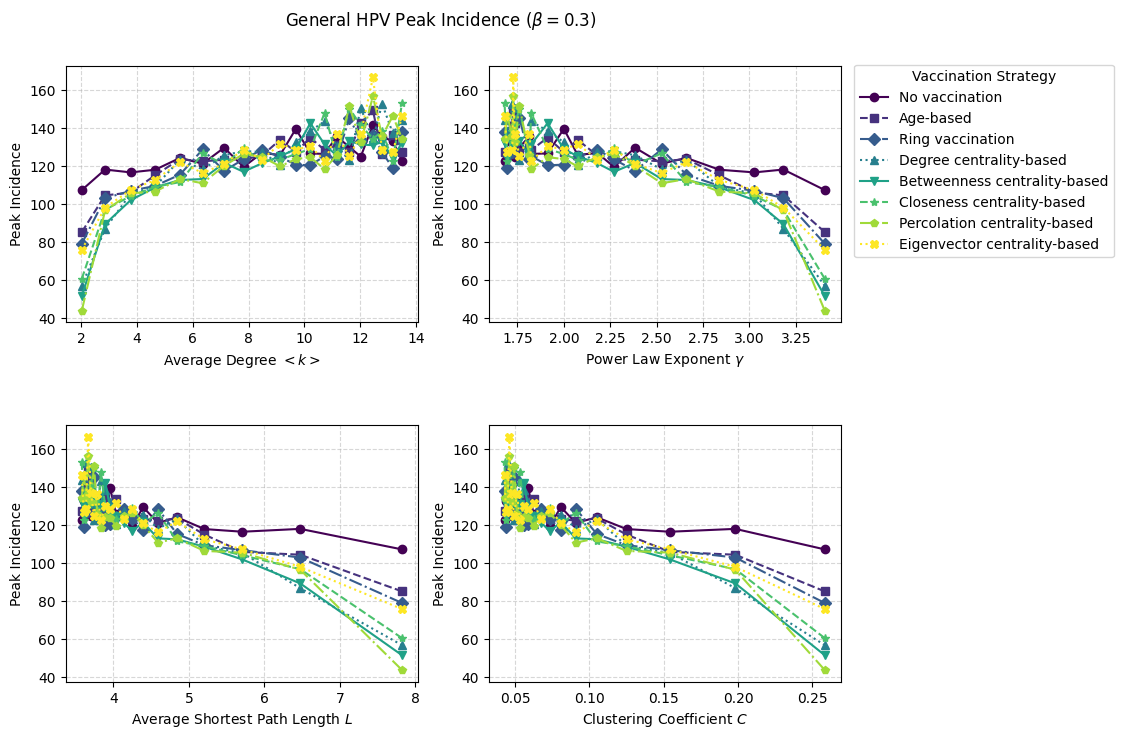}
\caption{Maximum Daily General HPV Incidence across Network Structures and Vaccination Strategies ($N = 3000$, $\beta = 0.3$, $\langle\delta\rangle = 100$, $c = 10\%$, $\langle \eta \rangle = 3.5$) - The elevated transmissibility yields a marked increase in peak general incidence across all network topology metrics. Percolation- and betweenness centrality-based strategies remain effective at lower link densities, though performance disparities across strategies narrow.}\label{21}
\end{figure}

Under the elevated transmissibility ($\beta = 0.3$), maximum daily general incidence increases substantially across all network configurations in Fig. \ref{21}. While percolation centrality-based and betweenness centrality-based strategies remain effective at lower link densities ($\langle k \rangle < 4$), performance disparities across different vaccination strategies narrow significantly as network connectivity increases.

\begin{figure}[htbp]
\centering
\includegraphics[width=\linewidth]{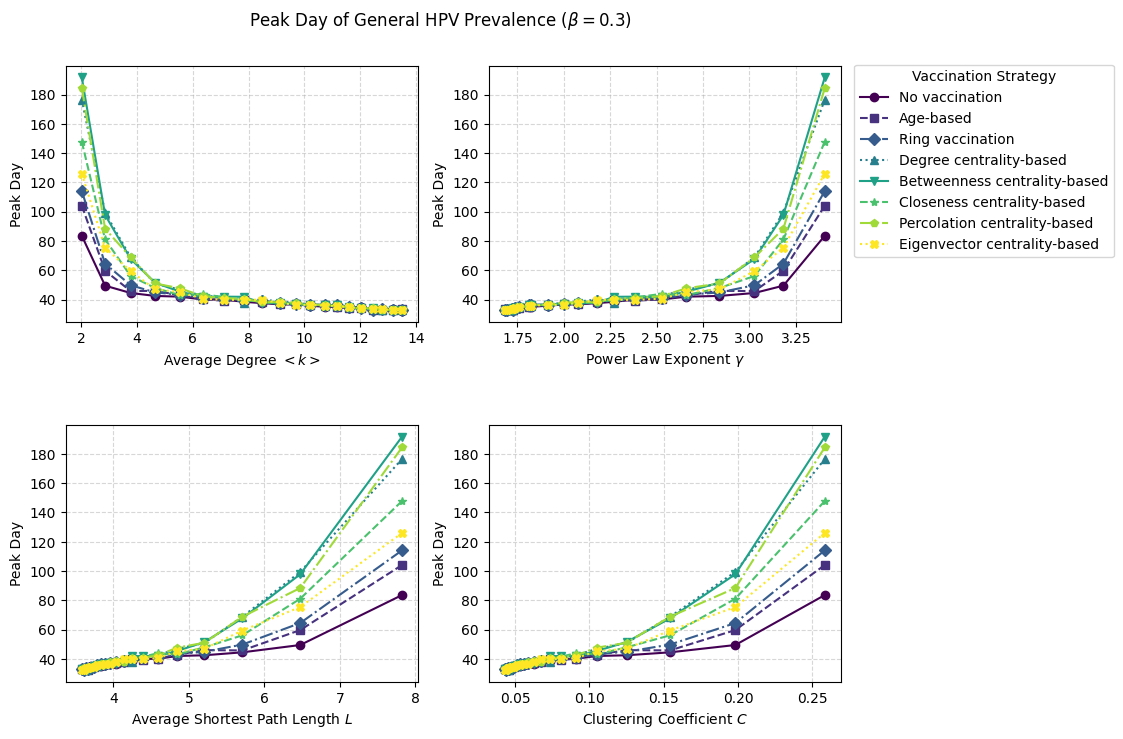}
\caption{Effect of Network Properties and Vaccination Strategies on Peak Day of General HPV Prevalence ($N = 3000$, $\beta = 0.3$, $\langle\delta\rangle = 100$, $c = 10\%$, $\langle \eta \rangle = 3.5$) - The higher transmissibility substantially compresses the epidemic timeline, advancing peak prevalence across all configurations. While percolation centrality-based strategy effectively delays peak prevalence in sparse networks, the performance gap among strategies shrinks as connectivity increases.}\label{22}
\end{figure}

Furthermore, higher transmissibility compresses the epidemic timeline, advancing the peak day of general HPV prevalence across all network structures in Fig. \ref{22}. 

\begin{figure}[htbp]
\centering
\includegraphics[width=\linewidth]{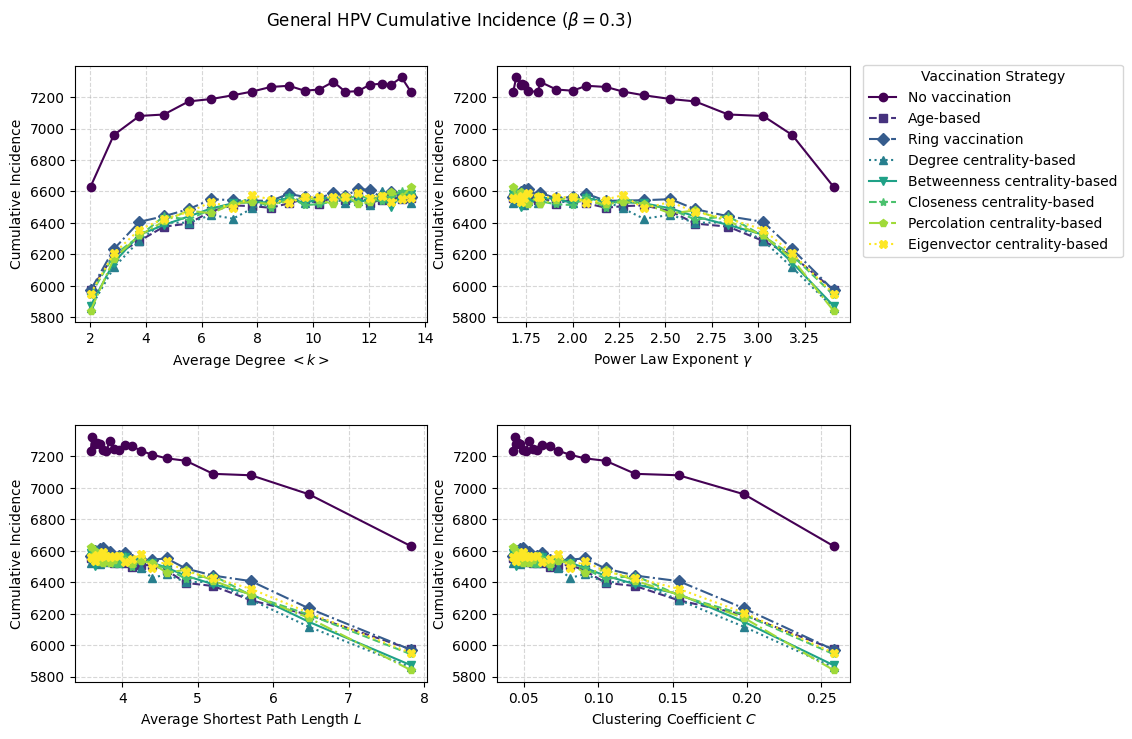}
\caption{Effect of Network Structure and Vaccination Strategy on General HPV Cumulative Incidence ($N = 3000$, $\beta = 0.3$, $\langle\delta\rangle = 100$, $c = 10\%$, $\langle \eta \rangle = 3.5$) - Although under high transmissibility, the general cumulative incidence increases modestly, the baseline structural trends persist, and the degree centrality-based strategy remains superior in suppressing overall case counts.}\label{23}
\end{figure}

Although overall cumulative incidence increases modestly under the high transmissibility, baseline structural patterns persist that degree centrality-based, betweenness centrality-based, and percolation centrality-based strategies consistently form a primary high-effectiveness cluster, which minimises total population-level case counts in Fig. \ref{23}.

\subsection{Simulation Results for the Female Cohort}

\begin{figure}[htbp]
\centering
\includegraphics[width=\linewidth]{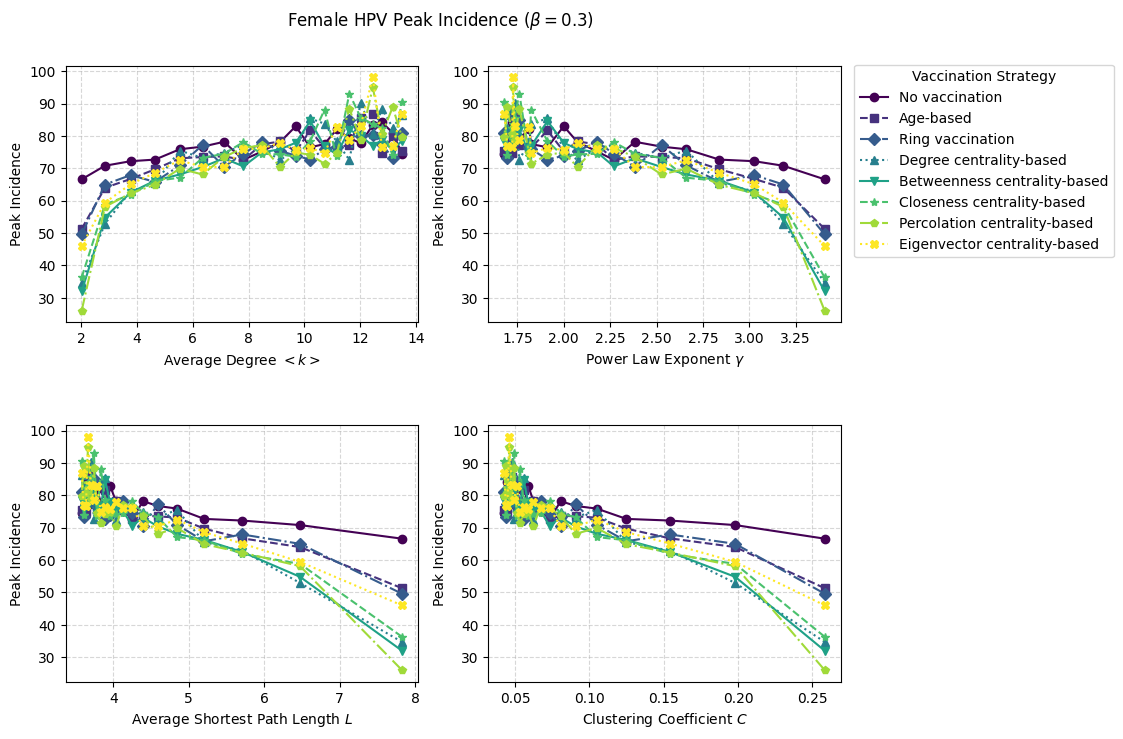}
\caption{Maximum Daily Female HPV Incidence across Network Structures and Vaccination Strategies ($N = 3000$, $\beta = 0.3$, $\langle\delta\rangle = 100$, $c = 10\%$, $\langle \eta \rangle = 3.5$) - The degree-, betweenness-, percolation centrality-based strategies remain pivotal in curbing female peak incidence; however, under heightened transmissibility, intervention performance becomes less stable, and strategy advantages become negligible in densely connected networks.}\label{24}
\end{figure}

For the female cohort, the degree-, betweenness-, percolation centrality-based strategies remain pivotal in curbing peak daily female incidence under elevated transmissibility in Fig. \ref{24}. However, as the viral spread intensifies, strategy performance becomes less stable, and relative intervention advantages become negligible in highly connected networks.

\begin{figure}[htbp]
\centering
\includegraphics[width=\linewidth]{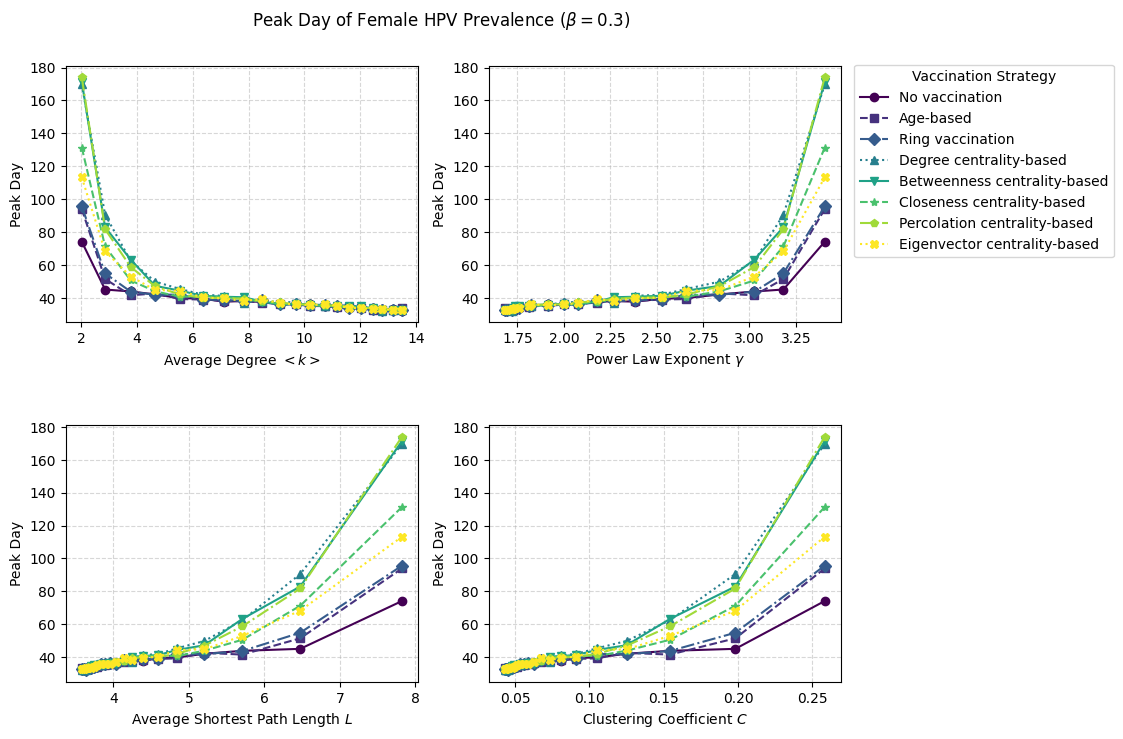}
\caption{Effect of Network Properties and Vaccination Strategies on Peak Day of Female HPV Prevalence ($N = 3000$, $\beta = 0.3$, $\langle\delta\rangle = 100$, $c = 10\%$, $\langle \eta \rangle = 3.5$) - The degree centrality-based vaccination strategy remains successful in delaying peak female prevalence. Because accelerated viral spread shortens the epidemic timeline, differences in relative strategy effectiveness shrink.}\label{25}
\end{figure}

In terms of epidemic timing, the degree centrality-based strategy remains effective in delaying peak female prevalence in Fig. \ref{25}, though the rapid viral spread shortens the temporal window overall, shrinking differences between active strategies. 

\begin{figure}[htbp]
\centering
\includegraphics[width=\linewidth]{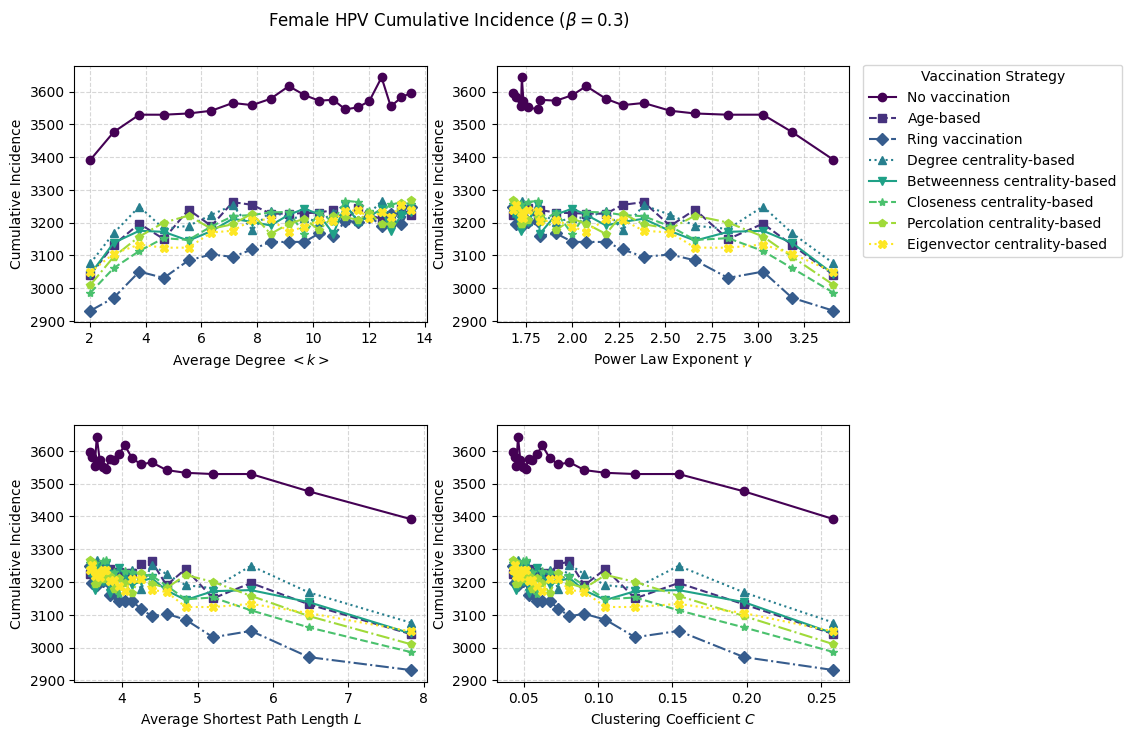}
\caption{Effect of Network Structure and Vaccination Strategy on Female HPV Cumulative Incidence ($N = 3000$, $\beta = 0.3$, $\langle\delta\rangle = 100$, $c = 10\%$, $\langle \eta \rangle = 3.5$) - Overall female cumulative incidence increases modestly, but the ring vaccination strategy consistently maintains its position as the most effective intervention for reducing infection burden in females.}\label{26}
\end{figure}

Crucially, despite elevated transmissibility raising overall case counts, the ring vaccination strategy consistently maintains its position as the most effective intervention for reducing cumulative incidence within the female cohort in Fig. \ref{26}, reinforcing its gender-specific advantage in containing localised female transmission chains.

\subsection{Simulation Results for the Male Cohort}
\begin{figure}[htbp]
\centering
\includegraphics[width=\linewidth]{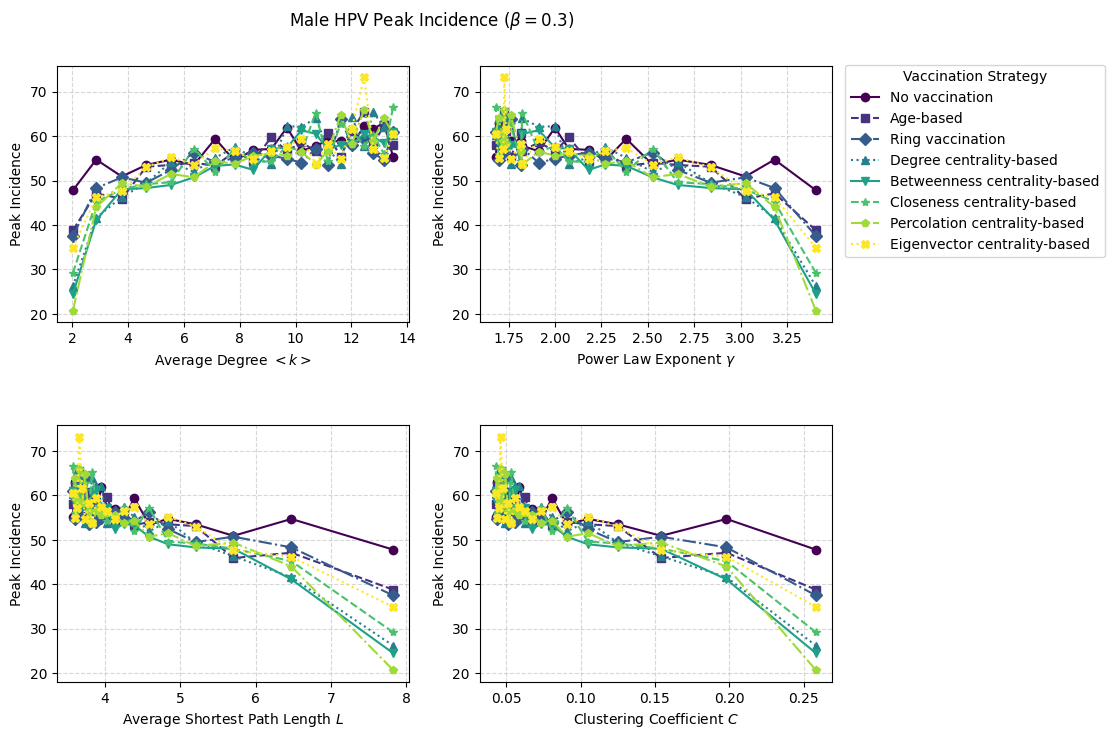}
\caption{Maximum Daily Male HPV Incidence across Network Structures and Vaccination Strategies ($N = 3000$, $\beta = 0.3$, $\langle\delta\rangle = 100$, $c = 10\%$, $\langle \eta \rangle = 3.5$) - The maximum daily male HPV incidence slightly increases, while degree-, betweenness- and percolation centrality-based vaccination strategies still work effectively in controlling new male cases.}\label{27}
\end{figure}

In the male cohort, peak daily incidence increases slightly under higher transmissibility in Fig. \ref{27}, but degree-, betweenness-, and percolation centrality-based strategies remain the top-performing schemes for controlling new cases.

\begin{figure}[htbp]
\centering
\includegraphics[width=\linewidth]{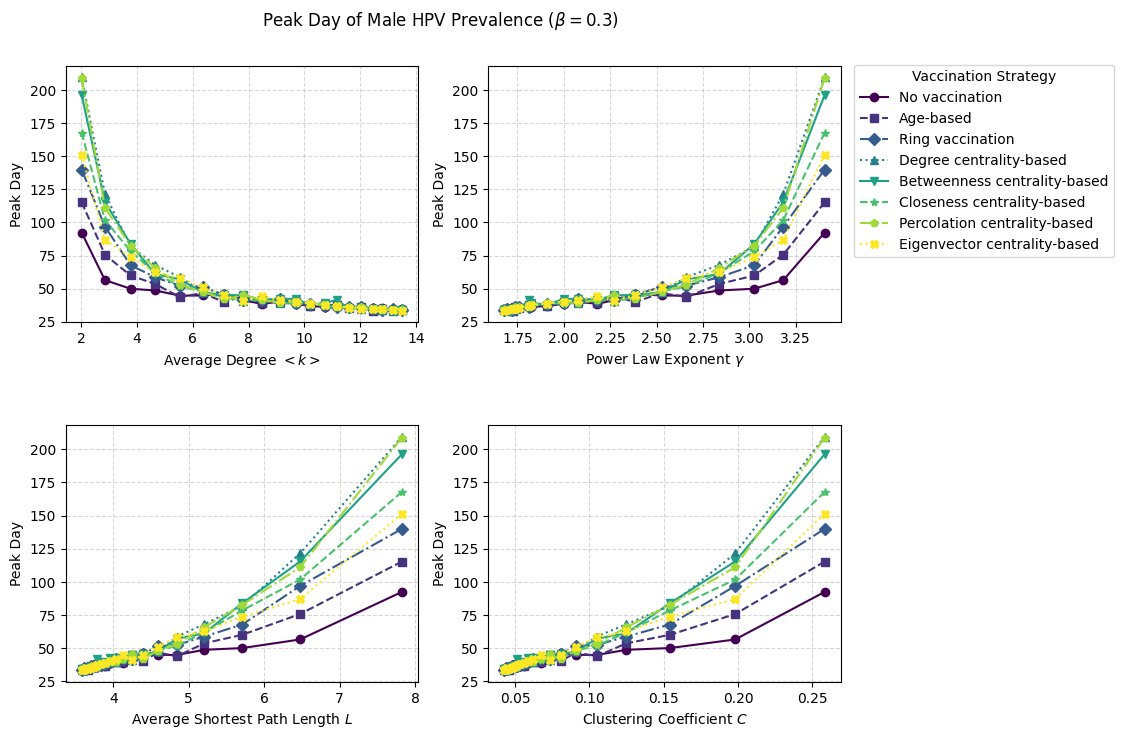}
\caption{Effect of Network Properties and Vaccination Strategies on Peak Day of Male HPV Prevalence ($N = 3000$, $\beta = 0.3$, $\langle\delta\rangle = 100$, $c = 10\%$, $\langle \eta \rangle = 3.5$) - Even as higher transmissibility advances the epidemic peak across all network, the overall structural trends stay robust.}\label{28}
\end{figure}

Temporal trends show that higher transmissibility advances the epidemic peak in males across all topological configurations in Fig. \ref{28}, though qualitative topological dependencies, such as peak delays driven by longer average path length $L$ and higher clustering $C$, remain robust. 

\begin{figure}[htbp]
\centering
\includegraphics[width=\linewidth]{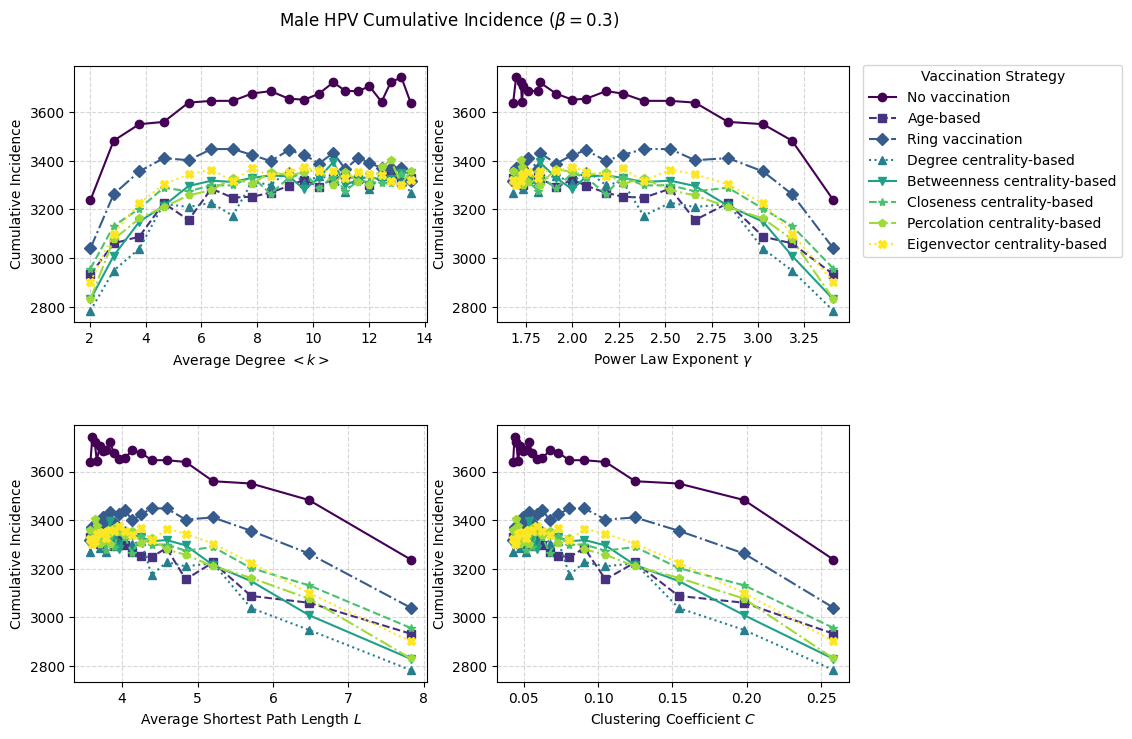}
\caption{Effect of Network Structure and Vaccination Strategy on Male HPV Cumulative Incidence ($N = 3000$, $\beta = 0.3$, $\langle\delta\rangle = 100$, $c = 10\%$, $\langle \eta \rangle = 3.5$)  - The degree centrality-based vaccination strategy remains superior in mitigating male cumulative incidence, as infection spread intensifies across the population.}\label{29}
\end{figure}

Finally, the degree centrality-based strategy remains superior in mitigating male cumulative disease burden as infection spread intensifies across the population in Fig. \ref{29}.

\section{Simulation Results with Varying Average Relationship Durations}\label{ac}

To investigate how relationship stability influences epidemic spread and intervention efficacy, we conducted sensitivity experiments across varying average partnership durations ($\langle\delta\rangle = 300, 500, \text{and } 700 \text{ days}$), compared to the baseline setting of $\langle\delta\rangle = 100 \text{ days}$. All other parameters ($N = 3000$, $\beta = 0.13$, vaccine coverage $c = 10\%$, and $\langle\eta\rangle = 3.5 \text{ years}$) were kept constant. Fig. \ref{30} - \ref{38} present the peak daily incidence, peak day of prevalence, and cumulative incidence across overall, female, and male cohorts under these extended partnership scenarios.

\subsection{Simulation Results for the Overall Cohort}

\begin{figure*}[htbp]
  \centering
  \begin{subfigure}[b]{0.32\textwidth}
    \centering
    \includegraphics[width=\linewidth]{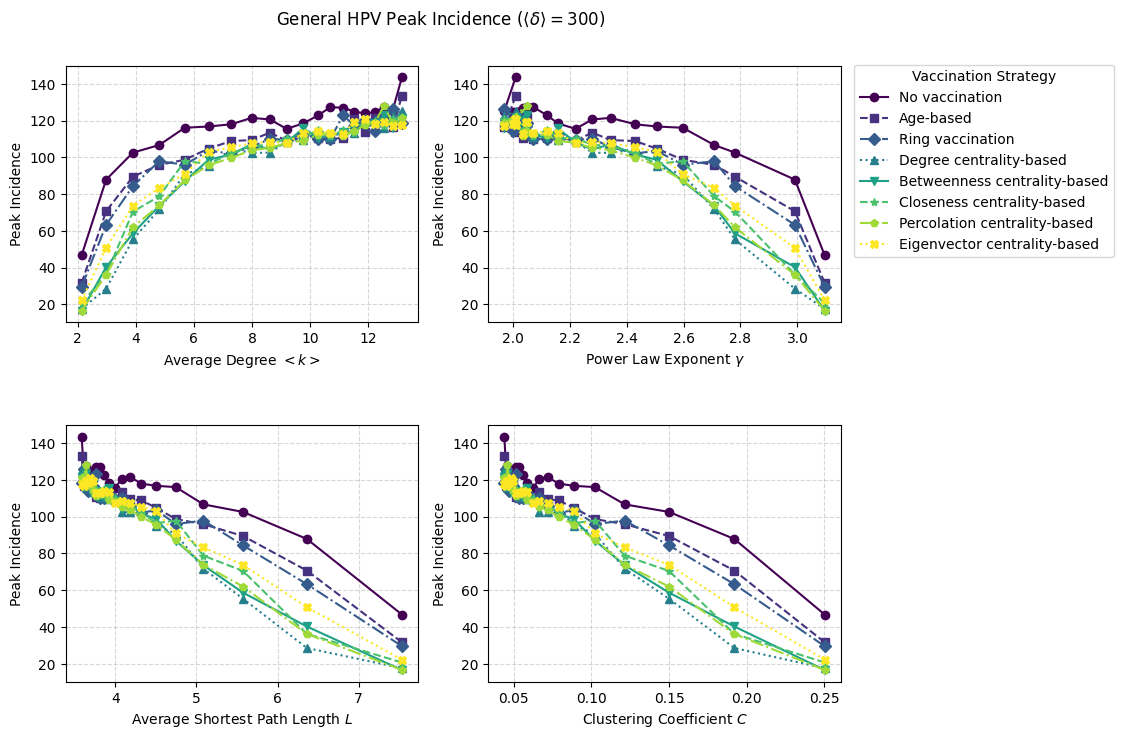}
    \caption{$\langle\delta\rangle = 300$}
  \end{subfigure}\hfill
  \begin{subfigure}[b]{0.32\textwidth}
    \centering
    \includegraphics[width=\linewidth]{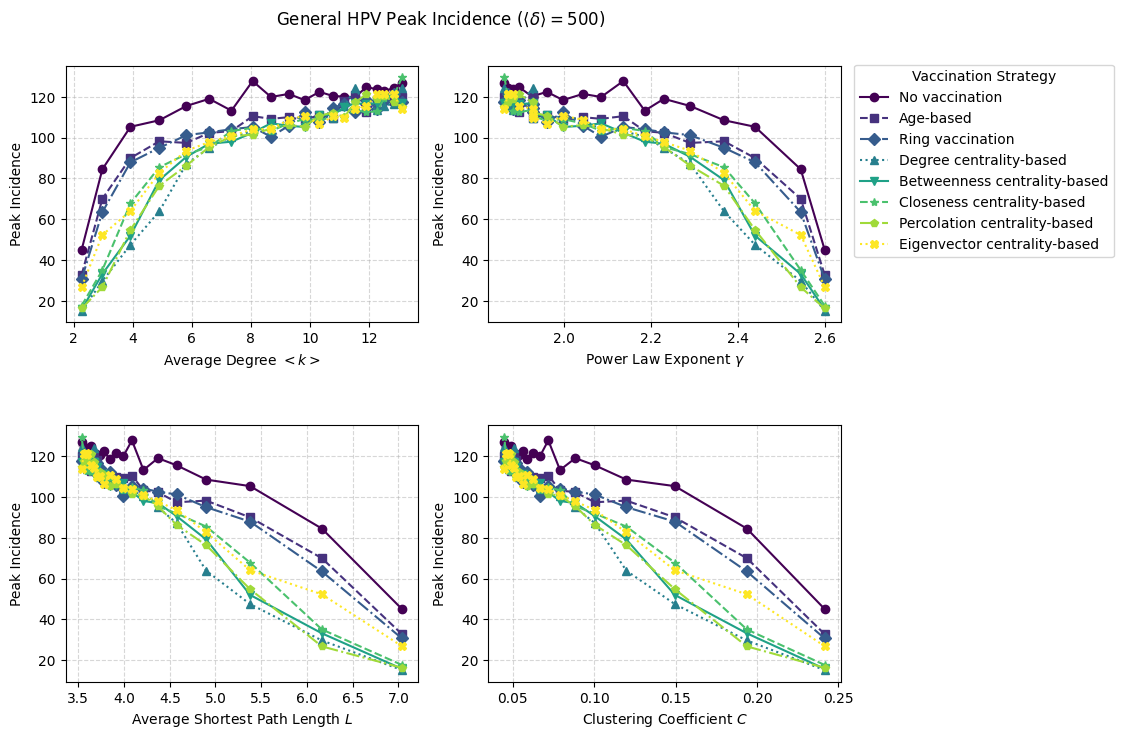}
    \caption{$\langle\delta\rangle = 500$}
  \end{subfigure}\hfill
  \begin{subfigure}[b]{0.32\textwidth}
    \centering
    \includegraphics[width=\linewidth]{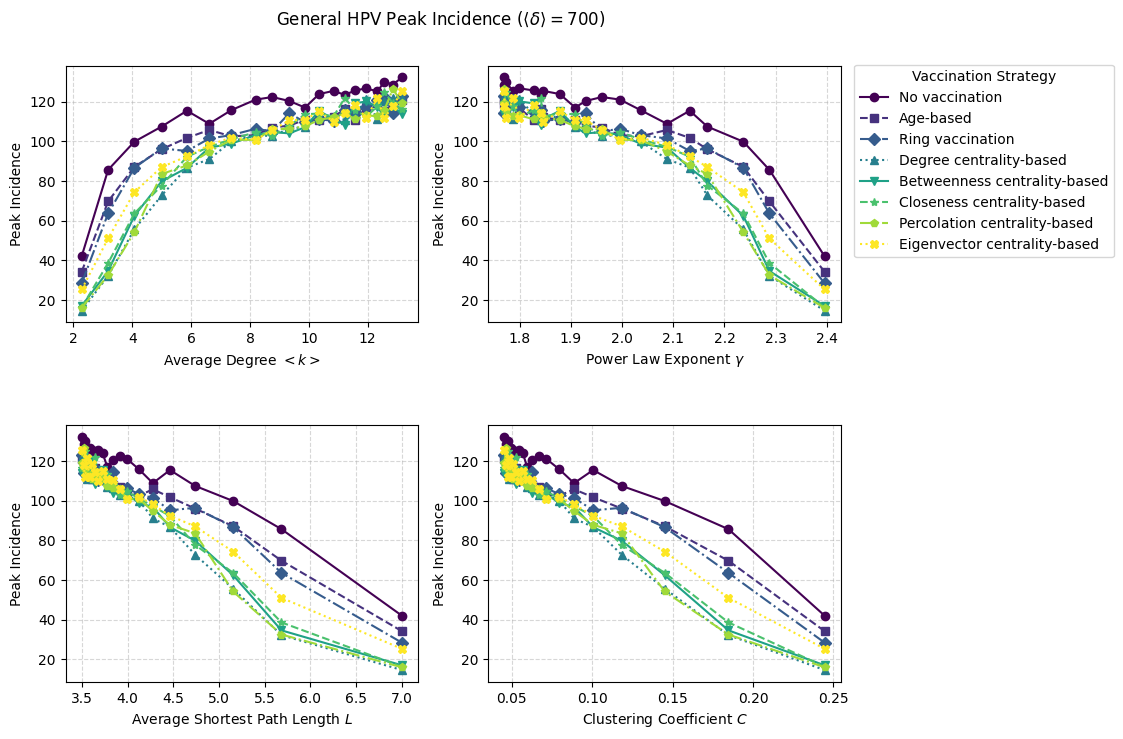}
    \caption{$\langle\delta\rangle = 700$}
  \end{subfigure}
  \caption{Maximum Daily General HPV Incidence across Network Structures and Vaccination Strategies under Varying Average Relationship Durations ($\langle\delta\rangle$) ($N = 3000$, $\beta = 0.13$, $c = 10\%$, $\langle \eta \rangle = 3.5$) - As average relationship duration increases ($\langle\delta\rangle = 300, 500, 700$), peak daily general incidence declines overall across sparse topologies. Degree- and percolation centrality-based strategies consistently maintain optimal transmission containment.}\label{30}
\end{figure*}

As average relationship duration increases ($\langle\delta\rangle = 300, 500, 700 \text{ days}$), maximum daily general incidence declines overall, particularly across sparse network topologies. Prolonged partnership stability restricts individual concurrency and link turnover, slowing infection turnover. Degree centrality-based and percolation centrality-based strategies consistently maintain optimal containment of peak daily incidence across all relationship duration regimes in Fig. \ref{30}.

\begin{figure*}[htbp]
  \centering
  \begin{subfigure}[b]{0.32\textwidth}
    \centering
    \includegraphics[width=\linewidth]{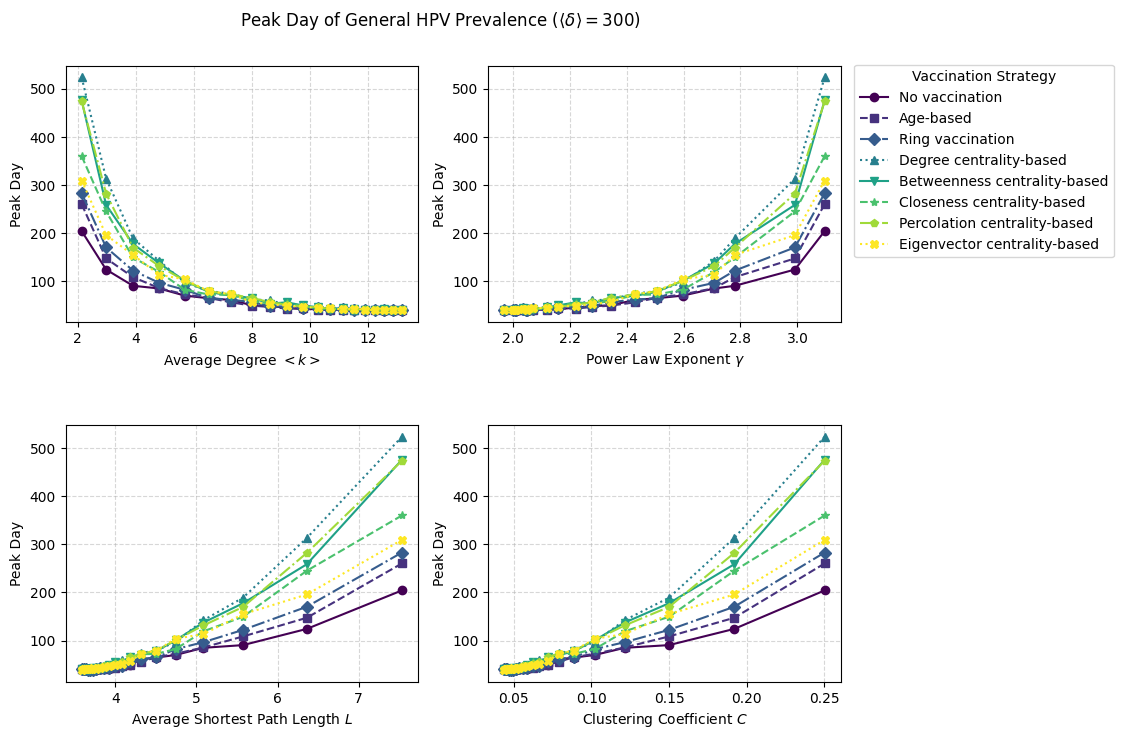}
    \caption{$\langle\delta\rangle = 300$}
  \end{subfigure}\hfill
  \begin{subfigure}[b]{0.32\textwidth}
    \centering
    \includegraphics[width=\linewidth]{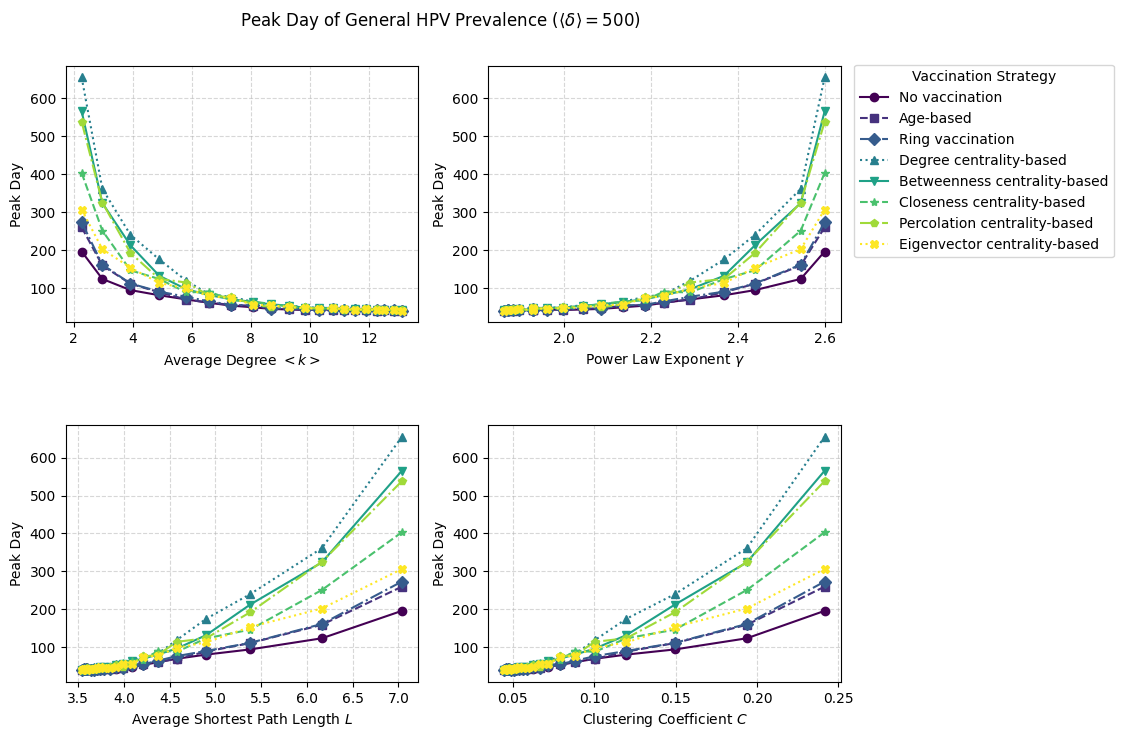}
    \caption{$\langle\delta\rangle = 500$}
  \end{subfigure}\hfill
  \begin{subfigure}[b]{0.32\textwidth}
    \centering
    \includegraphics[width=\linewidth]{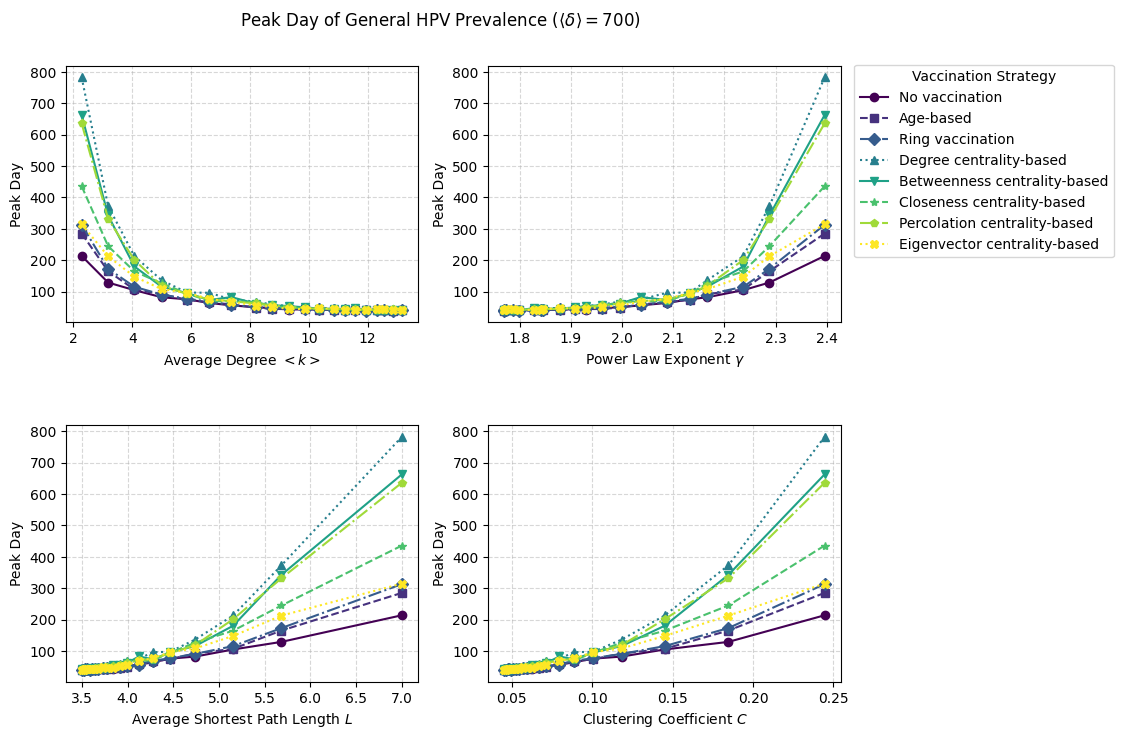}
    \caption{$\langle\delta\rangle = 700$}
  \end{subfigure}
  \caption{Effect of Network Properties and Vaccination Strategies on Peak Day of General HPV Prevalence under Varying Average Relationship Durations ($\langle\delta\rangle$) ($N = 3000$, $\beta = 0.13$, $c = 10\%$, $\langle \eta \rangle = 3.5$) - Prolonged relationship stability substantially delays the timing of peak general prevalence. Degree centrality-based targeting consistently offers the greatest temporal delay across all relationship duration scenarios.}\label{31}
\end{figure*}

In terms of epidemic timing, extended partnership stability substantially delays the peak day of general prevalence. Degree centrality-based targeting provides the greatest temporal buffer prior to peak prevalence in Fig. \ref{31}. 

\begin{figure*}[htbp]
  \centering
  \begin{subfigure}[b]{0.32\textwidth}
    \centering
    \includegraphics[width=\linewidth]{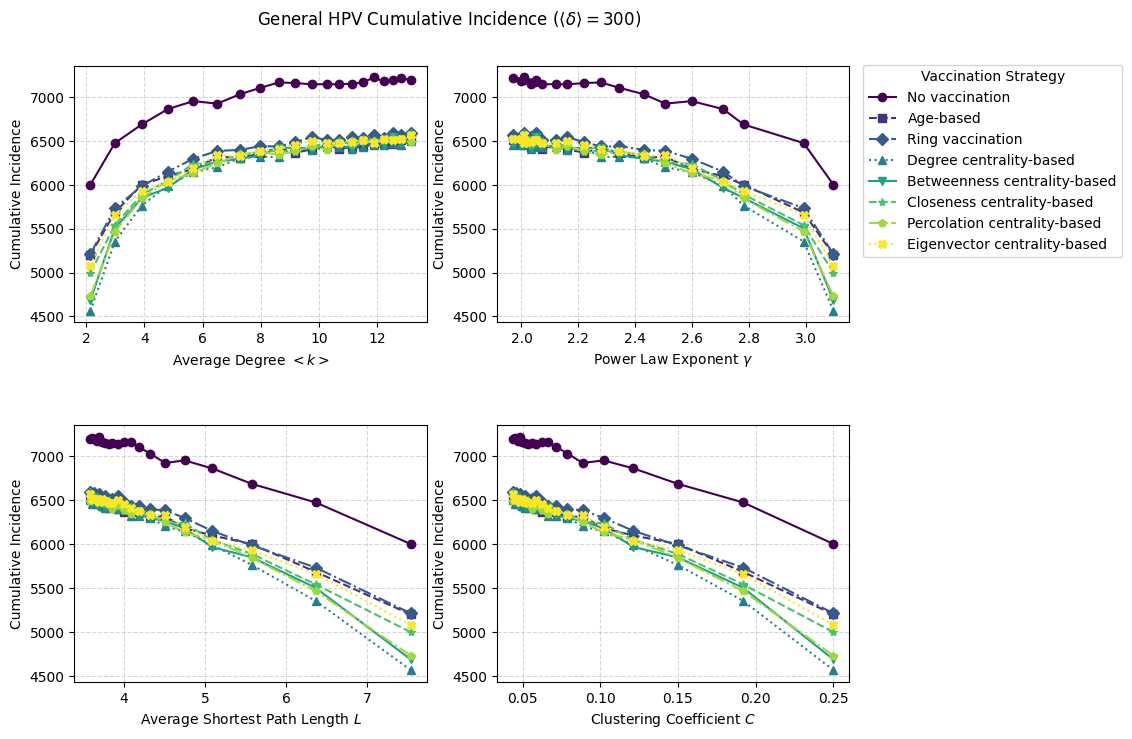}
    \caption{$\langle\delta\rangle = 300$}
  \end{subfigure}\hfill
  \begin{subfigure}[b]{0.32\textwidth}
    \centering
    \includegraphics[width=\linewidth]{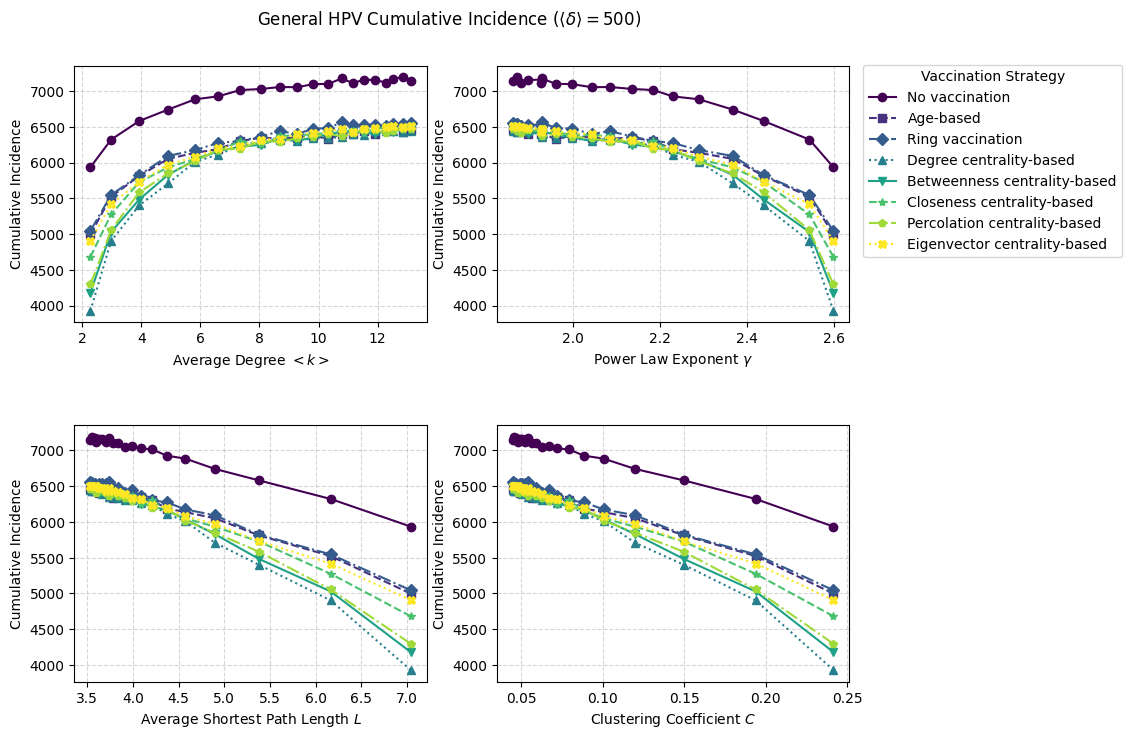}
    \caption{$\langle\delta\rangle = 500$}
  \end{subfigure}\hfill
  \begin{subfigure}[b]{0.32\textwidth}
    \centering
    \includegraphics[width=\linewidth]{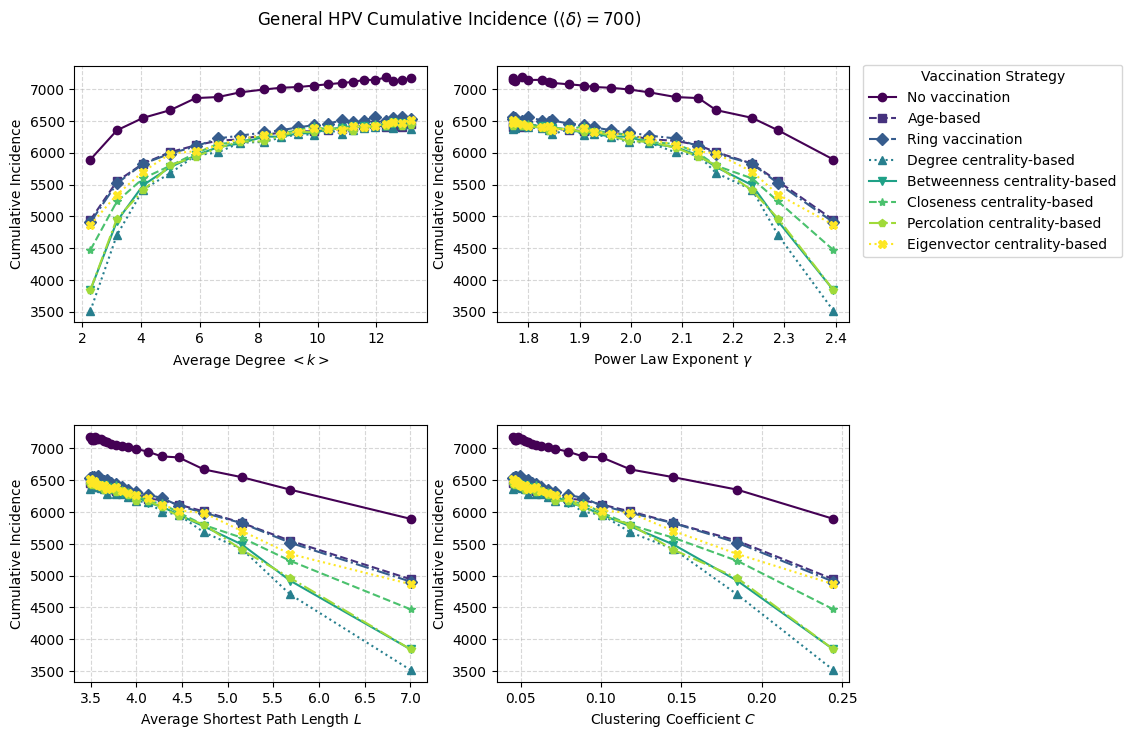}
    \caption{$\langle\delta\rangle = 700$}
  \end{subfigure}
  \caption{Effect of Network Structure and Vaccination Strategy on General HPV Cumulative Incidence under Varying Average Relationship Durations ($\langle\delta\rangle$) ($N = 3000$, $\beta = 0.13$, $c = 10\%$, $\langle \eta \rangle = 3.5$) - Degree-, betweenness- and percolation centrality-based strategies effectively suppress cumulative incidence, while increased partnership stability attenuates cumulative case growth as network connectivity increases.}\label{32}
\end{figure*}

Furthermore, increased partnership stability attenuates total cumulative case growth as network connectivity increases, with degree-, betweenness-, and percolation centrality-based strategies remaining most effective in curbing overall cumulative infections in Fig. \ref{32}.

\subsection{Simulation Results for the Female Cohort}

\begin{figure*}[htbp]
  \centering
  \begin{subfigure}[b]{0.32\textwidth}
    \centering
    \includegraphics[width=\linewidth]{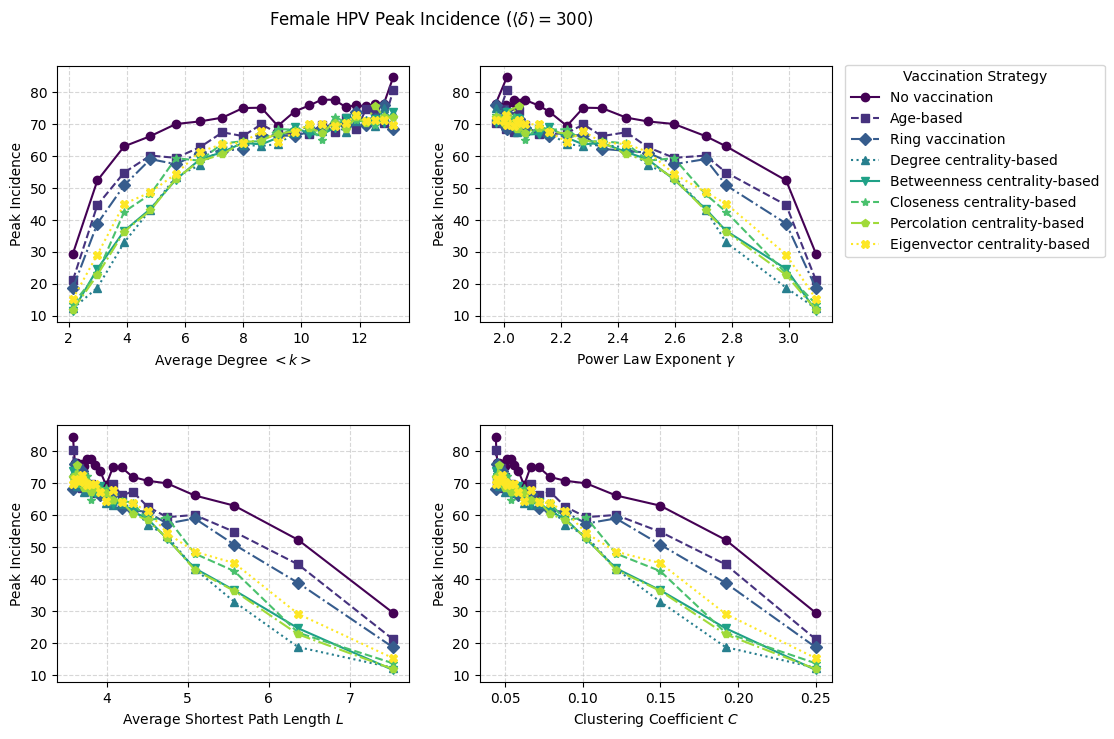}
    \caption{$\langle\delta\rangle = 300$}
  \end{subfigure}\hfill
  \begin{subfigure}[b]{0.32\textwidth}
    \centering
    \includegraphics[width=\linewidth]{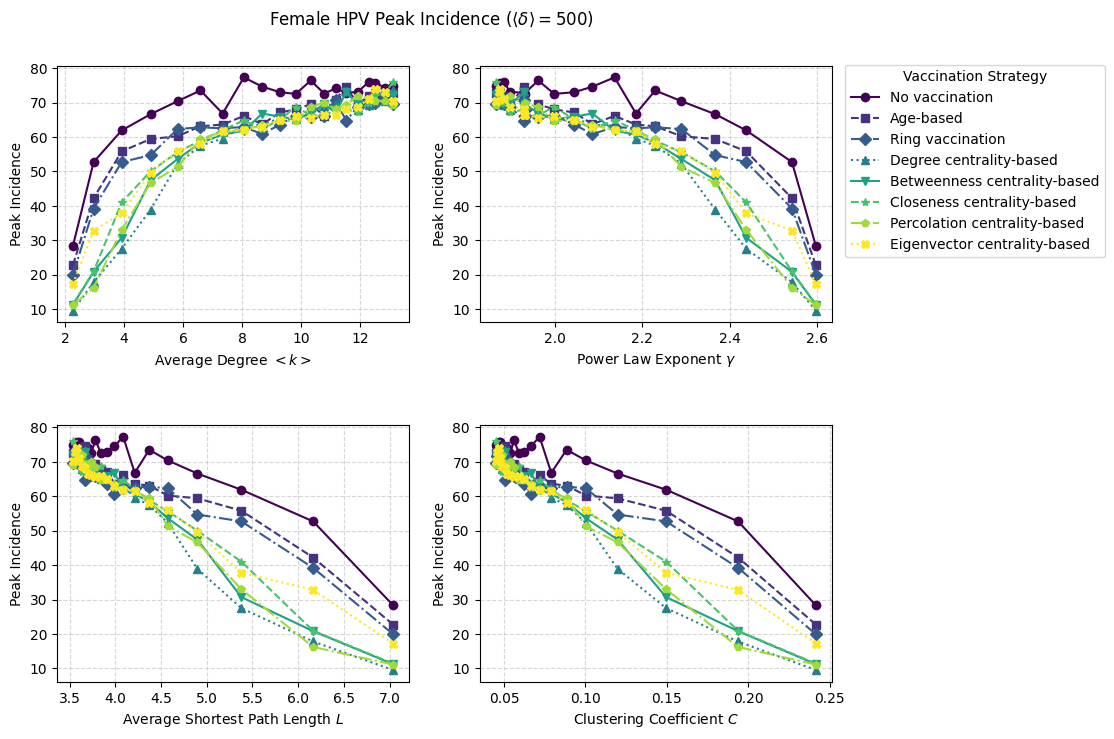}
    \caption{$\langle\delta\rangle = 500$}
  \end{subfigure}\hfill
  \begin{subfigure}[b]{0.32\textwidth}
    \centering
    \includegraphics[width=\linewidth]{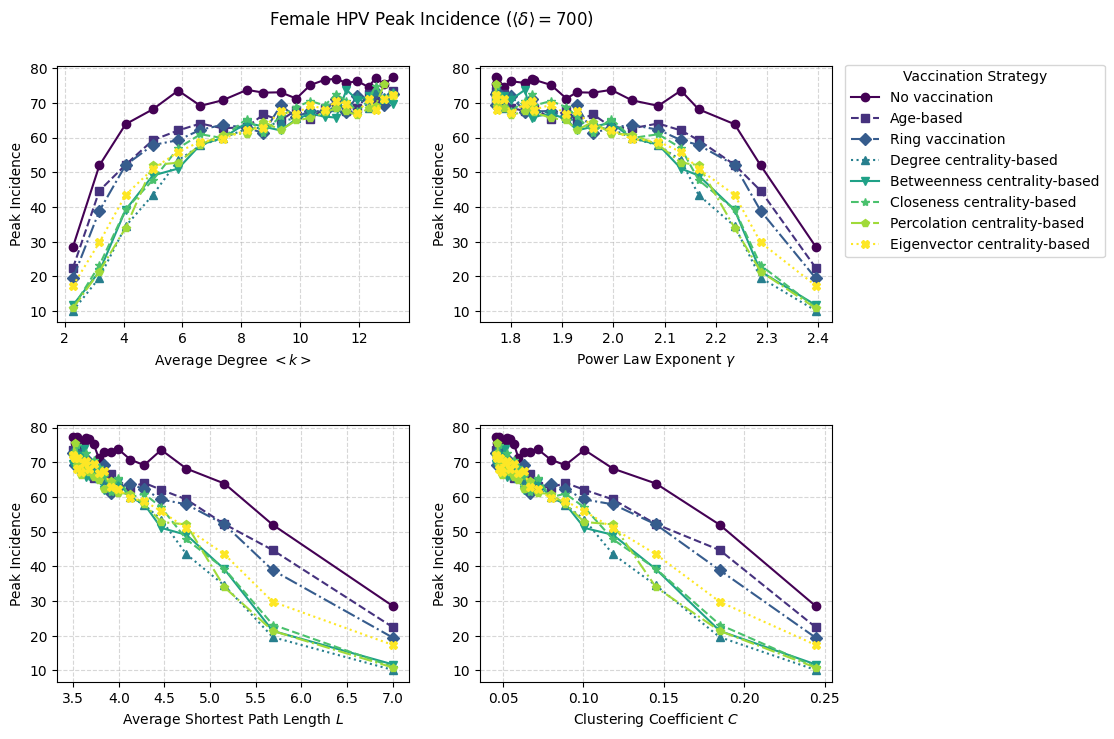}
    \caption{$\langle\delta\rangle = 700$}
  \end{subfigure}
  \caption{Maximum Daily Female HPV Incidence across Network Structures and Vaccination Strategies under Varying Average Relationship Durations ($\langle\delta\rangle$) ($N = 3000$, $\beta = 0.13$, $c = 10\%$, $\langle \eta \rangle = 3.5$) - Percolation centrality-based vaccination plays a key role in curbing peak female cases, with extended relationship duration providing additive reductions in daily transmission.}\label{33}
\end{figure*}

In the female cohort, percolation centrality-based vaccination continues to play a key role in curbing peak daily female cases, with extended partnership durations providing additive reductions in daily transmission rates in Fig. \ref{33}.

\begin{figure*}[htbp]
  \centering
  \begin{subfigure}[b]{0.32\textwidth}
    \centering
    \includegraphics[width=\linewidth]{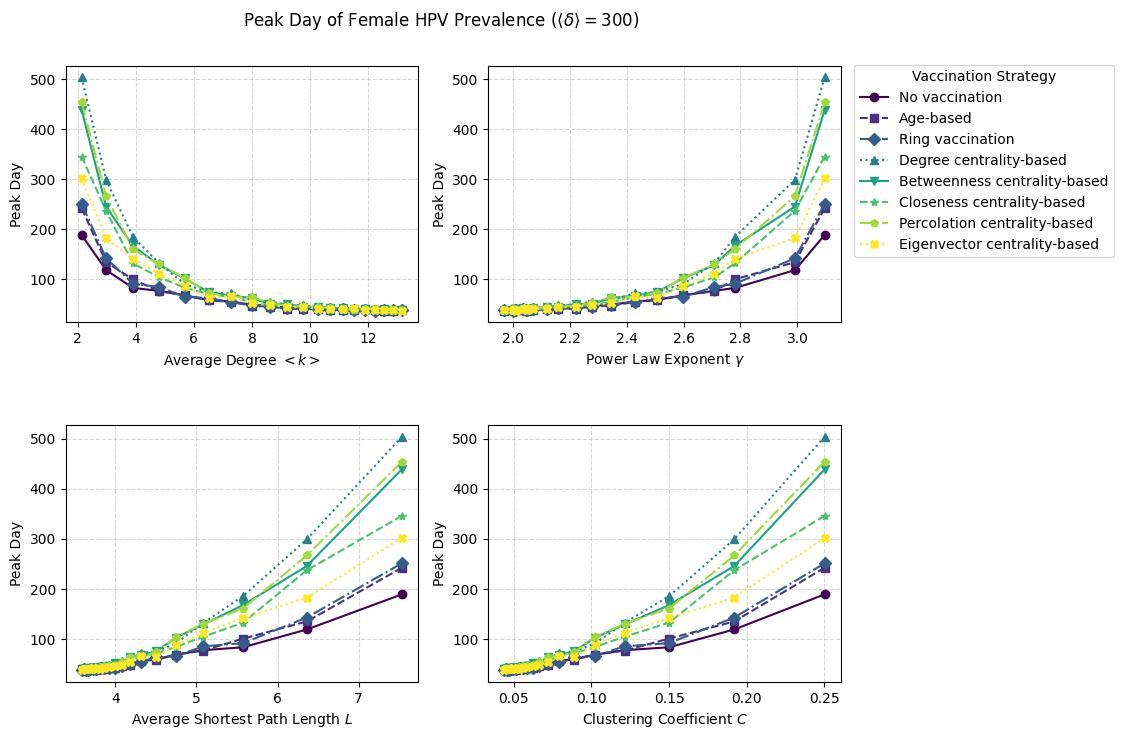}
    \caption{$\langle\delta\rangle = 300$}
  \end{subfigure}\hfill
  \begin{subfigure}[b]{0.32\textwidth}
    \centering
    \includegraphics[width=\linewidth]{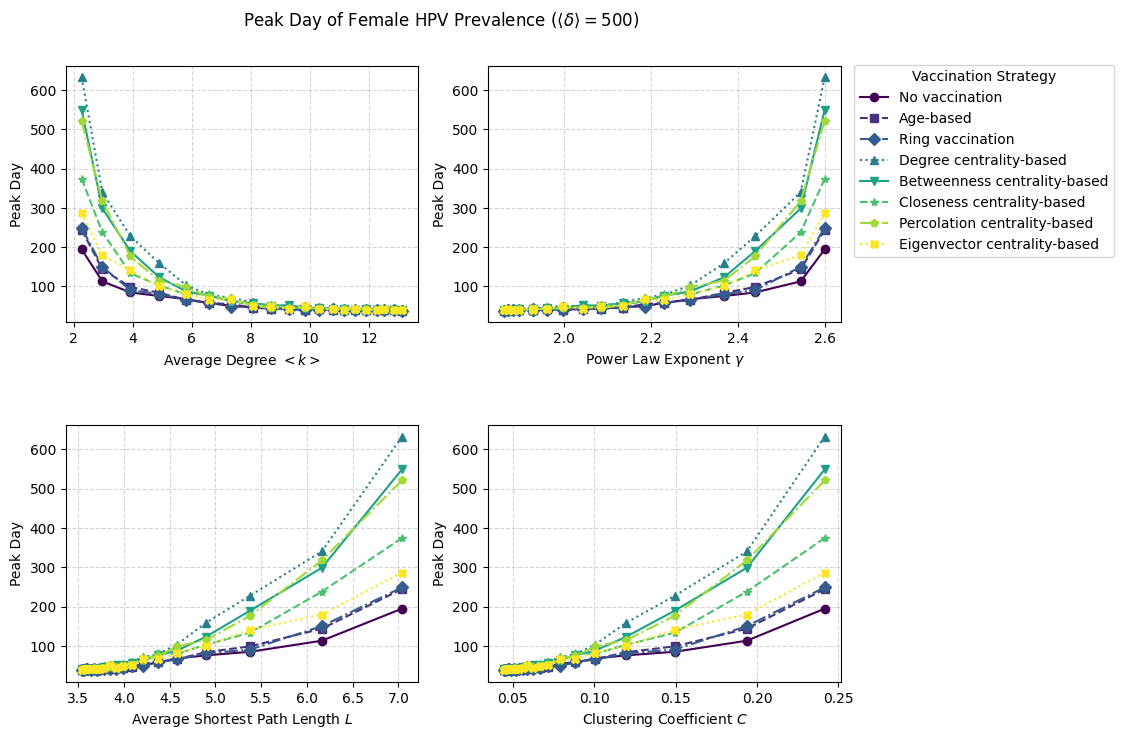}
    \caption{$\langle\delta\rangle = 500$}
  \end{subfigure}\hfill
  \begin{subfigure}[b]{0.32\textwidth}
    \centering
    \includegraphics[width=\linewidth]{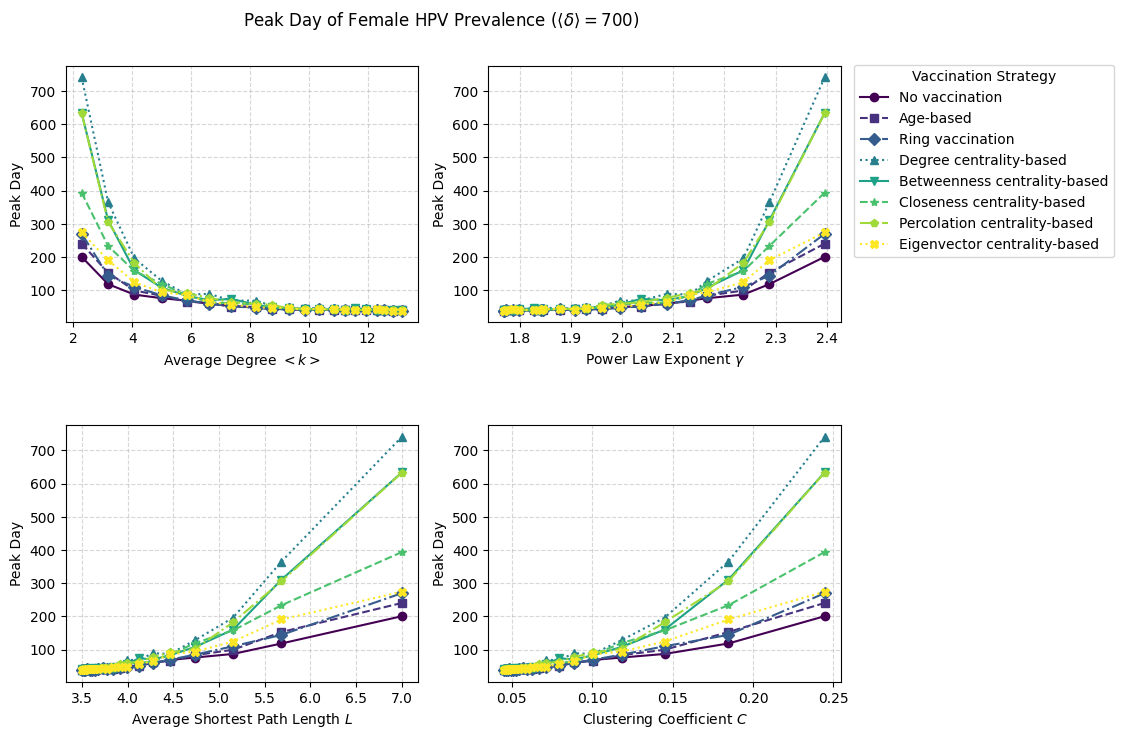}
    \caption{$\langle\delta\rangle = 700$}
  \end{subfigure}
  \caption{Effect of Network Properties and Vaccination Strategies on Peak Day of Female HPV Prevalence under Varying Average Relationship Durations ($\langle\delta\rangle$) ($N = 3000$, $\beta = 0.13$, $c = 10\%$, $\langle \eta \rangle = 3.5$) - Coupled with increased partnership stability, the degree centrality-based strategy maximises the temporal buffer prior to peak female infection prevalence.}\label{34}
\end{figure*}

When paired with increased partnership stability, degree centrality-based vaccination maximises the pre-peak temporal window before female prevalence reaches its maximum in Fig. \ref{34}. 

\begin{figure*}[htbp]
  \centering
  \begin{subfigure}[b]{0.32\textwidth}
    \centering
    \includegraphics[width=\linewidth]{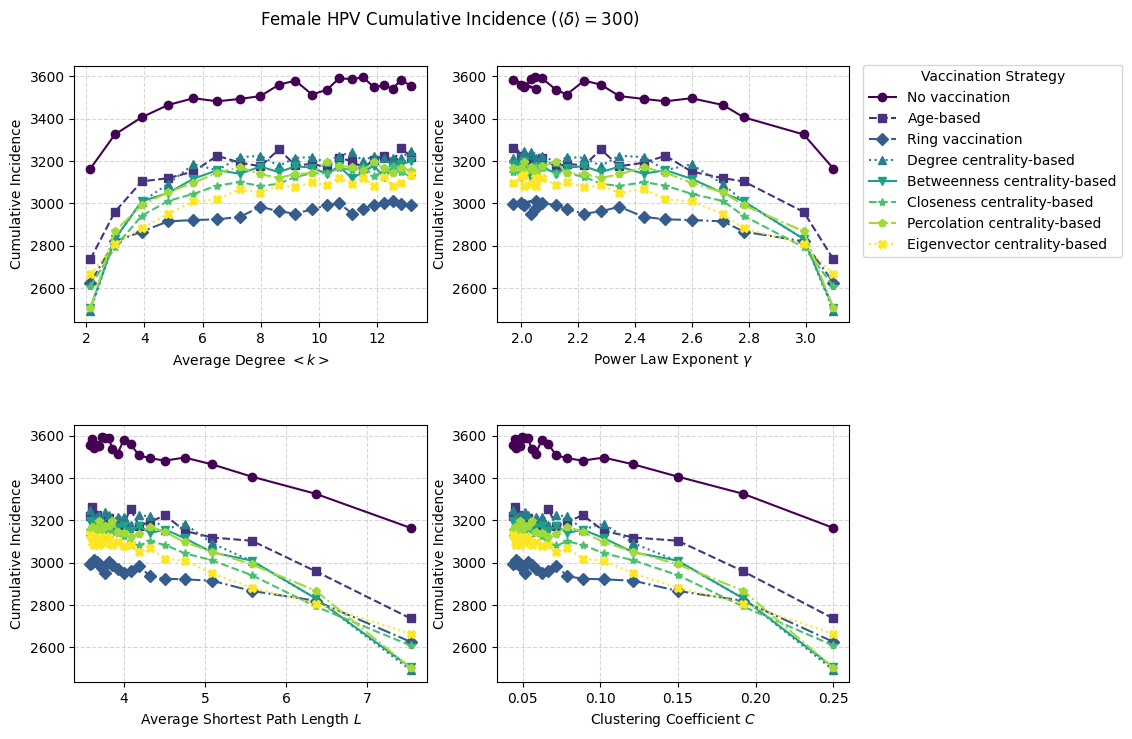}
    \caption{$\langle\delta\rangle = 300$}
  \end{subfigure}\hfill
  \begin{subfigure}[b]{0.32\textwidth}
    \centering
    \includegraphics[width=\linewidth]{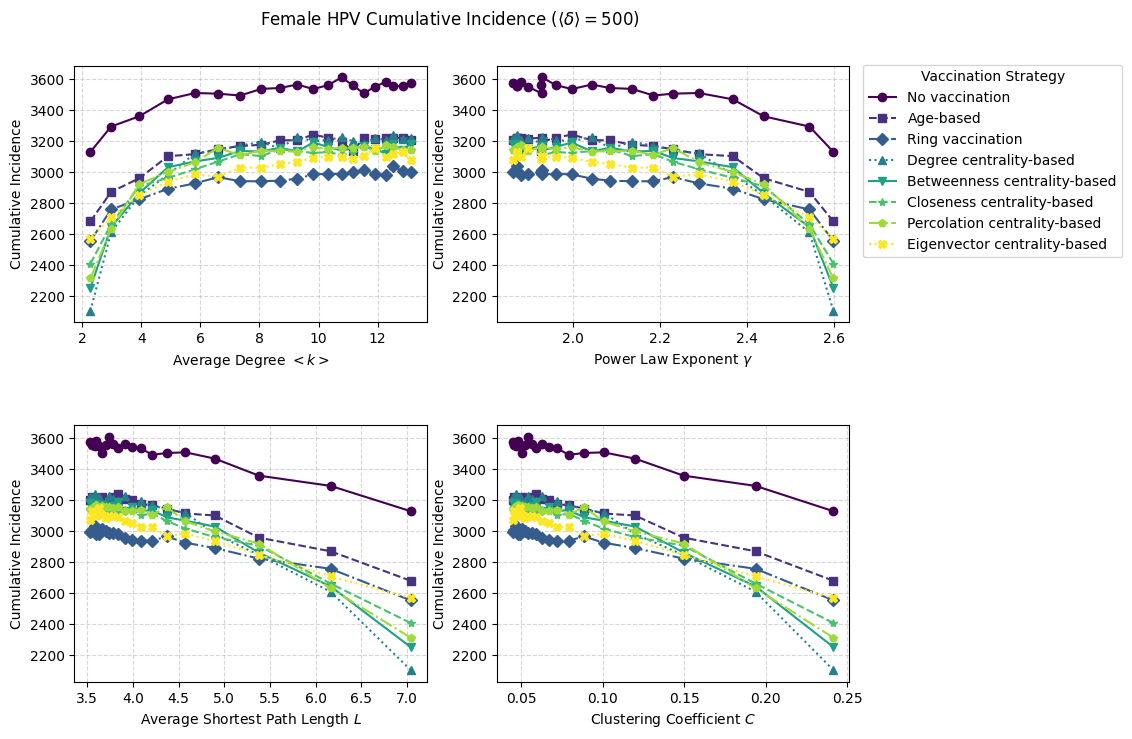}
    \caption{$\langle\delta\rangle = 500$}
  \end{subfigure}\hfill
  \begin{subfigure}[b]{0.32\textwidth}
    \centering
    \includegraphics[width=\linewidth]{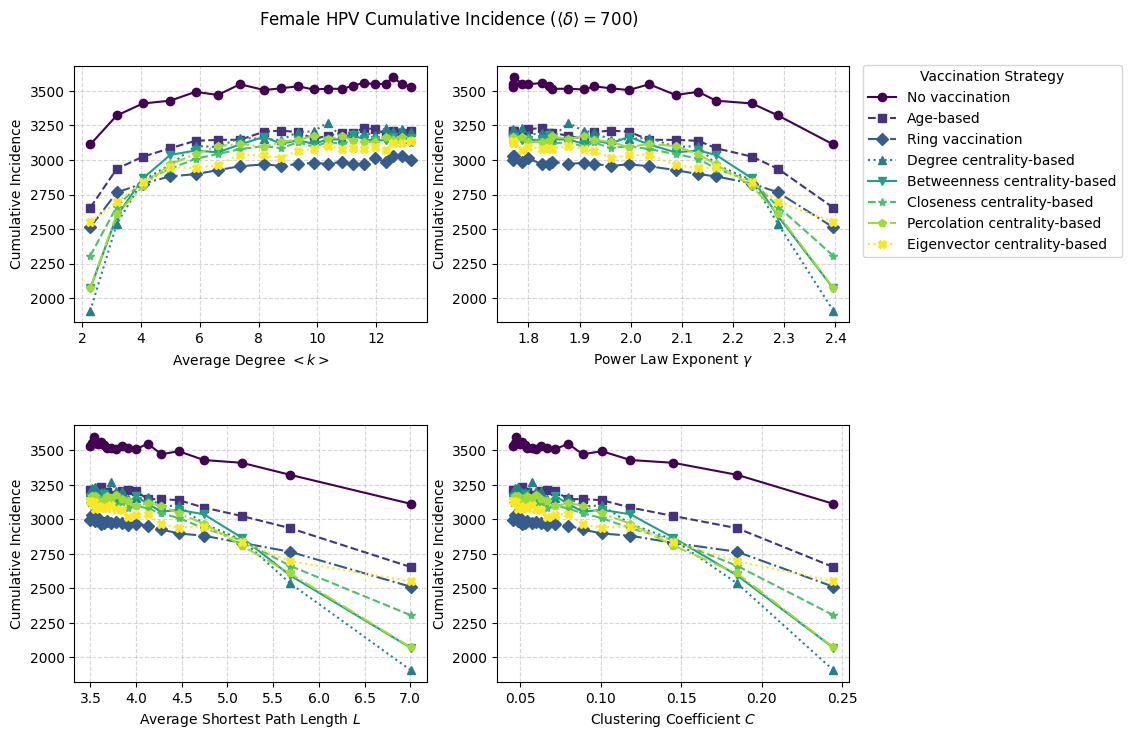}
    \caption{$\langle\delta\rangle = 700$}
  \end{subfigure}
  \caption{Effect of Network Structure and Vaccination Strategy on Female HPV Cumulative Incidence under Varying Average Relationship Durations ($\langle\delta\rangle$) ($N = 3000$, $\beta = 0.13$, $c = 10\%$, $\langle \eta \rangle = 3.5$) - Ring vaccination consistently achieves the lowest cumulative incidence in females; however, its relative advantage over centrality-based schemes shrinks as relationships become more stable and networks become sparse.}\label{35}
\end{figure*}

Crucially, ring vaccination consistently achieves the lowest cumulative incidence within the female cohort across all relationship durations. However, its relative performance advantage over centrality-based schemes gradually narrows as relationships become more stable ($\langle\delta\rangle = 700 \text{ days}$) and networks become sparse in Fig. \ref{35}.

\subsection{Simulation Results for the Male Cohort}

\begin{figure*}[htbp]
  \centering
  \begin{subfigure}[b]{0.32\textwidth}
    \centering
    \includegraphics[width=\linewidth]{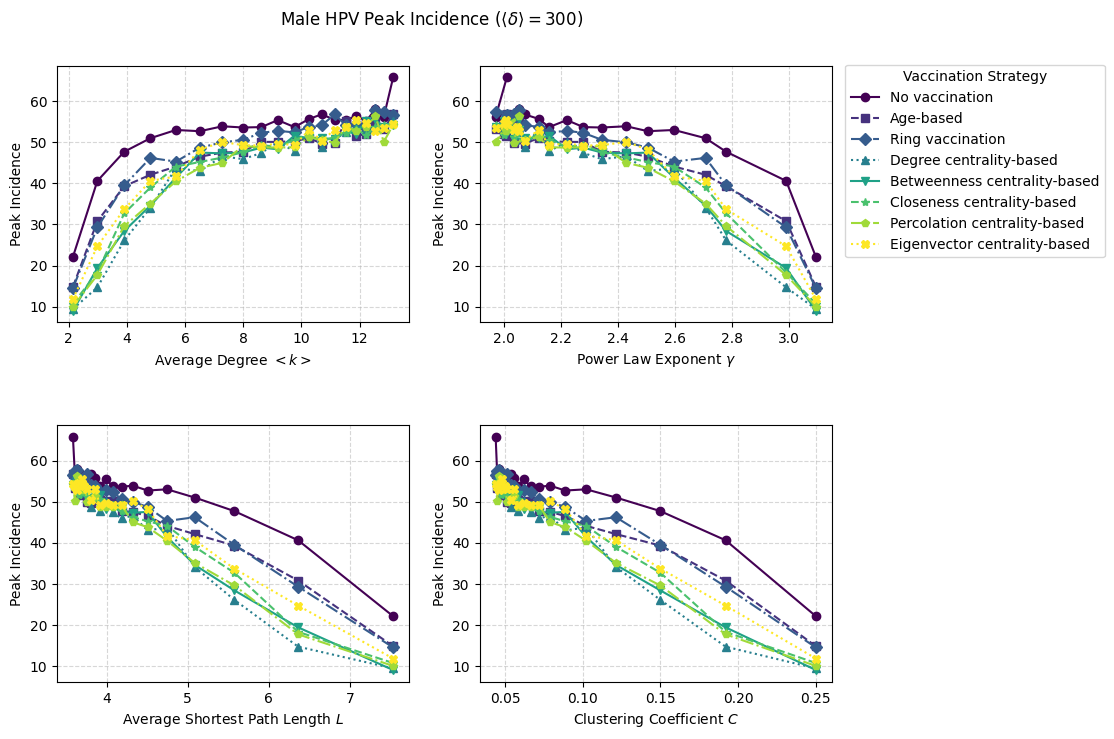}
    \caption{$\langle\delta\rangle = 300$}
  \end{subfigure}\hfill
  \begin{subfigure}[b]{0.32\textwidth}
    \centering
    \includegraphics[width=\linewidth]{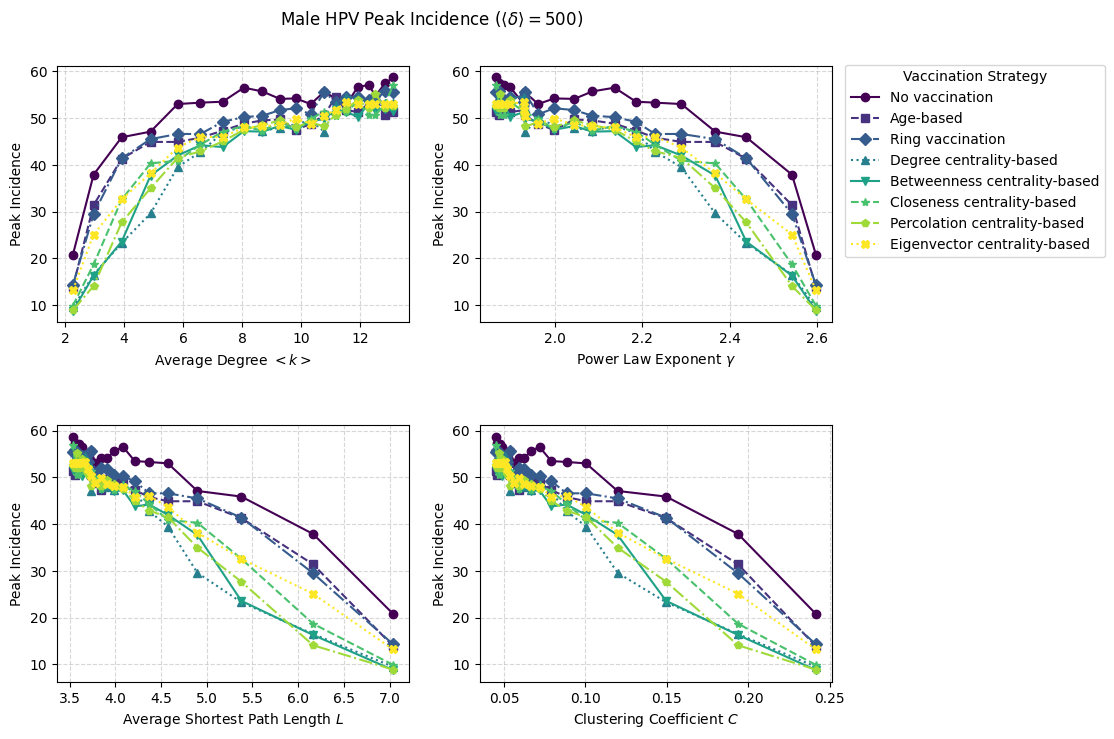}
    \caption{$\langle\delta\rangle = 500$}
  \end{subfigure}\hfill
  \begin{subfigure}[b]{0.32\textwidth}
    \centering
    \includegraphics[width=\linewidth]{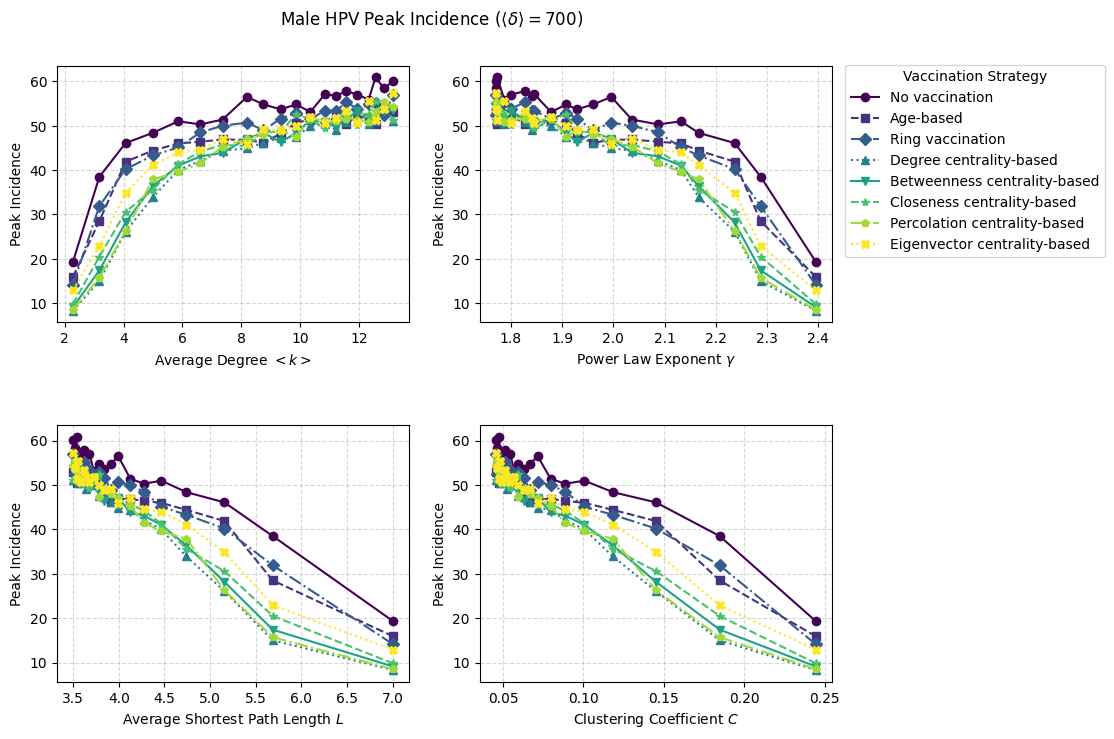}
    \caption{$\langle\delta\rangle = 700$}
  \end{subfigure}
  \caption{Maximum Daily Male HPV Incidence across Network Structures and Vaccination Strategies under Varying Average Relationship Durations ($\langle\delta\rangle$) ($N = 3000$, $\beta = 0.13$, $c = 10\%$, $\langle \eta \rangle = 3.5$) - Degree-, betweenness- and percolation centrality-based strategies effectively control peak male cases, with prolonged relationship duration lowering peak incidence overall.}\label{36}
\end{figure*}

For the male cohort, degree-, betweenness-, and percolation centrality-based strategies effectively suppress peak daily male incidence, while longer average relationship durations reduce overall peak case counts in Fig. \ref{36}.

\begin{figure*}[htbp]
  \centering
  \begin{subfigure}[b]{0.32\textwidth}
    \centering
    \includegraphics[width=\linewidth]{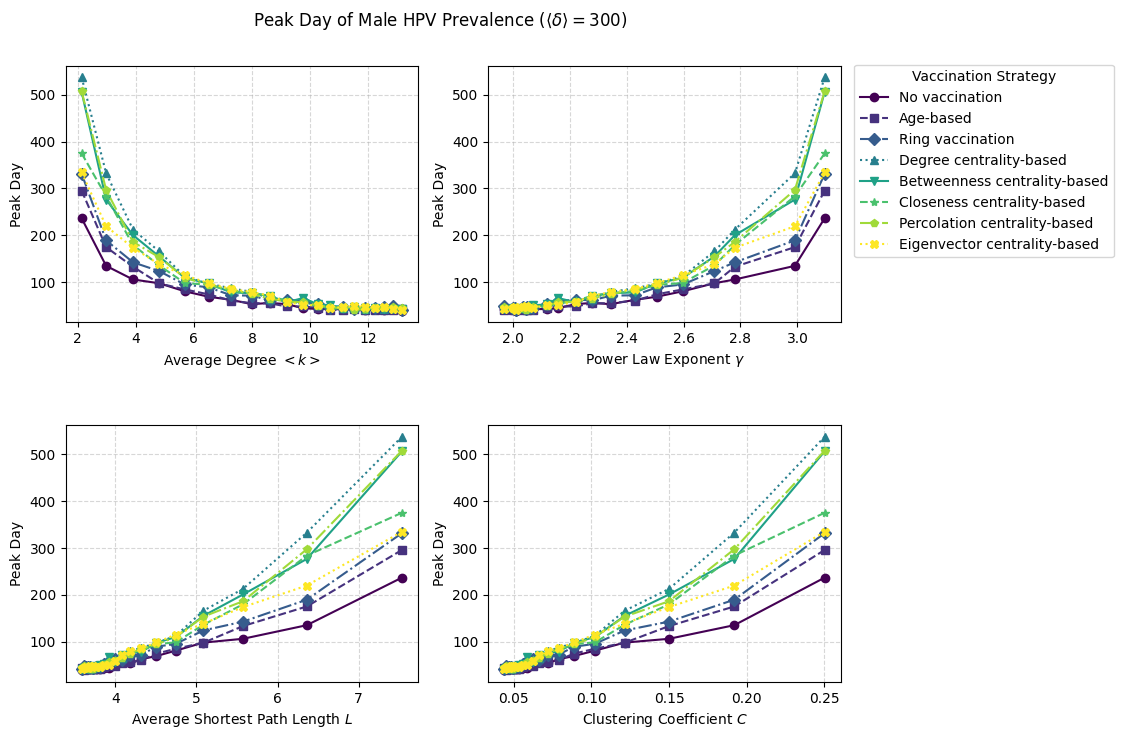}
    \caption{$\langle\delta\rangle = 300$}
  \end{subfigure}\hfill
  \begin{subfigure}[b]{0.32\textwidth}
    \centering
    \includegraphics[width=\linewidth]{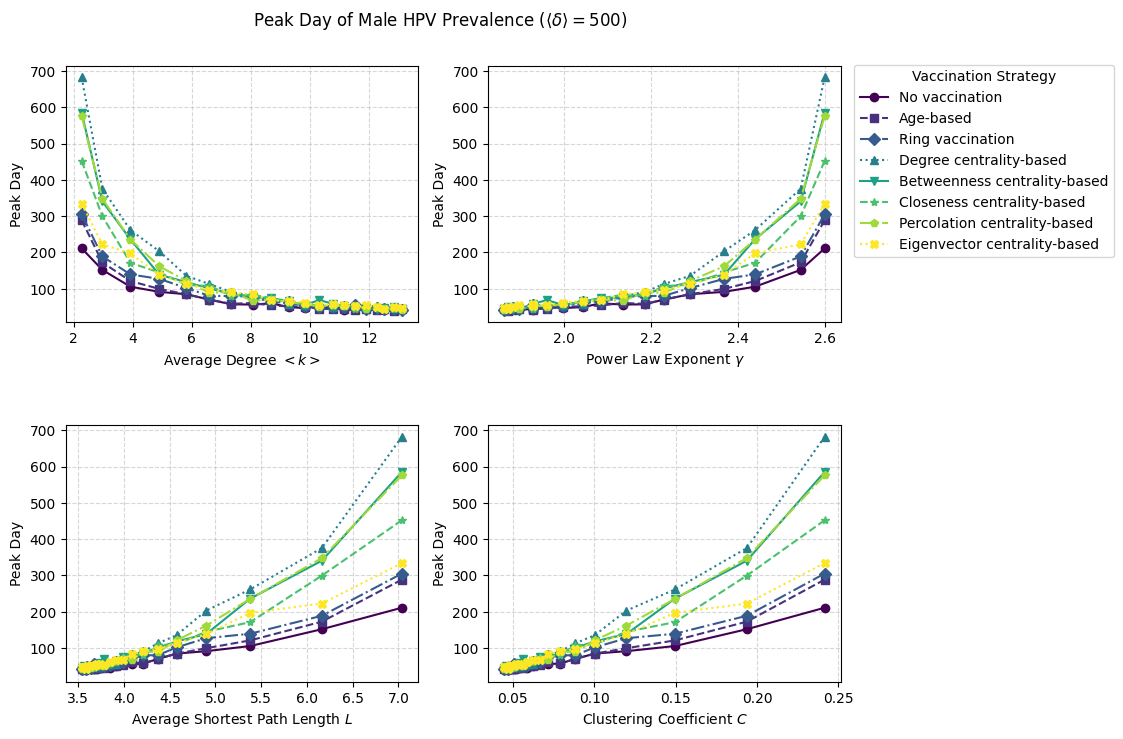}
    \caption{$\langle\delta\rangle = 500$}
  \end{subfigure}\hfill
  \begin{subfigure}[b]{0.32\textwidth}
    \centering
    \includegraphics[width=\linewidth]{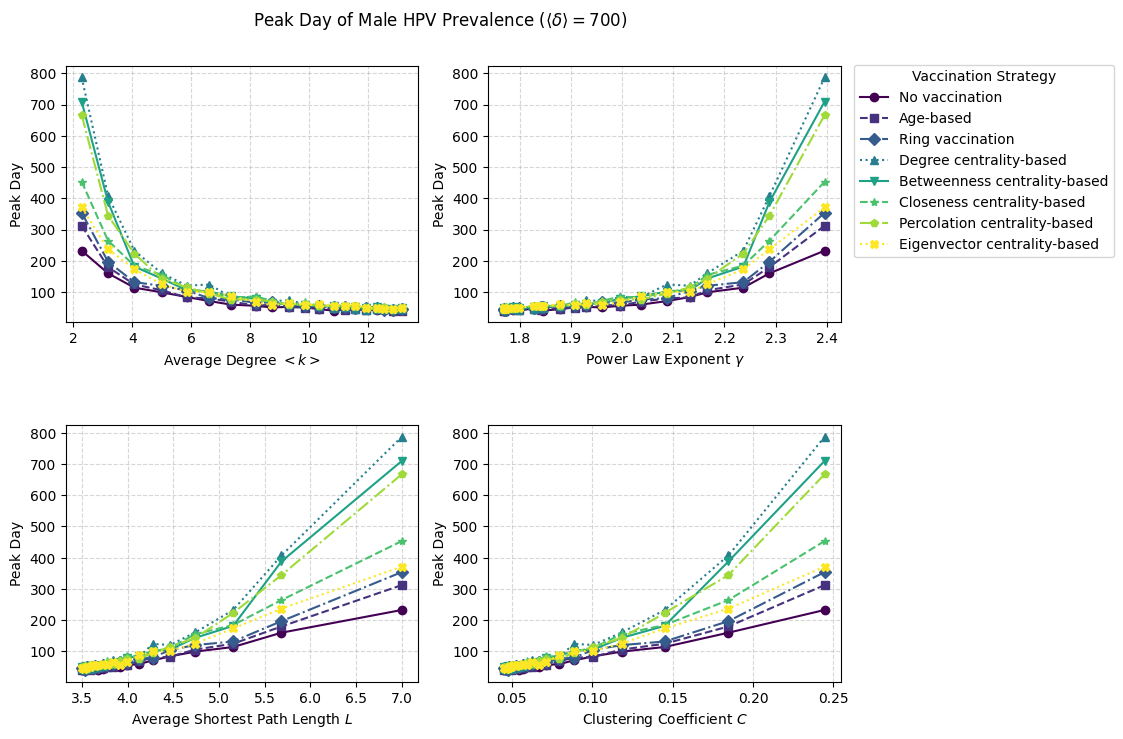}
    \caption{$\langle\delta\rangle = 700$}
  \end{subfigure}
  \caption{Effect of Network Properties and Vaccination Strategies on Peak Day of Male HPV Prevalence under Varying Average Relationship Durations ($\langle\delta\rangle$) ($N = 3000$, $\beta = 0.13$, $c = 10\%$, $\langle \eta \rangle = 3.5$) - Extended partnership durations postpone the peak of male prevalence, widening observable performance gaps among vaccination schemes in low-density networks.}\label{37}
\end{figure*}

Extended partnership durations postpone the peak of male prevalence across all topological configurations, widening observable performance gaps among active vaccination schemes in low-density networks in Fig. \ref{37}. 

\begin{figure*}[htbp]
  \centering
  \begin{subfigure}[b]{0.32\textwidth}
    \centering
    \includegraphics[width=\linewidth]{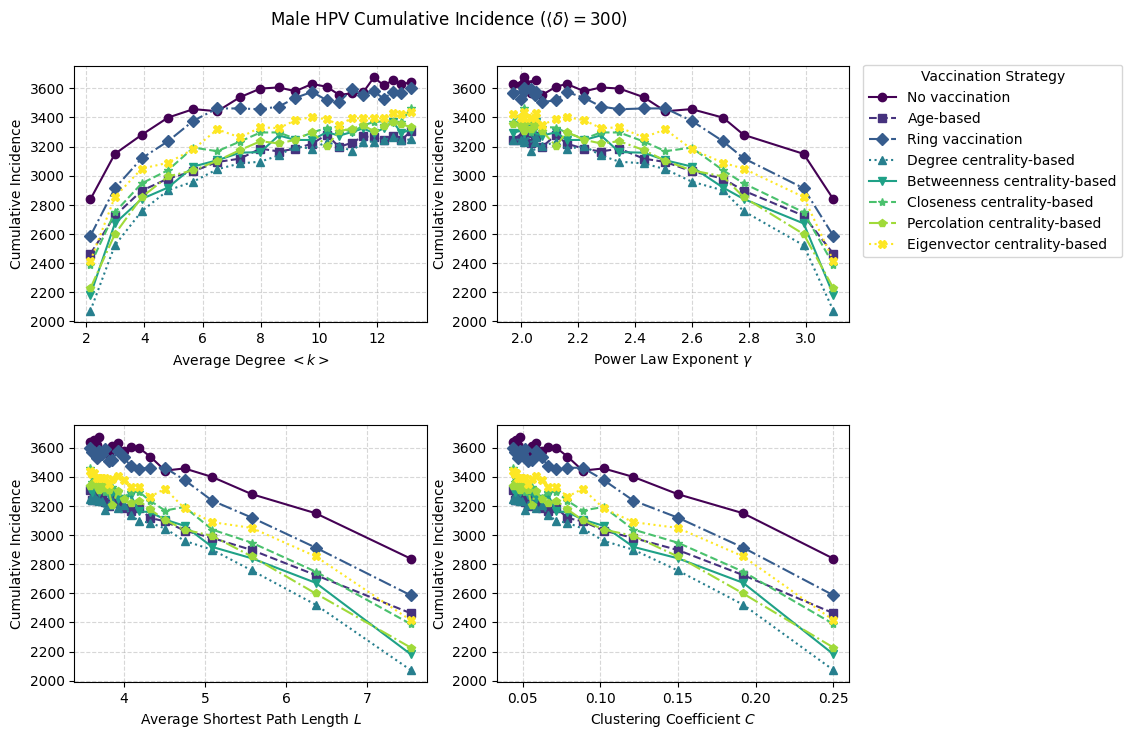}
    \caption{$\langle\delta\rangle = 300$}
  \end{subfigure}\hfill
  \begin{subfigure}[b]{0.32\textwidth}
    \centering
    \includegraphics[width=\linewidth]{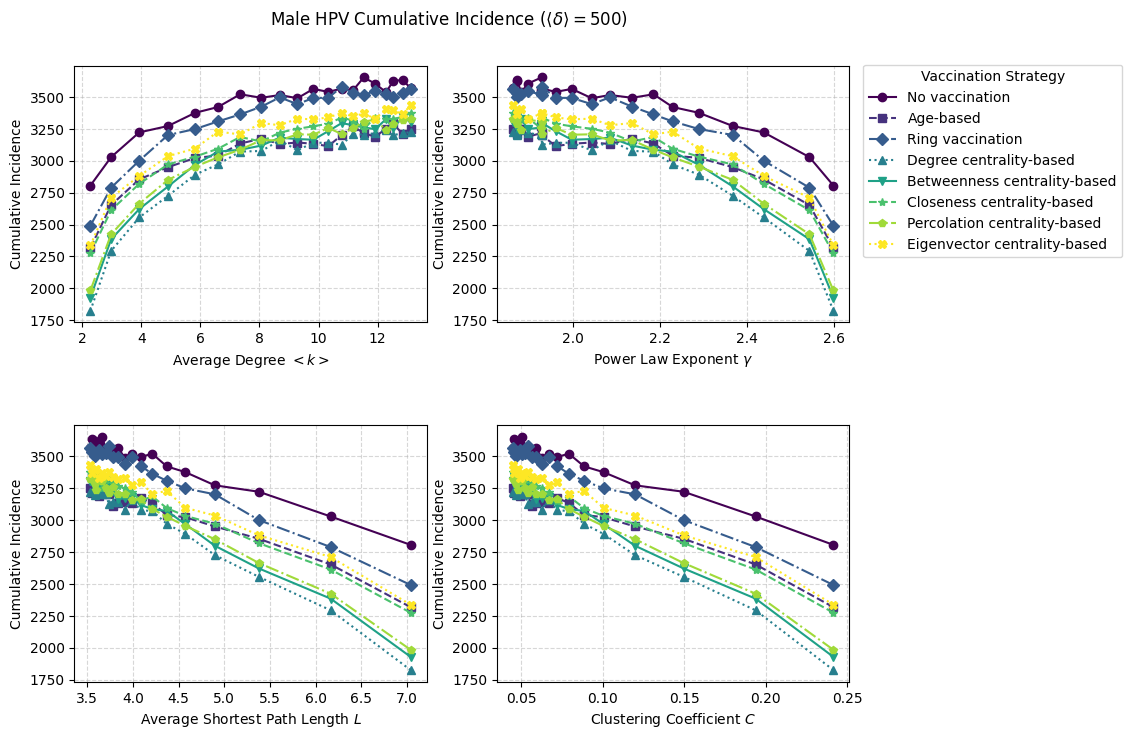}
    \caption{$\langle\delta\rangle = 500$}
  \end{subfigure}\hfill
  \begin{subfigure}[b]{0.32\textwidth}
    \centering
    \includegraphics[width=\linewidth]{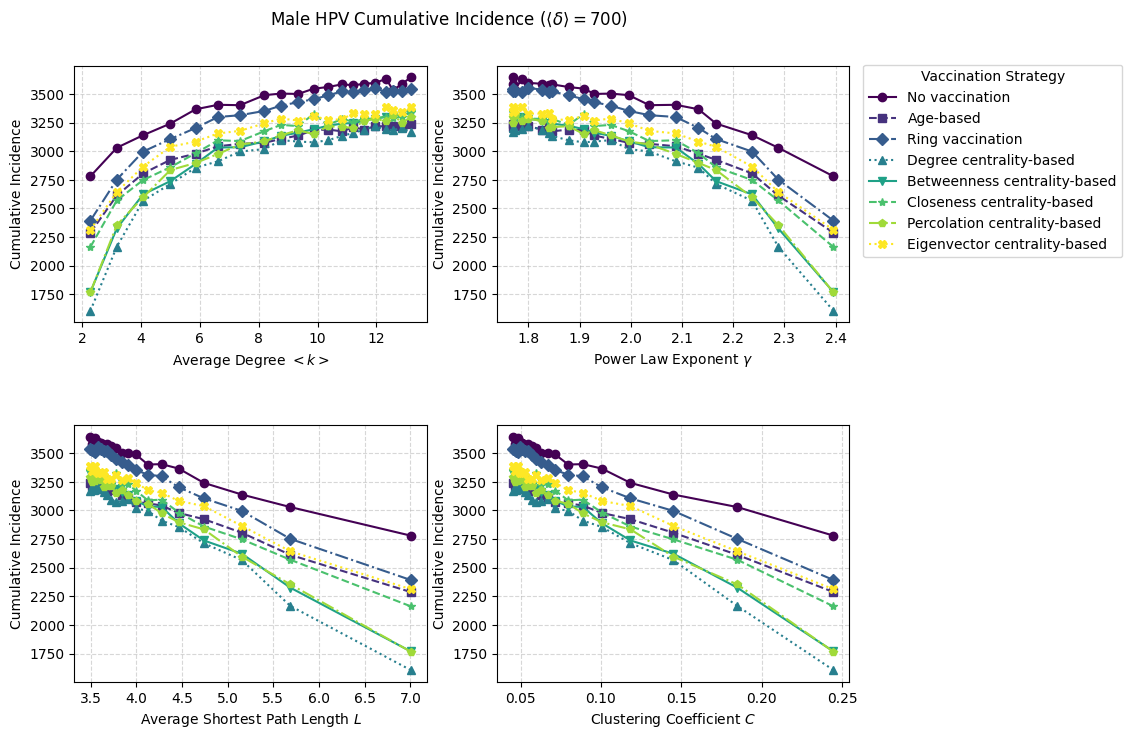}
    \caption{$\langle\delta\rangle = 700$}
  \end{subfigure}
  \caption{Effect of Network Structure and Vaccination Strategy on Male HPV Cumulative Incidence under Varying Average Relationship Durations ($\langle\delta\rangle$) ($N = 3000$, $\beta = 0.13$, $c = 10\%$, $\langle \eta \rangle = 3.5$) - Increased relationship stability markedly reduces male cumulative cases, with degree centrality-based targeting consistently delivering the lowest cumulative burden.}\label{38}
\end{figure*}

Finally, increased relationship stability markedly reduces cumulative male cases, with degree centrality-based targeting consistently delivering the lowest overall cumulative disease burden in Fig. \ref{38}.

\section{Simulation Results with Varying Vaccine Coverage For Individuals under 26}\label{ad}

To evaluate the impact of resource availability and population-level immunity, we conducted sensitivity experiments across varying vaccine coverage levels ($c = 20\%, 40\%, \text{and } 80\%$), compared to the baseline allocation of $c = 10\%$. All other network and epidemic parameters ($N = 3000$, $\beta = 0.13$, $\langle\delta\rangle = 100 \text{ days}$, and $\langle\eta\rangle = 3.5 \text{ years}$) were kept constant. Fig. \ref{39} - \ref{47} present the maximum daily incidence, peak day of prevalence, and cumulative incidence across overall, female, and male cohorts under these expanded coverage scenarios.

\subsection{Simulation Results for the Overall Cohort}

\begin{figure*}[htbp]
  \centering
  \begin{subfigure}[b]{0.32\textwidth}
    \centering
    \includegraphics[width=\linewidth]{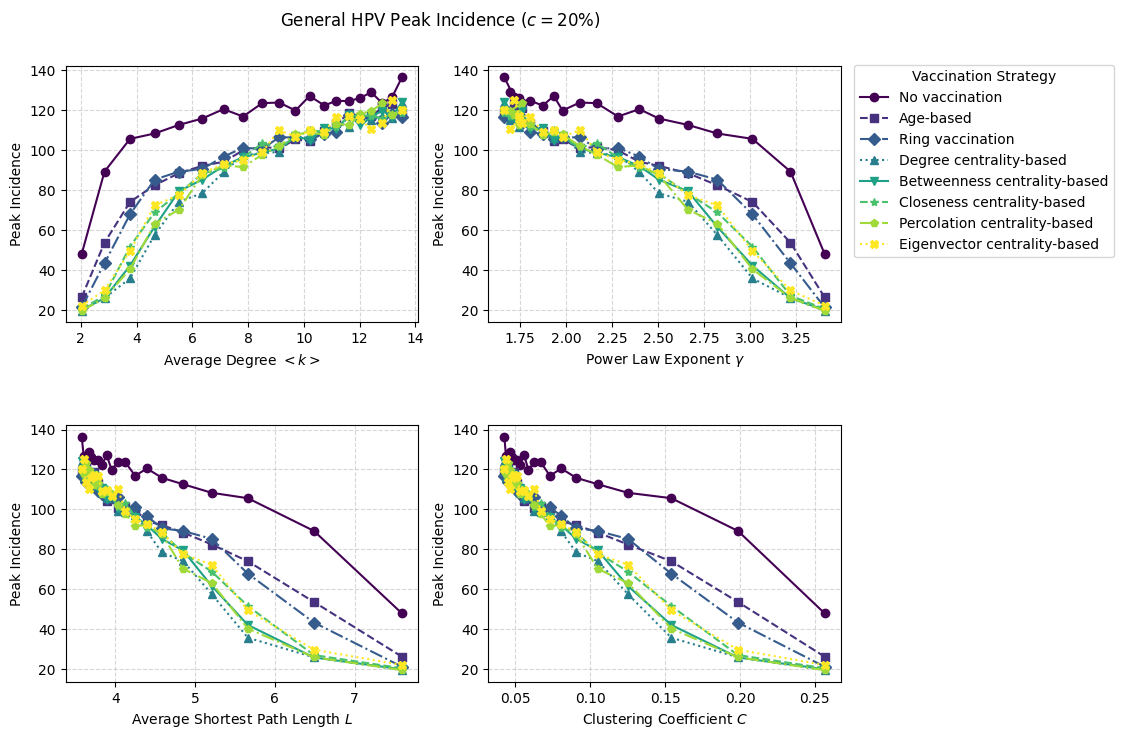}
    \caption{$c = 20\%$}
  \end{subfigure}\hfill
  \begin{subfigure}[b]{0.32\textwidth}
    \centering
    \includegraphics[width=\linewidth]{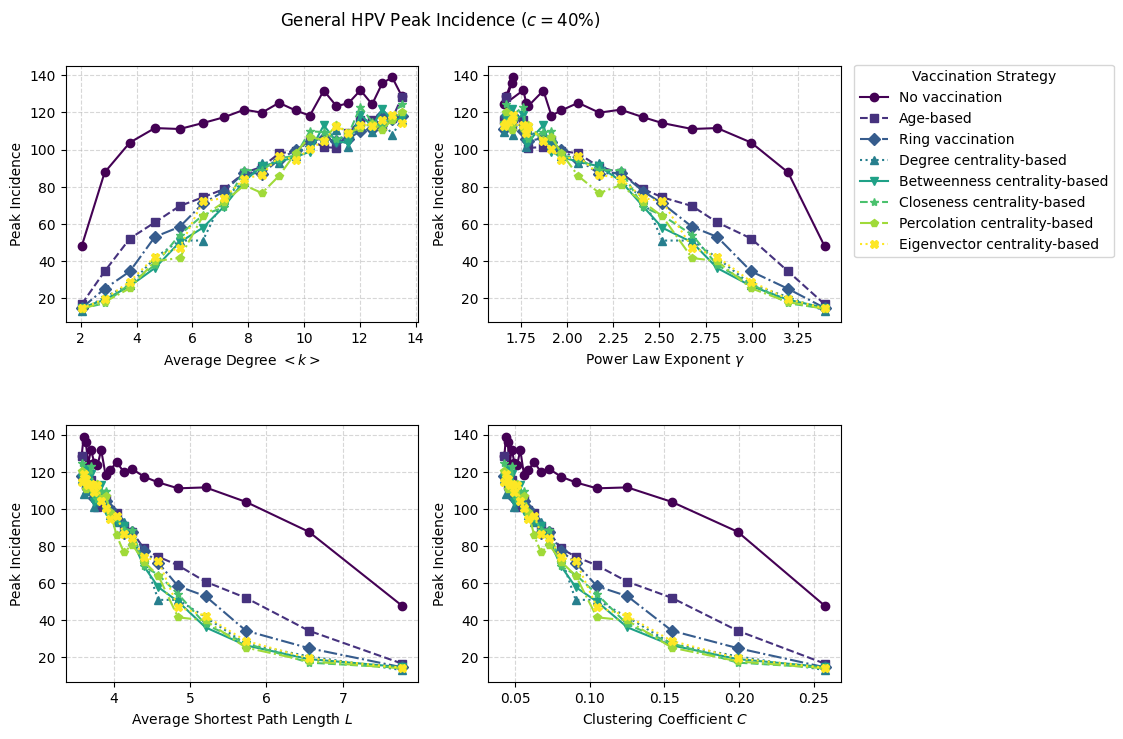}
    \caption{$c = 40\%$}
  \end{subfigure}\hfill
  \begin{subfigure}[b]{0.32\textwidth}
    \centering
    \includegraphics[width=\linewidth]{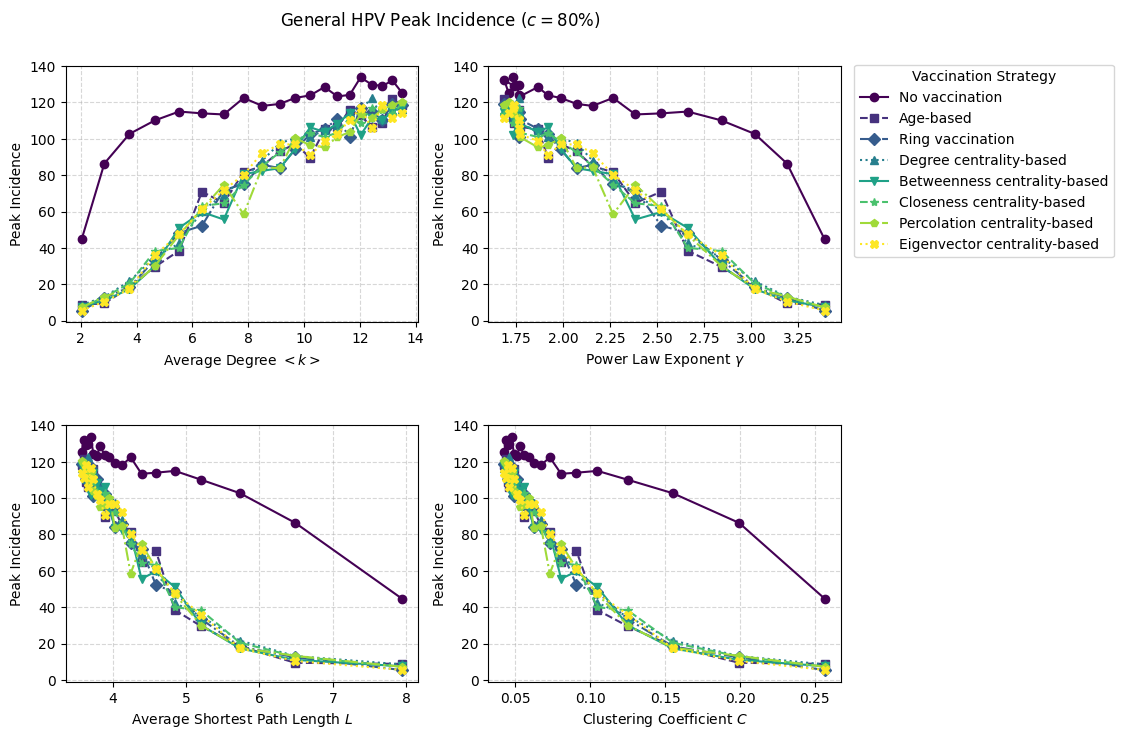}
    \caption{$c = 80\%$}
  \end{subfigure}
  \caption{Maximum Daily General HPV Incidence across Network Structures and Vaccination Strategies under Varying Vaccine Coverage $c$ For Individuals under 26 ($N = 3000$, $\beta = 0.13$, $\langle\delta\rangle = 100$, $\langle \eta \rangle = 3.5$) - Increasing vaccine coverage from 20\% to 80\% progressively suppresses peak general incidence. At 80\% coverage, all active vaccination strategies effectively flatten peak incidence across low and moderate network densities.}\label{39}
\end{figure*}

Increasing vaccine coverage from $20\%$ to $80\%$ progressively suppresses peak general incidence across all network configurations. At sub-saturating coverage levels ($20\%$ and $40\%$), degree-, betweenness-, and percolation centrality-based strategies maintain a distinct performance advantage in flattening the epidemic curve. However, at $80\%$ coverage, population-level herd immunity predominates, uniformly suppressing peak incidence across sparse and moderately connected networks in Fig. \ref{39}.

\begin{figure*}[htbp]
  \centering
  \begin{subfigure}[b]{0.32\textwidth}
    \centering
    \includegraphics[width=\linewidth]{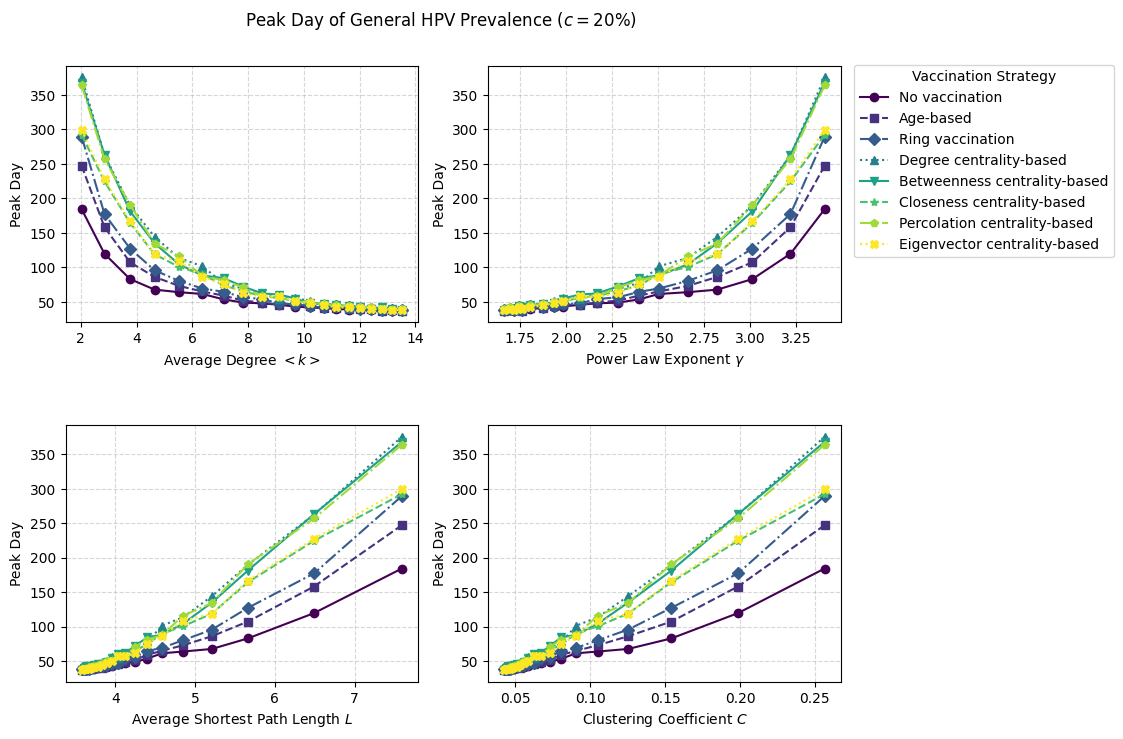}
    \caption{$c = 20\%$}
  \end{subfigure}\hfill
  \begin{subfigure}[b]{0.32\textwidth}
    \centering
    \includegraphics[width=\linewidth]{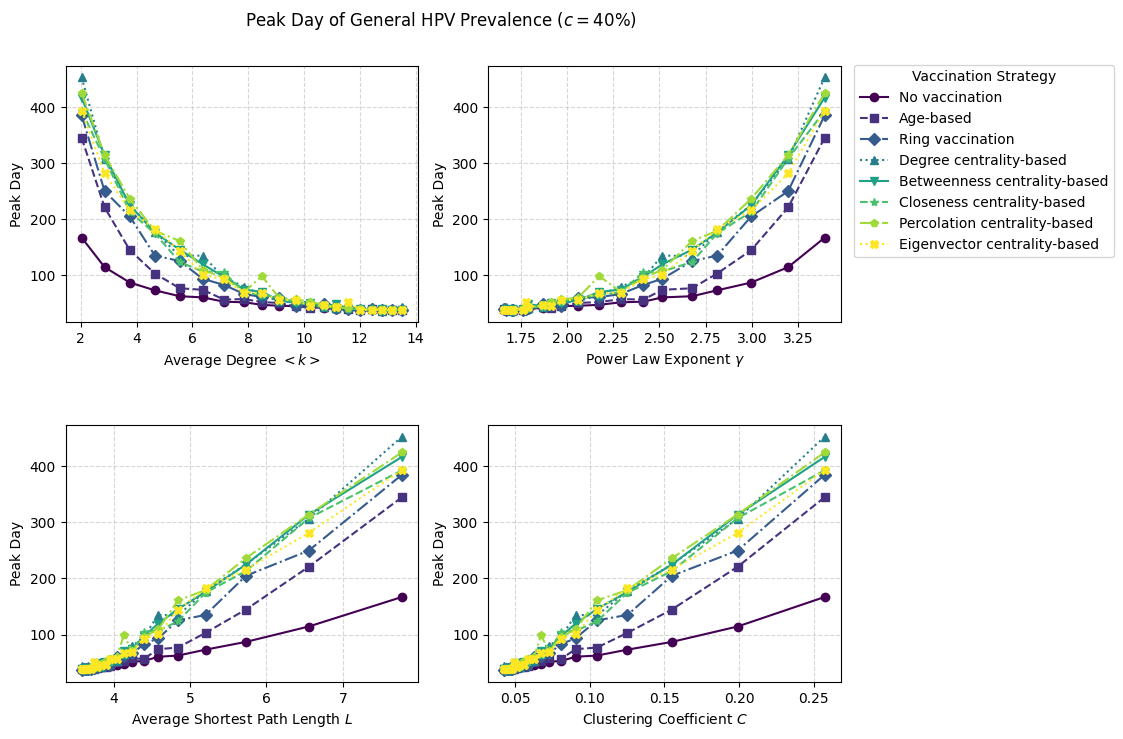}
    \caption{$c =40\%$}
  \end{subfigure}\hfill
  \begin{subfigure}[b]{0.32\textwidth}
    \centering
    \includegraphics[width=\linewidth]{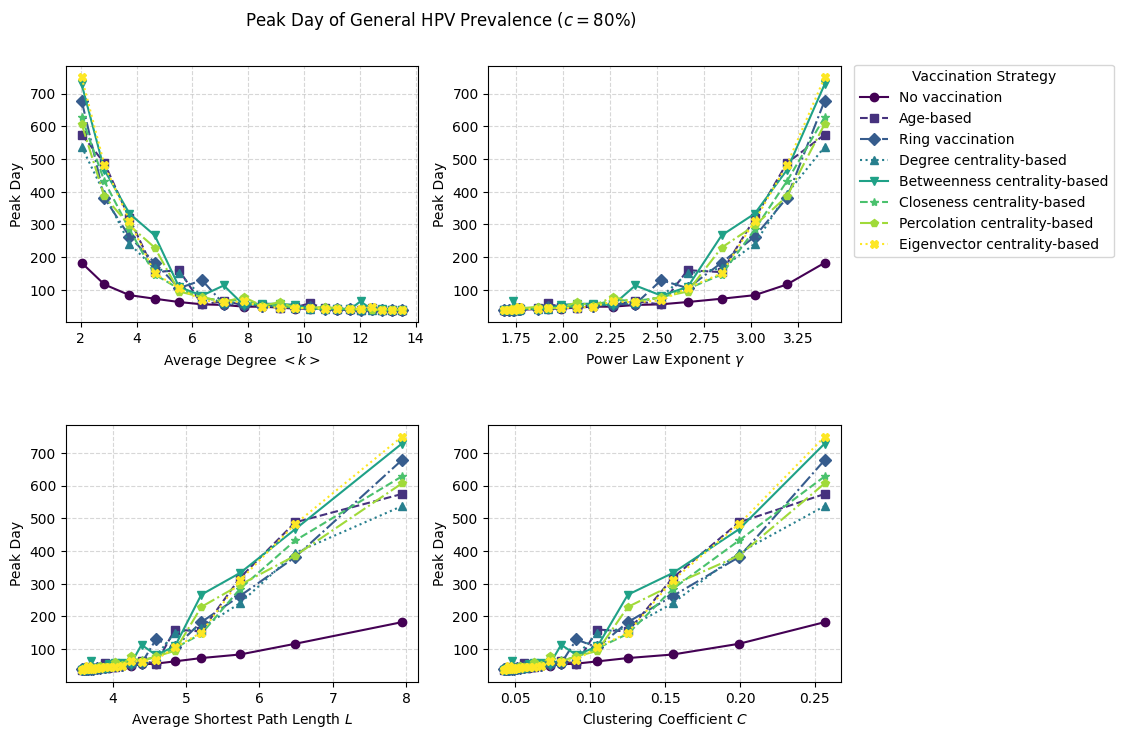}
    \caption{$c =80\%$}
  \end{subfigure}
  \caption{Effect of Network Properties and Vaccination Strategies on Peak Day of General HPV Prevalence under Varying Vaccine Coverage $c$ For Individuals under 26 ($N = 3000$, $\beta = 0.13$, $\langle\delta\rangle = 100$, $\langle \eta \rangle = 3.5$) - Higher vaccine coverage substantially delays peak general prevalence. At 80\% coverage, the epidemic peak is effectively postponed or prevented across sparse network configurations.}\label{40}
\end{figure*}

Higher vaccine coverage substantially extends the temporal window prior to peak general prevalence. At $80\%$ coverage, the epidemic peak is effectively postponed or prevented across sparse network configurations in Fig. \ref{40}. 

\begin{figure*}[htbp]
  \centering
  \begin{subfigure}[b]{0.32\textwidth}
    \centering
    \includegraphics[width=\linewidth]{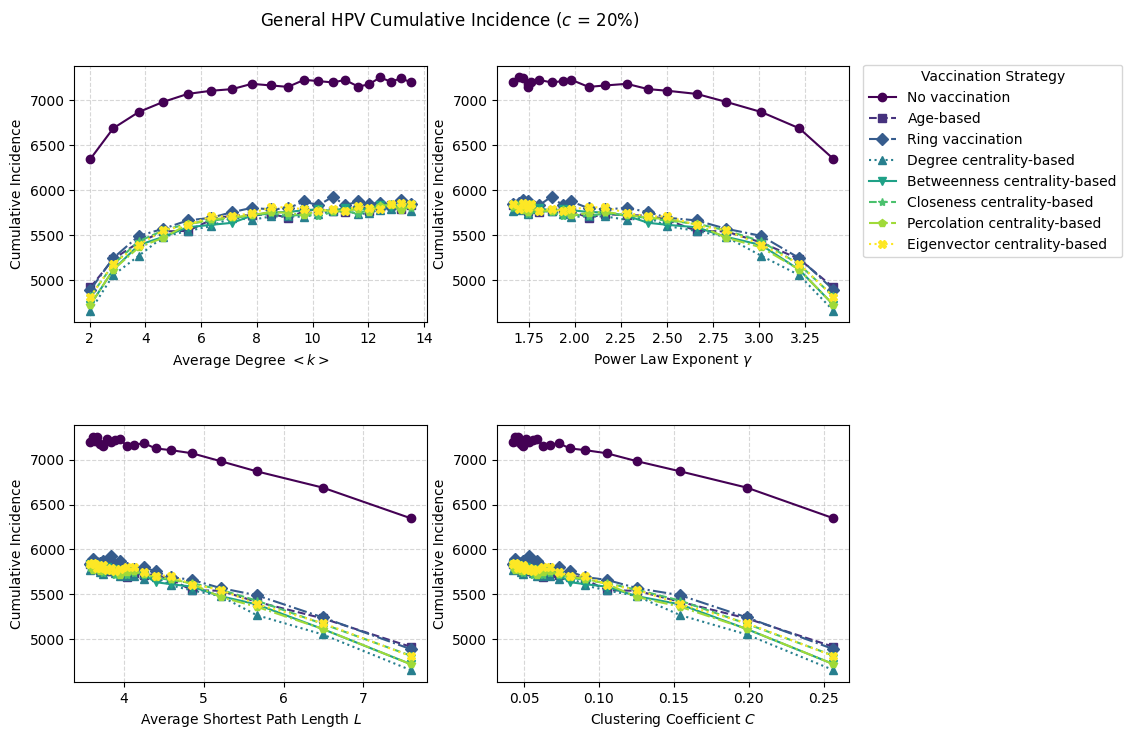}
    \caption{$c =20\%$}
  \end{subfigure}\hfill
  \begin{subfigure}[b]{0.32\textwidth}
    \centering
    \includegraphics[width=\linewidth]{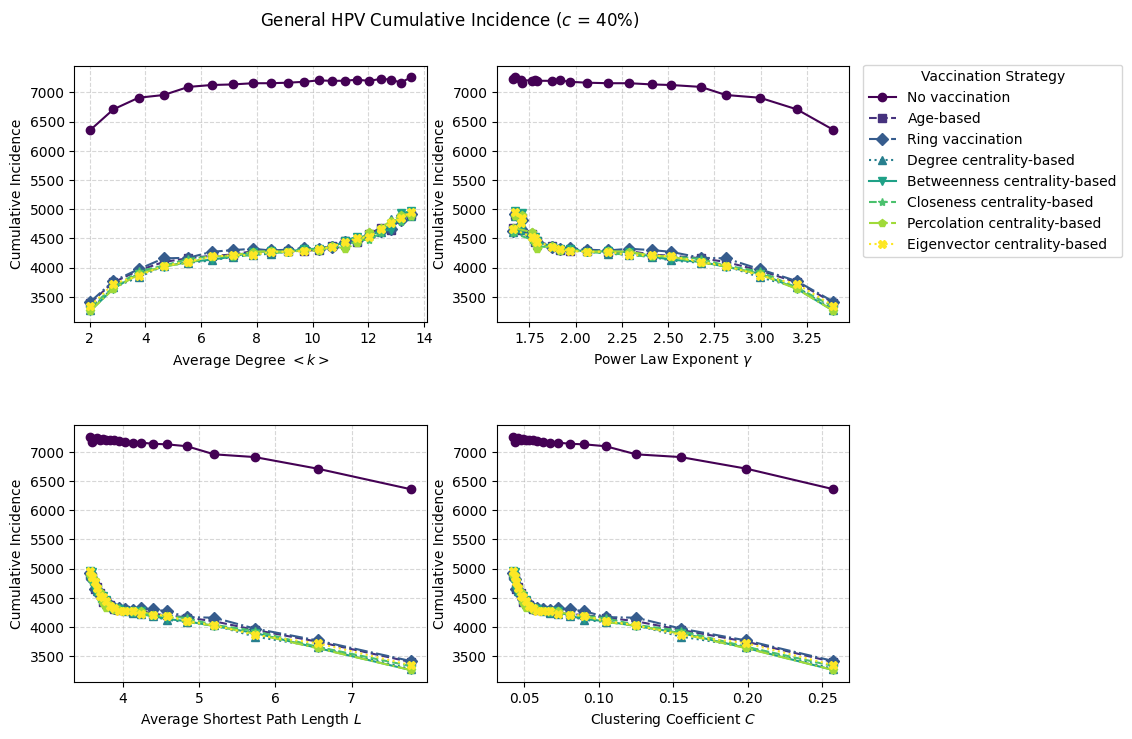}
    \caption{$c = 40\%$}
  \end{subfigure}\hfill
  \begin{subfigure}[b]{0.32\textwidth}
    \centering
    \includegraphics[width=\linewidth]{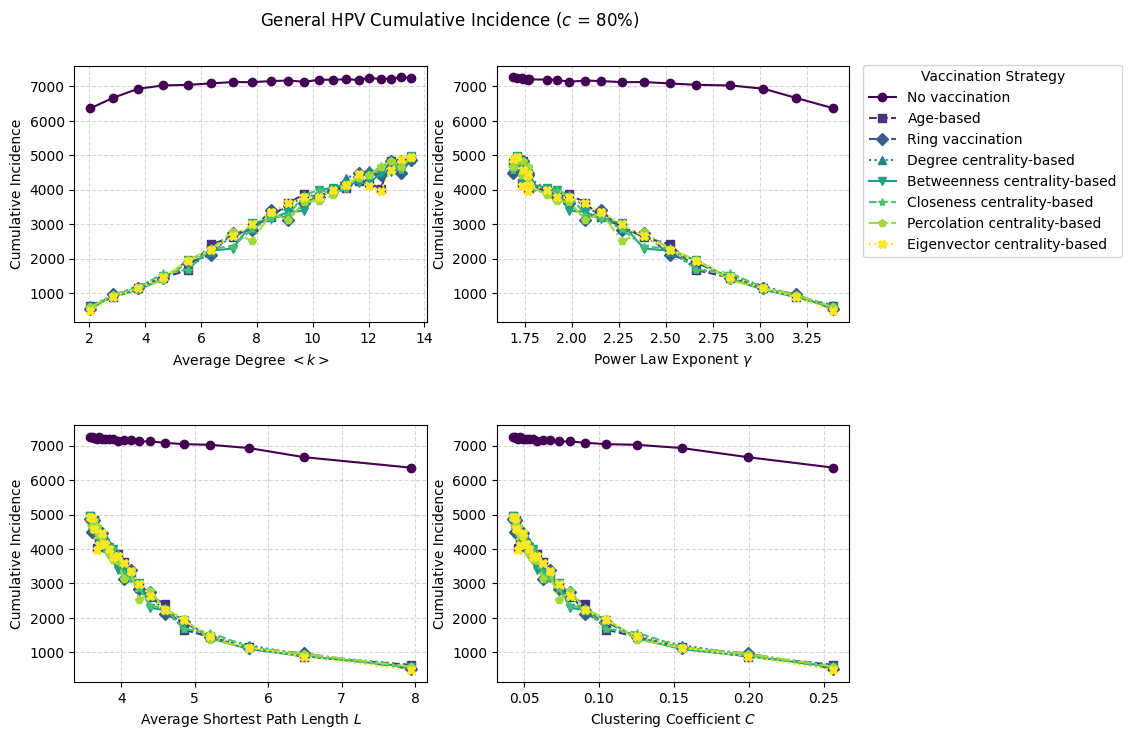}
    \caption{$c = 80\%$}
  \end{subfigure}
  \caption{Effect of Network Structure and Vaccination Strategy on General HPV Cumulative Incidence under Varying Vaccine Coverage $c$ For Individuals under 26 ($N = 3000$, $\beta = 0.13$, $\langle\delta\rangle = 100$, $\langle \eta \rangle = 3.5$) - Cumulative disease burden drops sharply as coverage increases. At 80\% coverage, performance distinctions among individual vaccination strategies vanish, as population-level immunity predominates.}\label{41}
\end{figure*}

Furthermore, cumulative general disease burden drops sharply as coverage increases; at $80\%$ coverage, strategy-specific performance distinctions vanish as high vaccination coverage suppresses overall population transmission in Fig. \ref{41}.

\subsection{Simulation Results for the Female Cohort}

\begin{figure*}[htbp]
  \centering
  \begin{subfigure}[b]{0.32\textwidth}
    \centering
    \includegraphics[width=\linewidth]{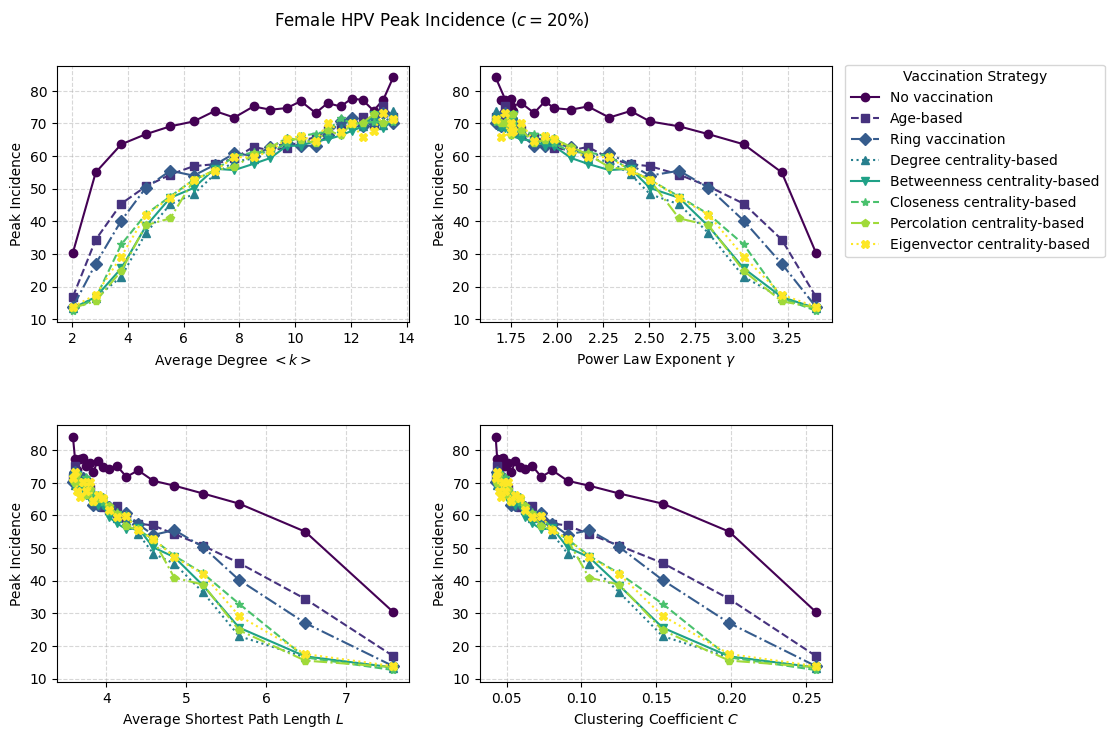}
    \caption{$c = 20\%$}
  \end{subfigure}\hfill
  \begin{subfigure}[b]{0.32\textwidth}
    \centering
    \includegraphics[width=\linewidth]{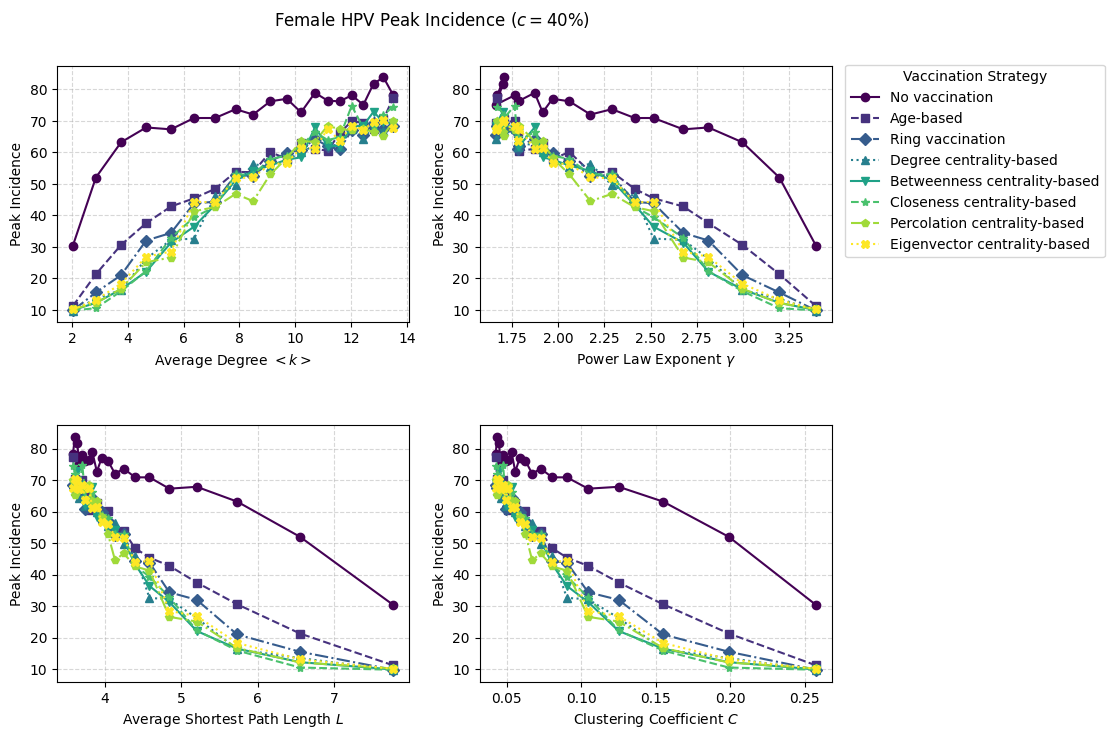}
    \caption{$c = 40\%$}
  \end{subfigure}\hfill
  \begin{subfigure}[b]{0.32\textwidth}
    \centering
    \includegraphics[width=\linewidth]{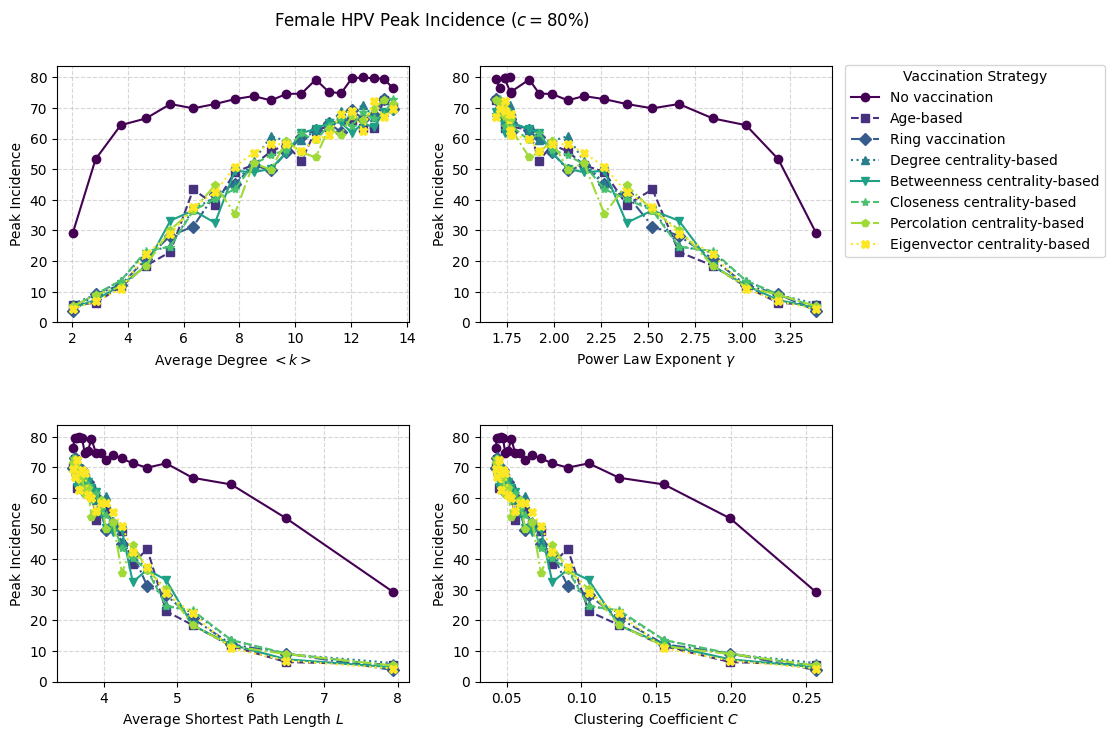}
    \caption{$c = 80\%$}
  \end{subfigure}
  \caption{Maximum Female HPV Incidence across Network Structures and Vaccination Strategies under Varying Vaccine Coverage $c$ For Individuals under 26 ($N = 3000$, $\beta = 0.13$, $\langle\delta\rangle = 100$, $\langle \eta \rangle = 3.5$) - Higher vaccine coverage markedly suppresses female peak incidence. Percolation- and degree centrality-based strategies provide superior control at lower coverage levels (20–40\%), whereas high coverage (80\%) uniformly suppresses peak female cases.}\label{42}
\end{figure*}

In the female cohort, higher vaccine coverage markedly suppresses peak daily incidence. Percolation centrality-based and degree centrality-based strategies provide superior control at lower coverage levels ($20\%–40\%$), whereas high coverage ($80\%$) uniformly suppresses peak female cases across all topologies in Fig. \ref{42}.

\begin{figure*}[htbp]
  \centering
  \begin{subfigure}[b]{0.32\textwidth}
    \centering
    \includegraphics[width=\linewidth]{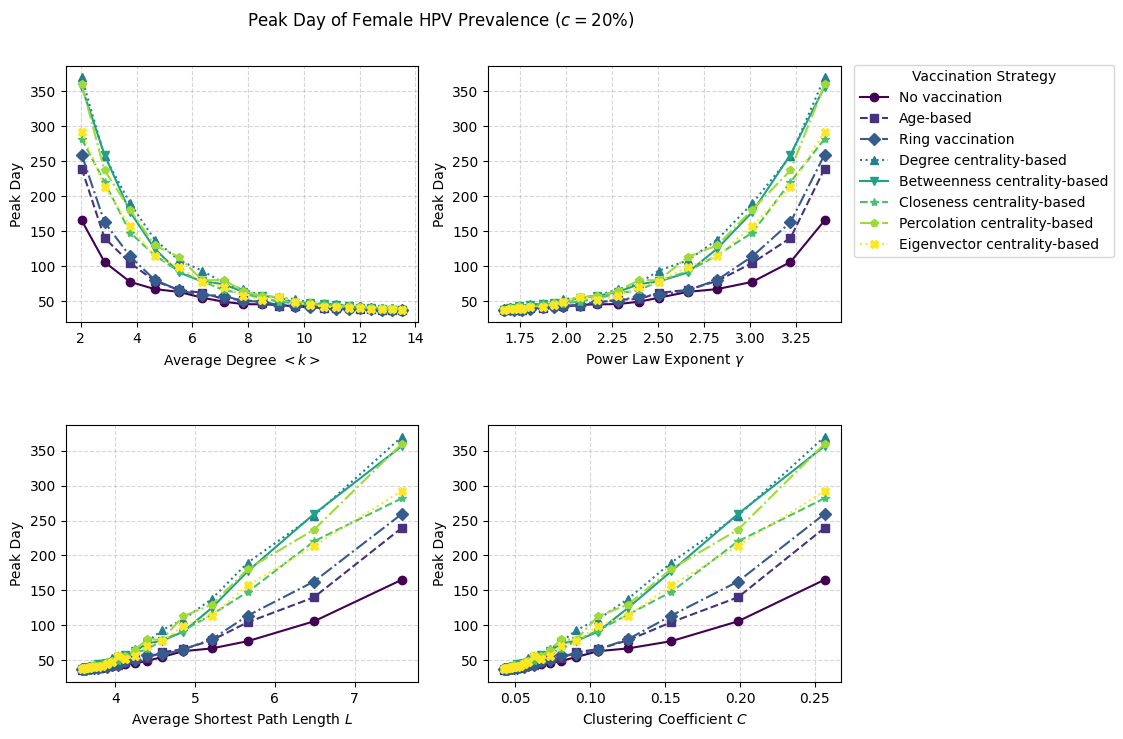}
    \caption{$c =20\%$}
  \end{subfigure}\hfill
  \begin{subfigure}[b]{0.32\textwidth}
    \centering
    \includegraphics[width=\linewidth]{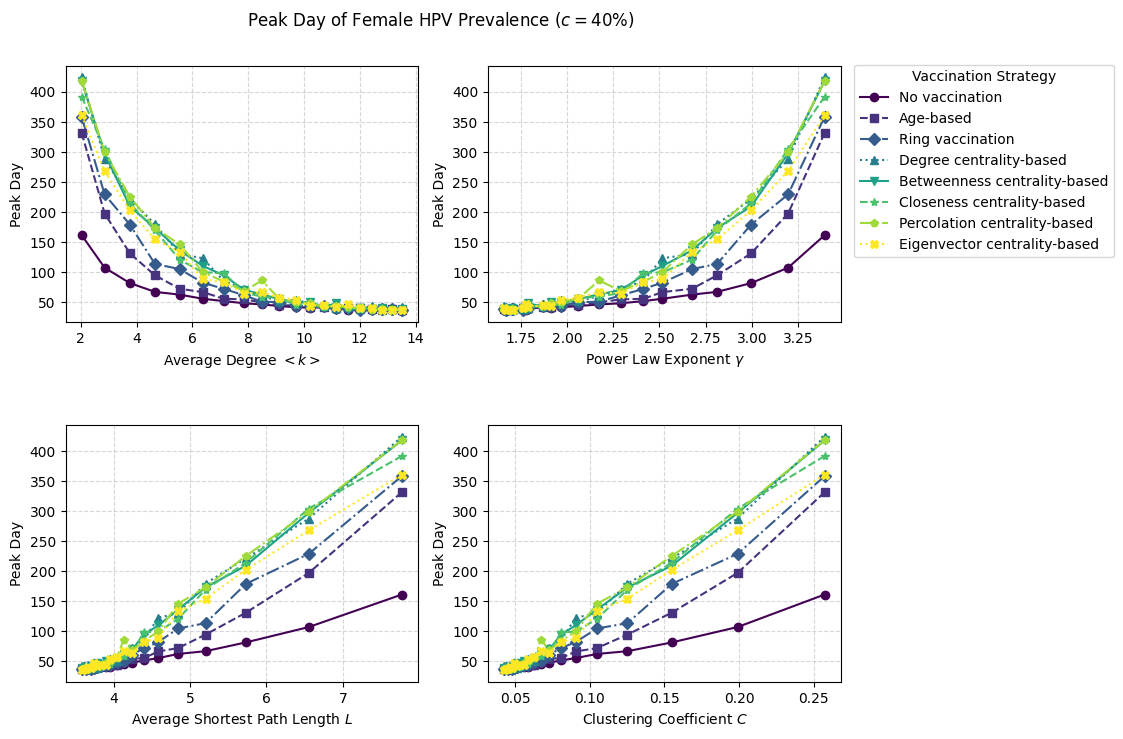}
    \caption{$c =  40\%$}
  \end{subfigure}\hfill
  \begin{subfigure}[b]{0.32\textwidth}
    \centering
    \includegraphics[width=\linewidth]{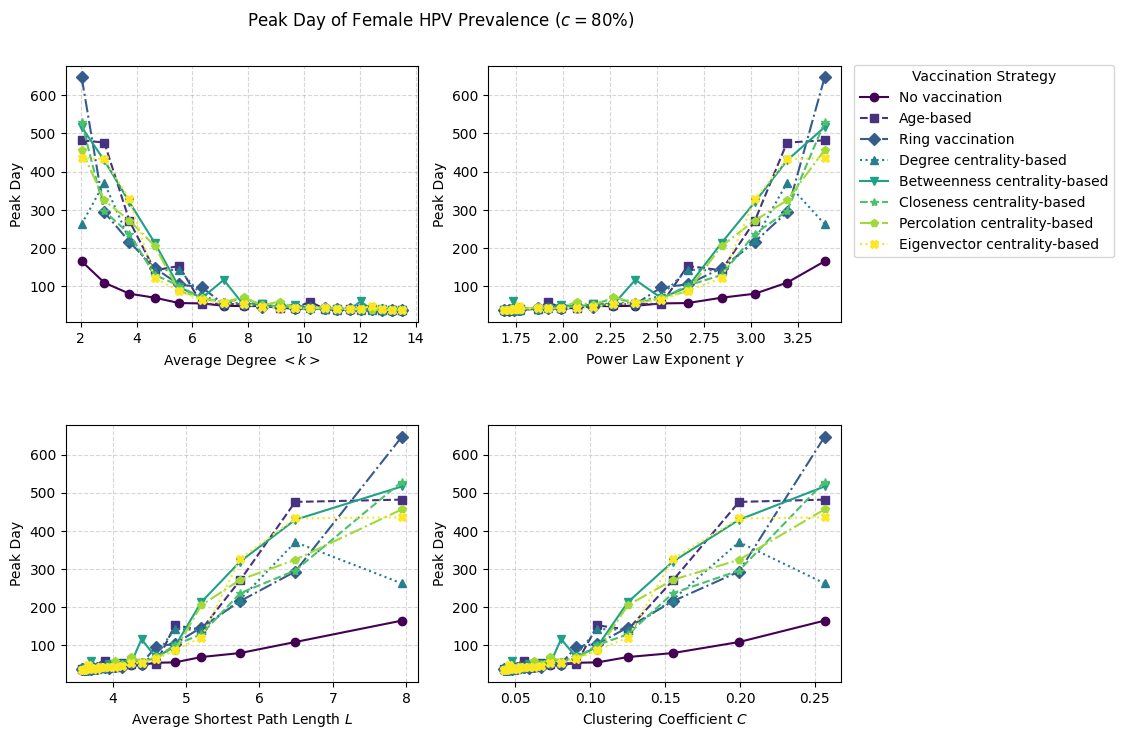}
    \caption{$c =  80\%$}
  \end{subfigure}
  \caption{Effect of Network Properties and Vaccination Strategies on Peak Day of Female HPV Prevalence under Varying Vaccine Coverage $c$ For Individuals under 26 ($N = 3000$, $\beta = 0.13$, $\langle\delta\rangle = 100$, $\langle \eta \rangle = 3.5$) - Increasing coverage extends the pre-peak temporal window in females, with centrality targeted schemes offering optimal delays under limited-resource scenarios.}\label{43}
\end{figure*}

Increasing coverage extends the pre-peak temporal window in females, with centrality-targeted schemes offering optimal delays under limited-resource scenarios in Fig. \ref{43}. 

\begin{figure*}[htbp]
  \centering
  \begin{subfigure}[b]{0.32\textwidth}
    \centering
    \includegraphics[width=\linewidth]{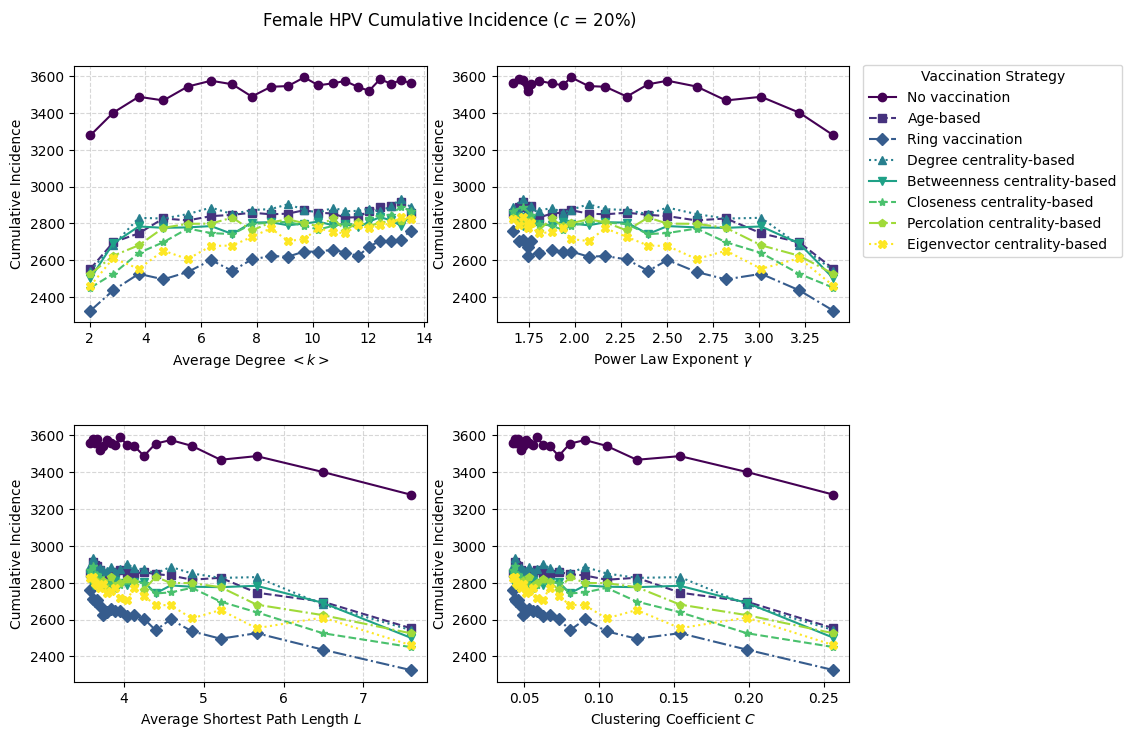}
    \caption{$c = 20\%$}
  \end{subfigure}\hfill
  \begin{subfigure}[b]{0.32\textwidth}
    \centering
    \includegraphics[width=\linewidth]{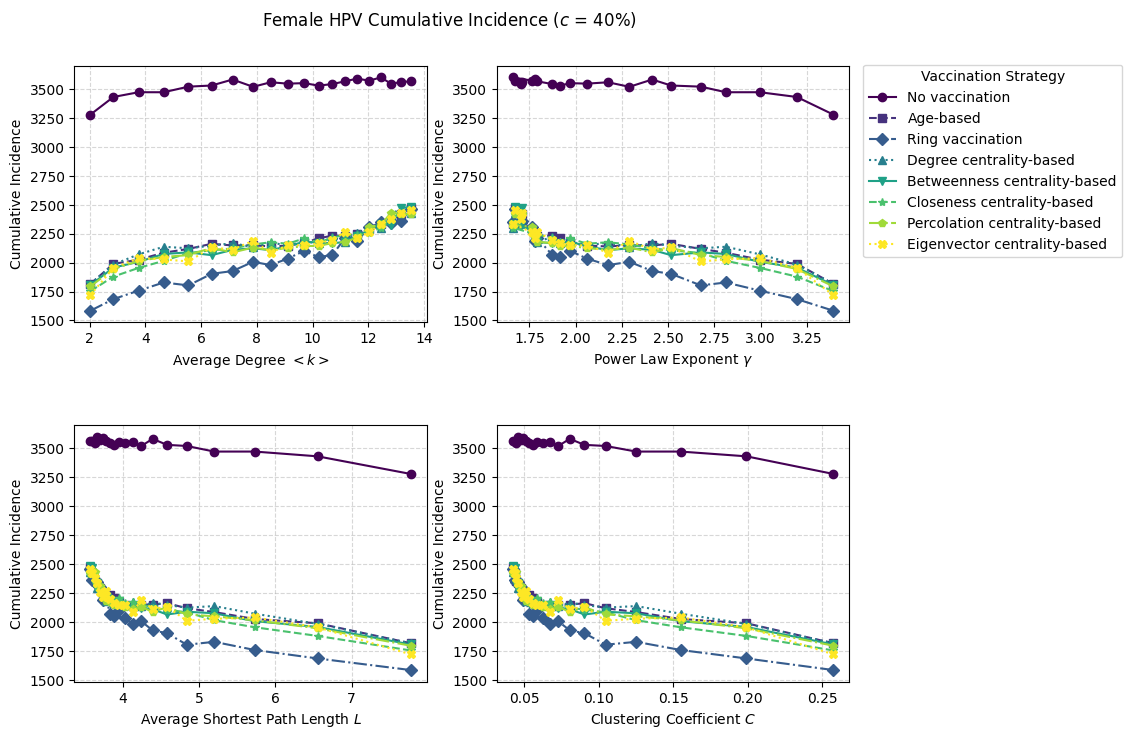}
    \caption{$c = 40\%$}
  \end{subfigure}\hfill
  \begin{subfigure}[b]{0.32\textwidth}
    \centering
    \includegraphics[width=\linewidth]{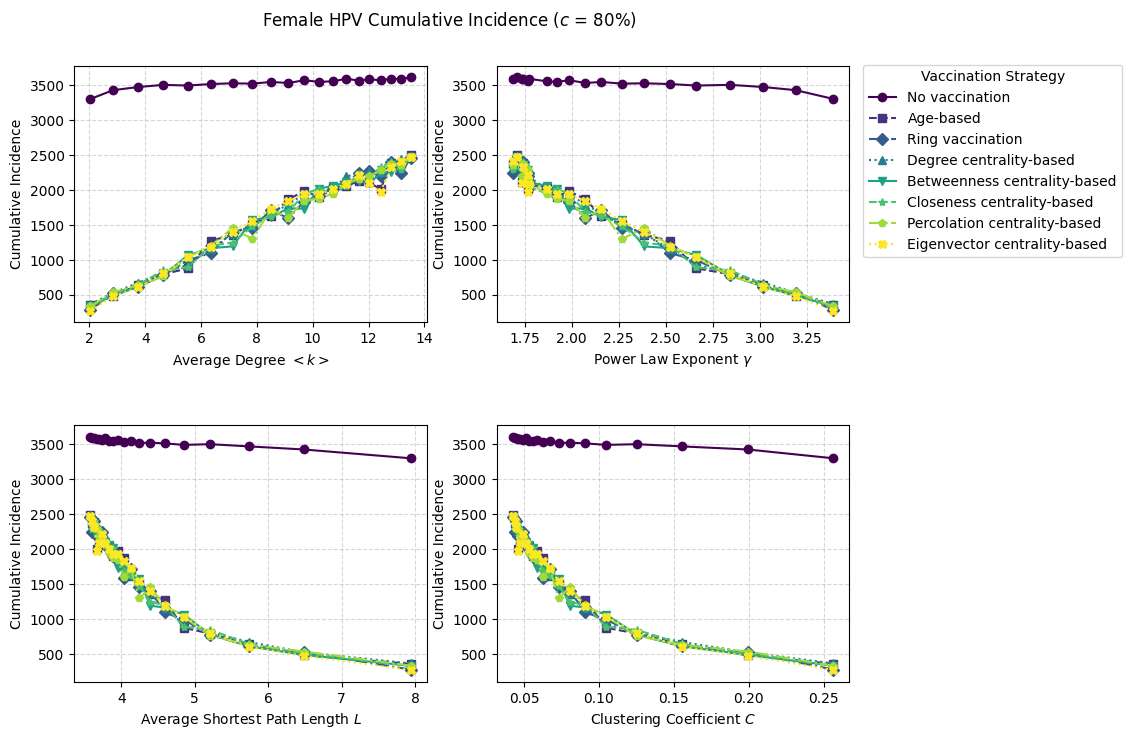}
    \caption{$c =80\%$}
  \end{subfigure}
  \caption{Effect of Network Structure and Vaccination Strategy on Female HPV Cumulative Incidence under Varying Vaccine Coverage $c$ For Individuals under 26 ($N = 3000$, $\beta = 0.13$, $\langle\delta\rangle = 100$, $\langle \eta \rangle = 3.5$) - Ring vaccination maintains a clear advantage in reducing female cumulative burden at 20\% and 40\% coverage, but strategy-specific performance converges at 80\% coverage.}\label{44}
\end{figure*}

Most notably, ring vaccination maintains a clear advantage in reducing the female cumulative burden at $20\%$ and $40\%$ coverage, but strategy-specific performance converges at $80\%$ coverage as population-wide protection takes over in Fig. \ref{44}.

\subsection{Simulation Results for the Male Cohort}

\begin{figure*}[htbp]
  \centering
  \begin{subfigure}[b]{0.32\textwidth}
    \centering
    \includegraphics[width=\linewidth]{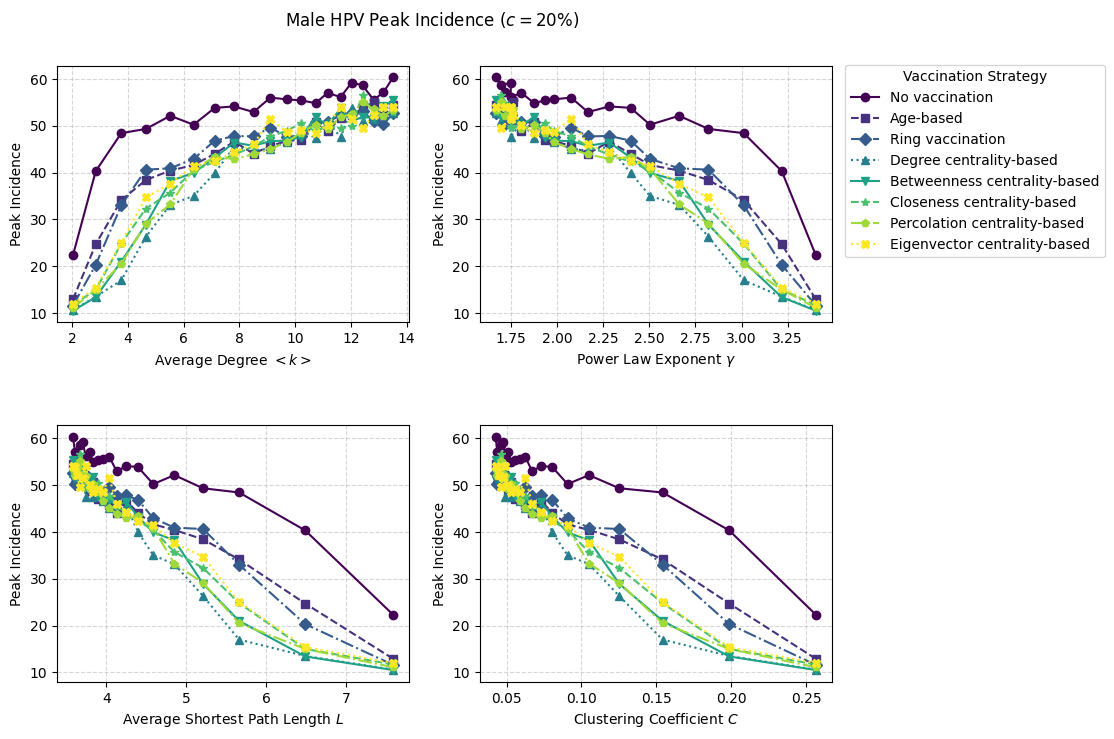}
    \caption{$c = 20\%$}
  \end{subfigure}\hfill
  \begin{subfigure}[b]{0.32\textwidth}
    \centering
    \includegraphics[width=\linewidth]{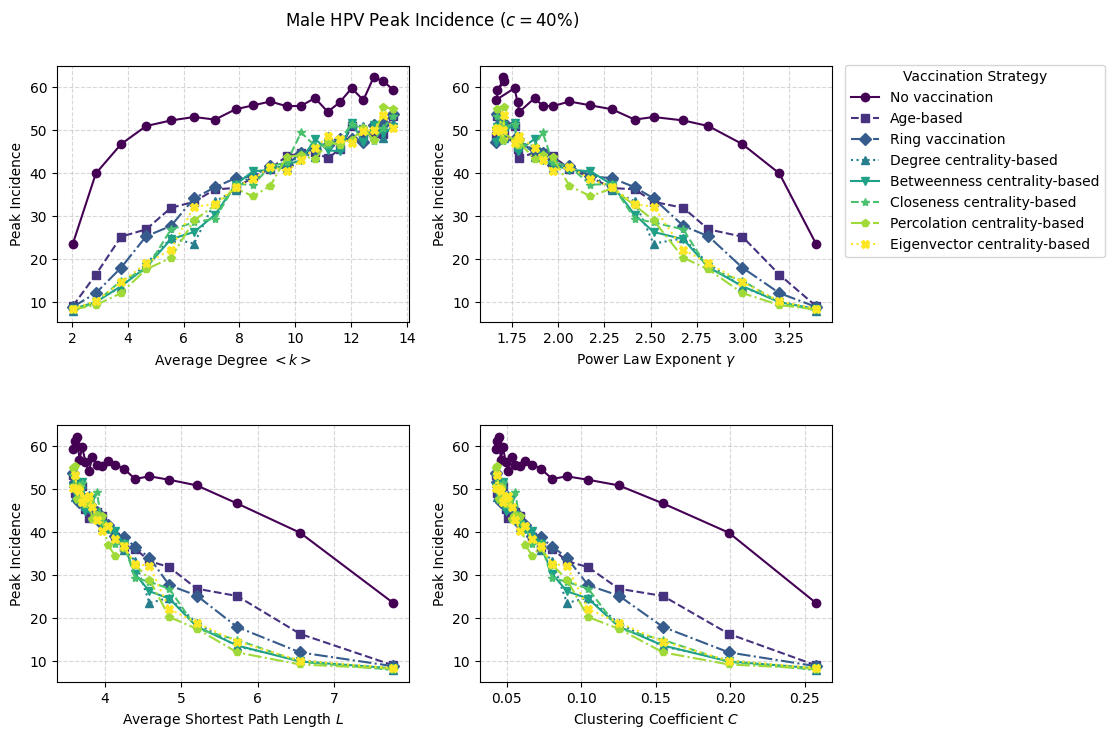}
    \caption{$c = 40\%$}
  \end{subfigure}\hfill
  \begin{subfigure}[b]{0.32\textwidth}
    \centering
    \includegraphics[width=\linewidth]{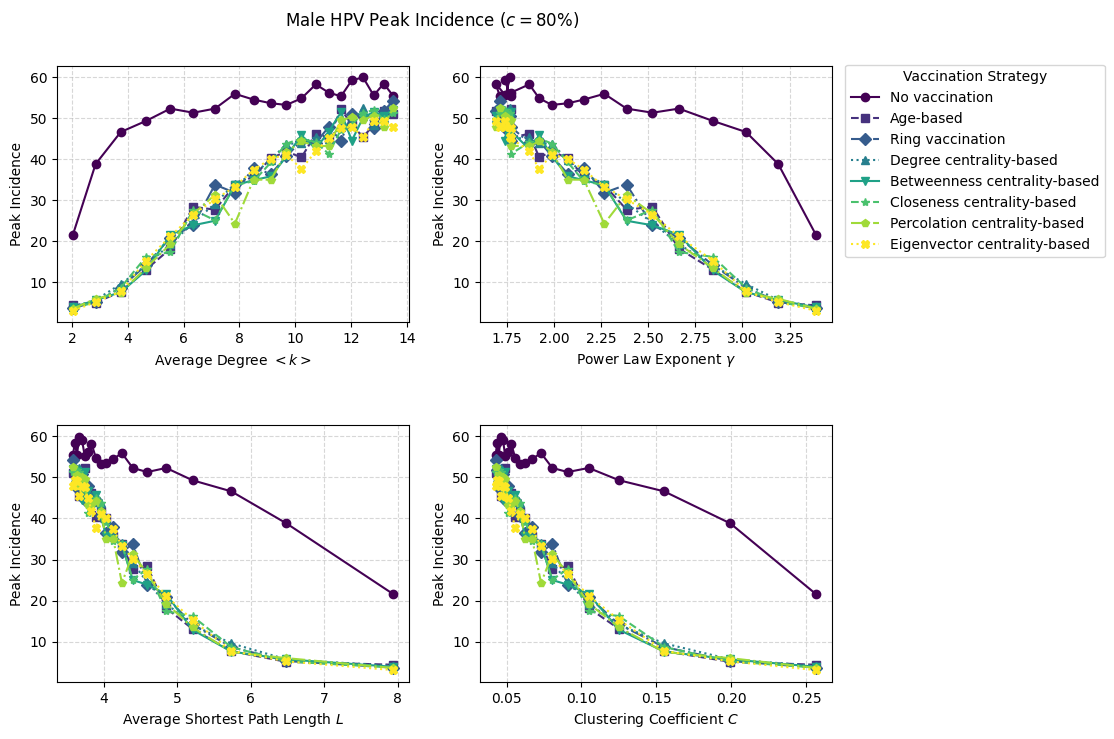}
    \caption{$c =80\%$}
  \end{subfigure}
  \caption{Maximum Daily Male HPV Incidence across Network Structures and Vaccination Strategies under Varying Vaccine Coverage $c$ For Individuals under 26 ($N = 3000$, $\beta = 0.13$, $\langle\delta\rangle = 100$, $\langle \eta \rangle = 3.5$) - Peak male incidence decreases monotonically with higher vaccine coverage, with degree-, betweenness- and percolation centrality-based strategies maintaining superior performance until high coverage (80\%) eliminates strategy variance.}\label{45}
\end{figure*}

For the male cohort, peak daily incidence decreases monotonically with higher vaccine coverage. Degree-, betweenness-, and percolation centrality-based strategies maintain superior performance until high coverage ($80\%$) eliminates strategy variance in Fig. \ref{45}.

\begin{figure*}[htbp]
  \centering
  \begin{subfigure}[b]{0.32\textwidth}
    \centering
    \includegraphics[width=\linewidth]{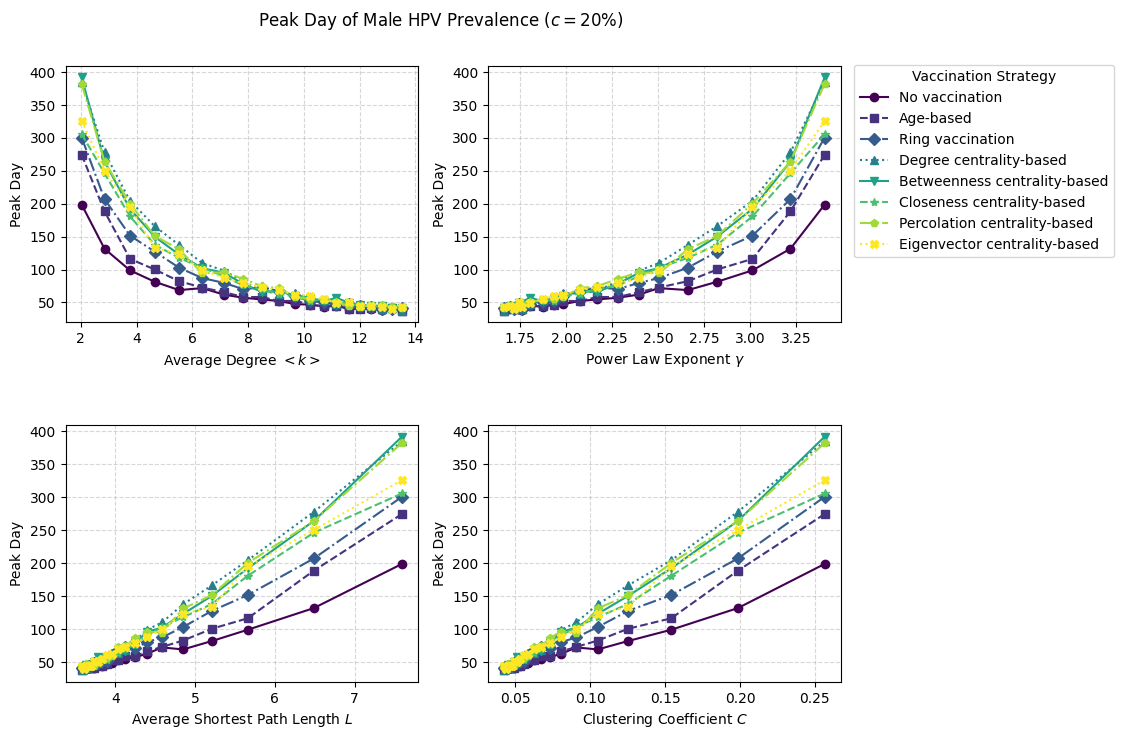}
    \caption{$c =  20\%$}
  \end{subfigure}\hfill
  \begin{subfigure}[b]{0.32\textwidth}
    \centering
    \includegraphics[width=\linewidth]{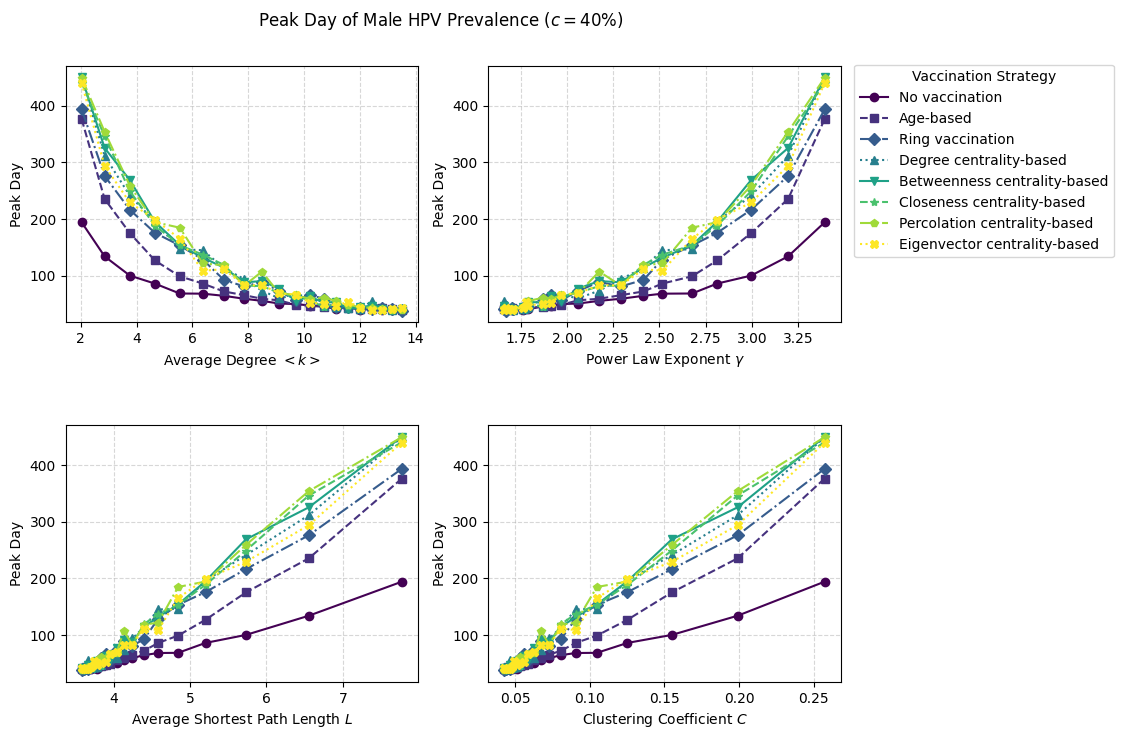}
    \caption{$c = 40\%$}
  \end{subfigure}\hfill
  \begin{subfigure}[b]{0.32\textwidth}
    \centering
    \includegraphics[width=\linewidth]{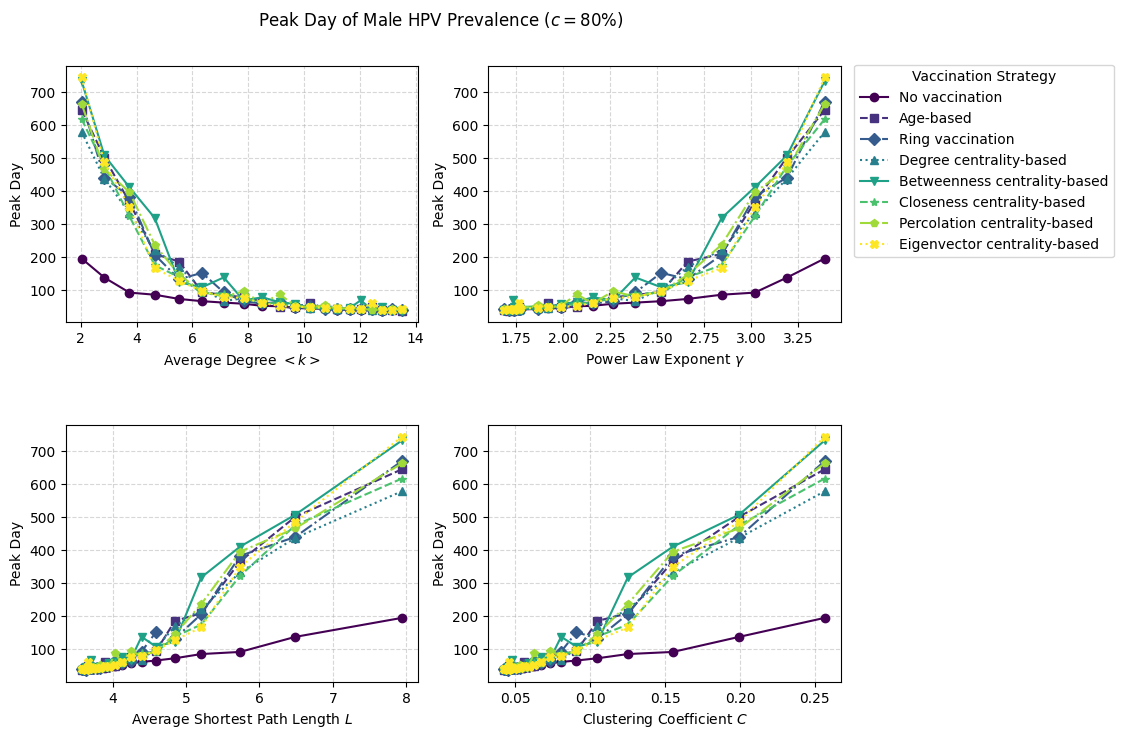}
    \caption{$c =80\%$}
  \end{subfigure}
  \caption{Effect of Network Properties and Vaccination Strategies on Peak Day of Male HPV Prevalence under Varying Vaccine Coverage $c$ For Individuals under 26 ($N = 3000$, $\beta = 0.13$, $\langle\delta\rangle = 100$, $\langle \eta \rangle = 3.5$) - Higher coverage delays peak male prevalence significantly across all network topologies, with centrality strategies delivering the greatest postponement at sub-saturating coverage levels.}\label{46}
\end{figure*}

Higher coverage delays peak male prevalence significantly across all network topologies, with centrality strategies delivering the greatest postponement at sub-saturating coverage levels in Fig. \ref{46}. 

\begin{figure*}[htbp]
  \centering
  \begin{subfigure}[b]{0.32\textwidth}
    \centering
    \includegraphics[width=\linewidth]{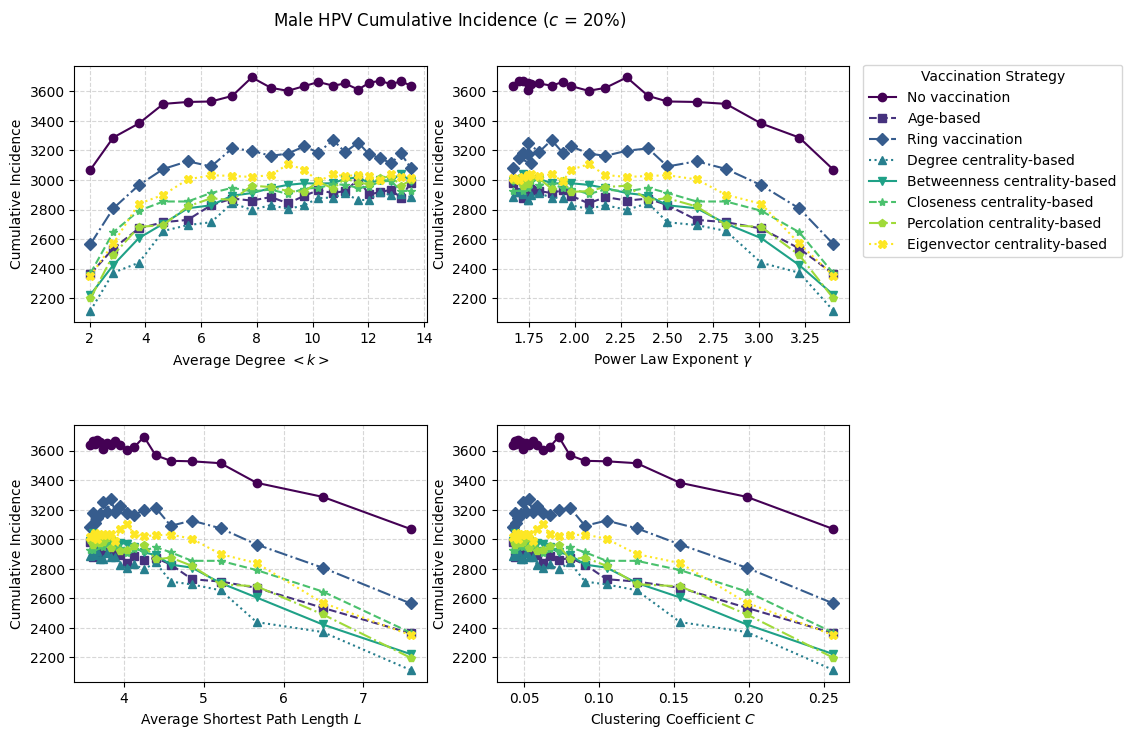}
    \caption{$c = 20\%$}
  \end{subfigure}\hfill
  \begin{subfigure}[b]{0.32\textwidth}
    \centering
    \includegraphics[width=\linewidth]{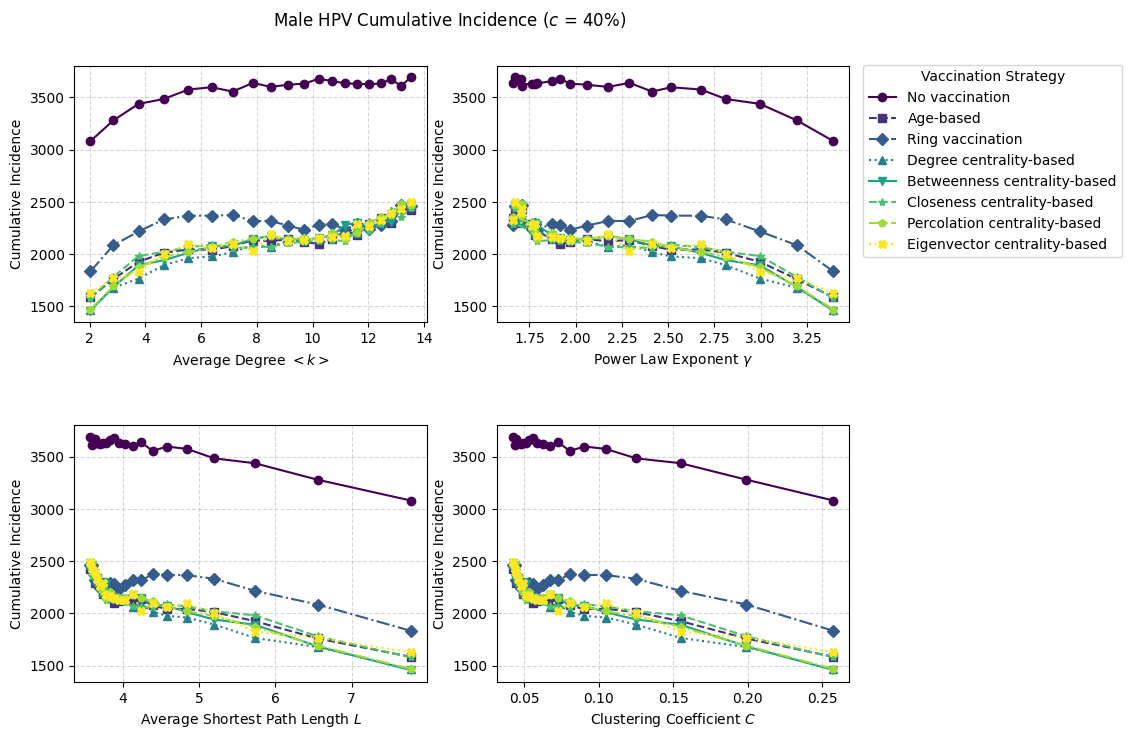}
    \caption{$c = 40\%$}
  \end{subfigure}\hfill
  \begin{subfigure}[b]{0.32\textwidth}
    \centering
    \includegraphics[width=\linewidth]{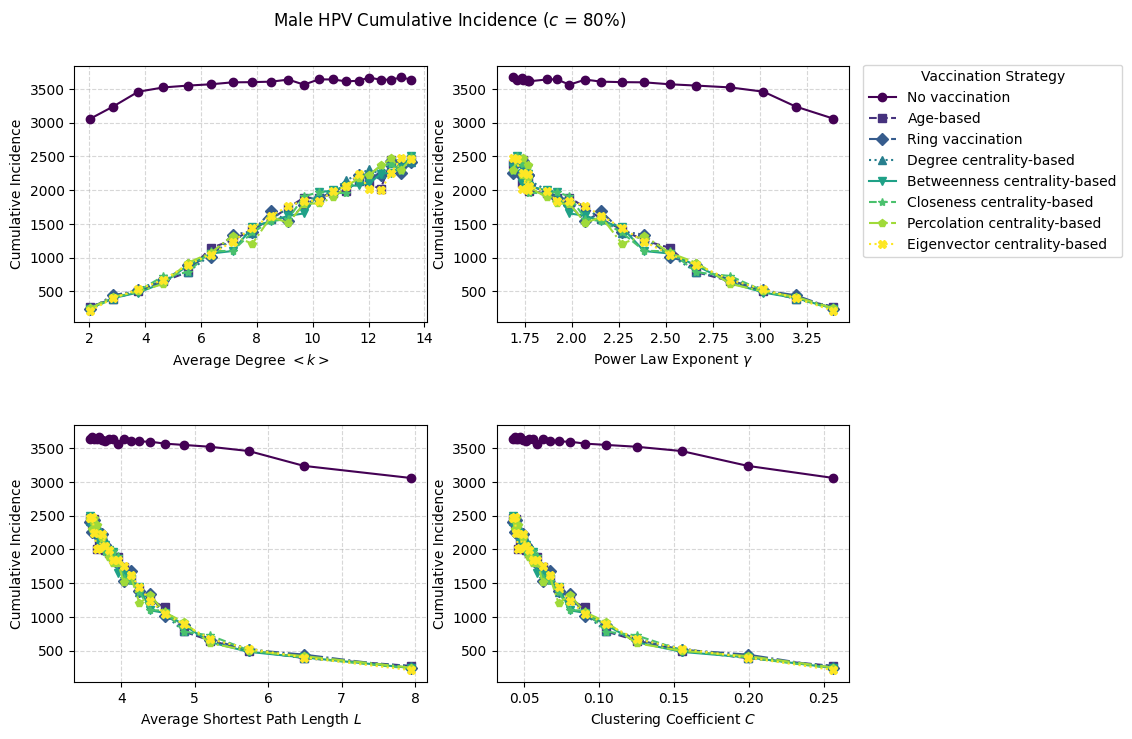}
    \caption{$c = 80\%$}
  \end{subfigure}
  \caption{Effect of Network Structure and Vaccination Strategy on Male HPV Cumulative Incidence under Varying Vaccine Coverage $c$ For Individuals under 26 ($N = 3000$, $\beta = 0.13$, $\langle\delta\rangle = 100$, $\langle \eta \rangle = 3.5$) - Male cumulative cases drop substantially as vaccine coverage increases, with degree centrality-based targeting remaining the most efficient scheme at 20\% and 40\% coverage.}\label{47}
\end{figure*}

Finally, male cumulative cases drop substantially as vaccine coverage increases, with degree centrality-based targeting remaining the most efficient scheme at $20\%$ and $40\%$ coverage in Fig. \ref{47}.

\section{Simulation Results for a Larger Population}\label{ae}

To evaluate the scalability of the model and determine whether core topological findings hold as population size increases, we conducted additional simulation experiments by scaling the network size from $N = 3000$ to $N = 5000$. All epidemic parameters ($\beta = 0.13$, $\langle\delta\rangle = 100 \text{ days}$), vaccination coverage ($c = 10\%$), and age preference ($\langle \eta \rangle = 3.5 \text{ years}$) were kept identical to the baseline setting. Fig. \ref{48} - \ref{56} present the maximum daily incidence, peak day of prevalence, and cumulative incidence across the general, female, and male cohorts for $N = 5000$.

\subsection{Simulation Results for the Overall Cohort}

\begin{figure}[htbp]
\centering
\includegraphics[width=\linewidth]{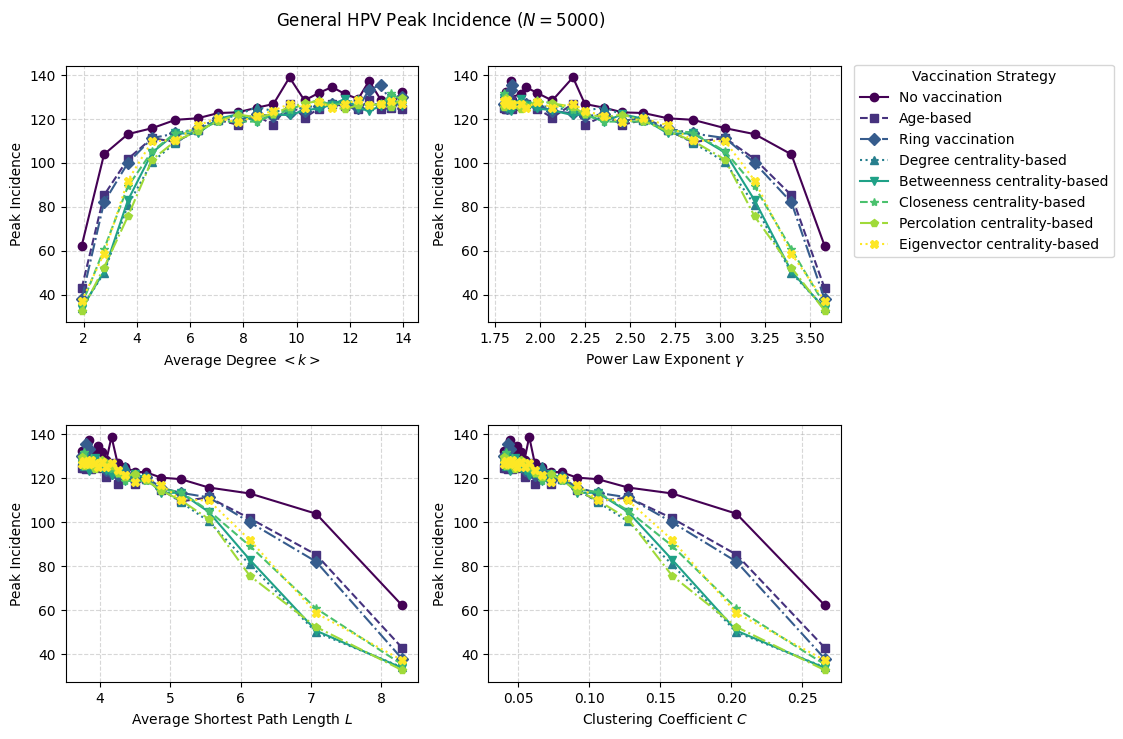}
\caption{Maximum Daily General HPV Incidence across Network Structures and Vaccination Strategies ($N = 5000$, $\beta = 0.13$, $\langle\delta\rangle = 100$, $c =10\%$, $\langle \eta \rangle = 3.5$) - Scaling the population size to $N = 5000$ proportionally increases absolute peak incidence while maintaining baseline topological trends. Degree- and percolation centrality-based strategies remain most effective in curbing new daily infections, with performance gaps narrowing at higher link densities.}\label{48}
\end{figure}

Scaling the population size to $N = 5000$, leads to a proportional increase in absolute case numbers, reflecting the $66.7\%$ expansion in population volume. Maximum daily general incidence peaks at higher absolute levels (reaching $\sim 200$ cases compared to $\sim 120$ cases in the baseline model), while strategy performance disparities narrow at higher link densities in Fig. \ref{48}.

\begin{figure}[htbp]
\centering
\includegraphics[width=\linewidth]{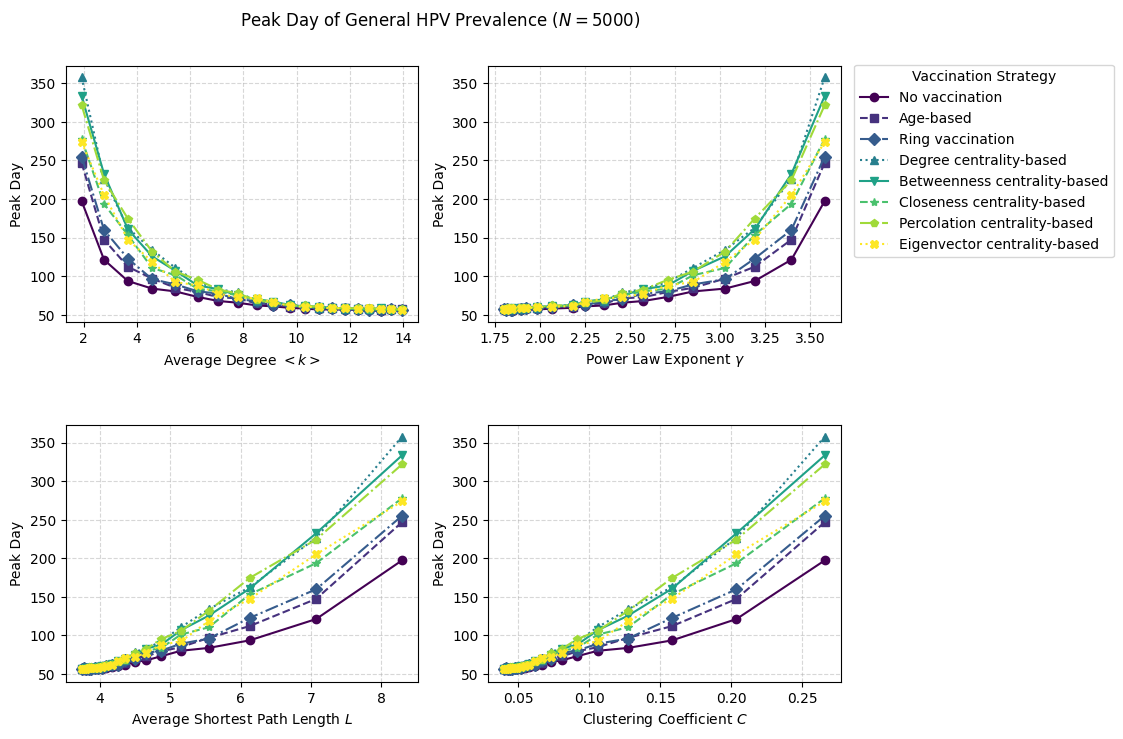}
\caption{Effect of Network Properties and Vaccination Strategies on Peak Day of General HPV Prevalence ($N = 5000$, $\beta = 0.13$, $\langle\delta\rangle = 100$, $c =10\%$, $\langle \eta \rangle = 3.5$) - The temporal progression to peak general prevalence remains invariant to population scale. Degree centrality-based targeting consistently delivers the greatest delay in epidemic peak timing across all topological metrics.}\label{49}
\end{figure}

However, the temporal dynamics of epidemic progression remain scale-invariant: the peak day of general prevalence across all network structures ranges from $\sim 50$ days in dense networks to $\sim 350+$ days in sparse networks, with degree centrality-based targeting delivering the greatest delay in epidemic timing in Fig. \ref{49}. 

\begin{figure}[htbp]
\centering
\includegraphics[width=\linewidth]{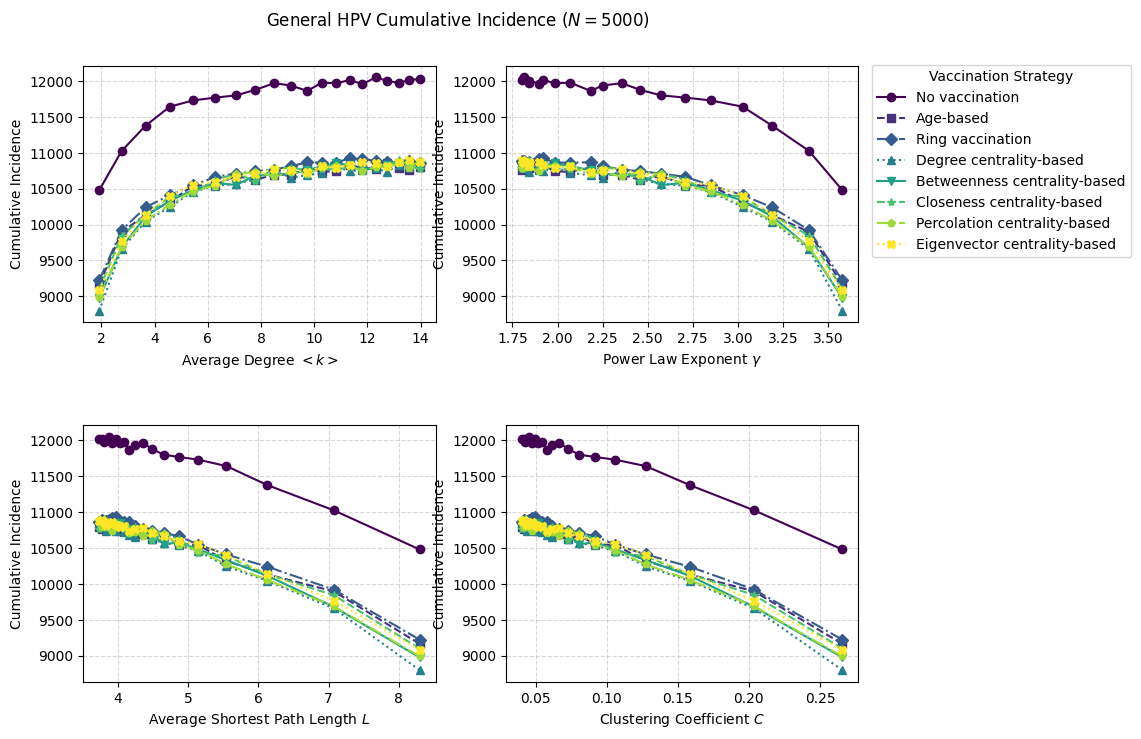}
\caption{Effect of Network Structure and Vaccination Strategy on General HPV Cumulative Incidence ($N = 5000$, $\beta = 0.13$, $\langle\delta\rangle = 100$, $c =10\%$, $\langle \eta \rangle = 3.5$) - Population-wide cumulative infection burden scales with population size, but strategy efficacy hierarchies persist, degree, betweenness, and percolation centrality-based strategies form a primary high-performance cluster that minimises overall cases.} \label{50}
\end{figure}

Cumulative general incidence scales up to $\sim 9,000–12,000$ cases, yet the relative strategy ranking persists, degree-, betweenness-, and percolation centrality-based strategies form a primary high-efficacy cluster that minimises overall infection burden in Fig. \ref{50}.

\subsection{Simulation Results for the Female Cohort}
\begin{figure}[htbp]
\centering
\includegraphics[width=\linewidth]{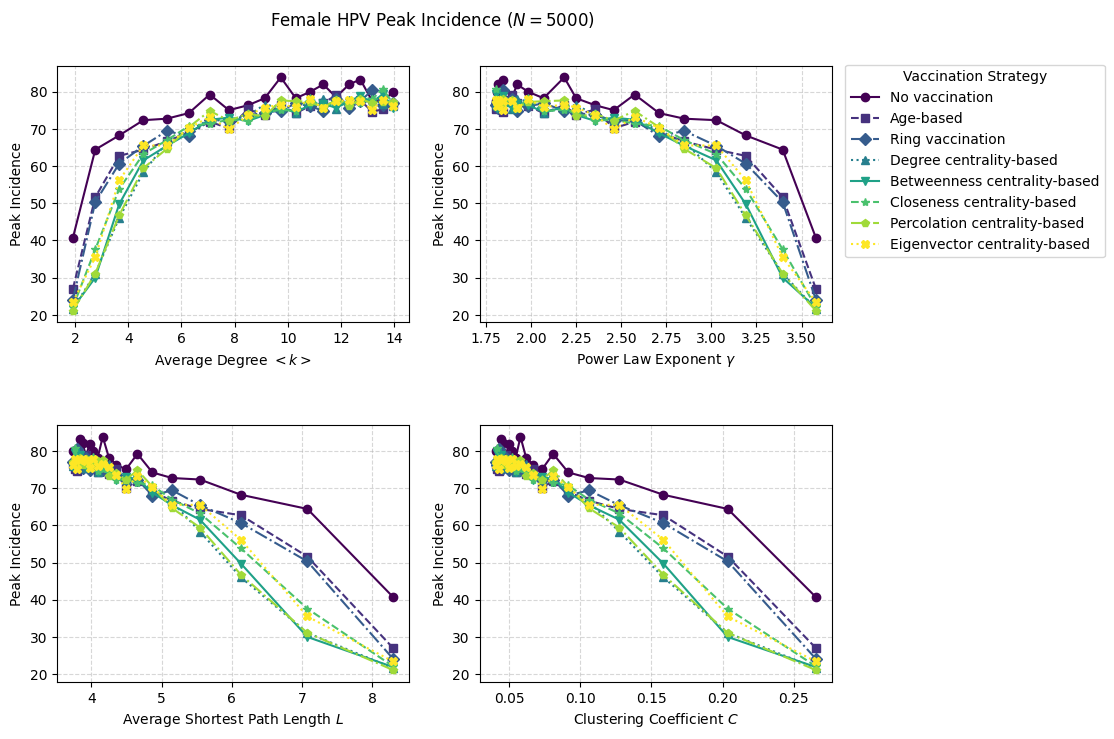}
\caption{Maximum Daily Female HPV Incidence across Network Structures and Vaccination Strategies ($N = 5000$, $\beta = 0.13$, $\langle\delta\rangle = 100$, $c =10\%$, $\langle \eta \rangle = 3.5$) - Percolation centrality-based vaccination plays a pivotal role in controlling peak daily female incidence in larger populations, with lower link density and higher network clustering providing strong additive protective benefits.}\label{51}
\end{figure}

In the female cohort, percolation centrality-based vaccination plays a pivotal role in controlling peak daily female incidence in the larger populations, with lower link density and higher network clustering providing strong additive protective benefits in Fig. \ref{51}.

\begin{figure}[htbp]
\centering
\includegraphics[width=\linewidth]{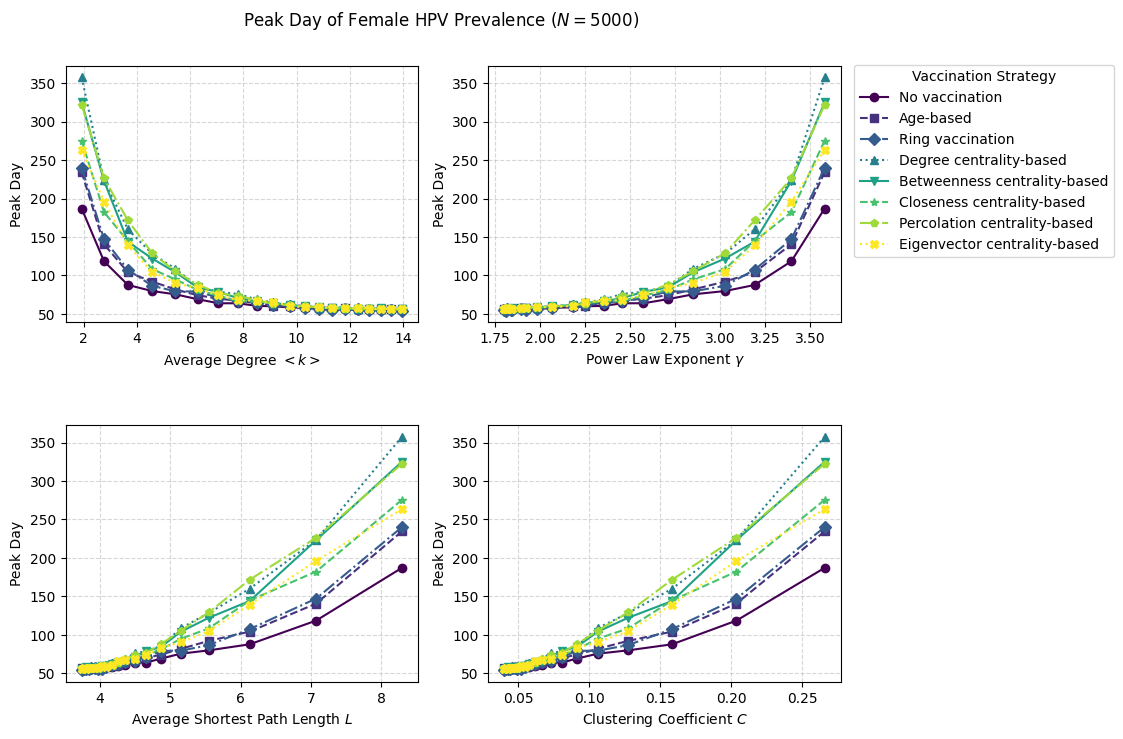}
\caption{Effect of Network Properties and Vaccination Strategies on Peak Day of Female HPV Prevalence ($N = 5000$, $\beta = 0.13$, $\langle\delta\rangle = 100$, $c =10\%$, $\langle \eta \rangle = 3.5$) - The degree centrality-based strategy demonstrates optimal efficacy in delaying peak female prevalence, matching baseline timing dynamics.}\label{52}
\end{figure}

The timeline to peak female prevalence matches baseline dynamics, where degree centrality-based targeting consistently achieves the longest postponement of the epidemic peak across topological configurations in Fig. \ref{52}. 

\begin{figure}[htbp]
\centering
\includegraphics[width=\linewidth]{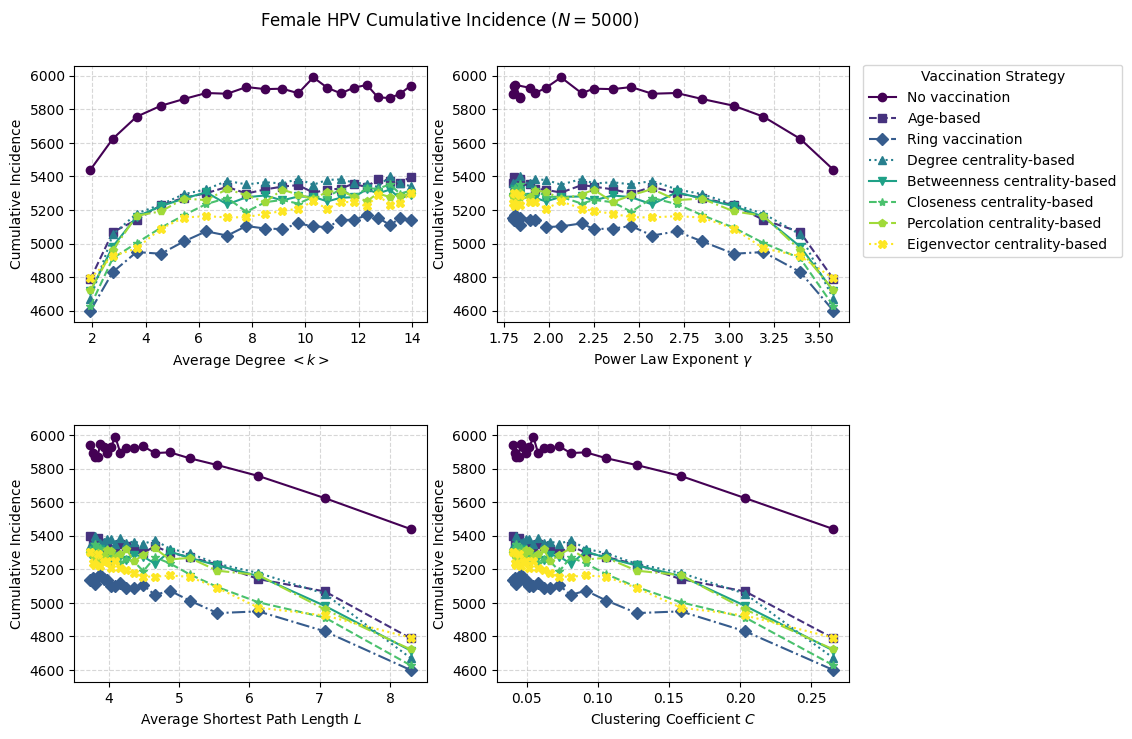}
\caption{Effect of Network Structure and Vaccination Strategy on Female HPV Cumulative Incidence ($N = 5000$, $\beta = 0.13$, $\langle\delta\rangle = 100$, $c =10\%$, $\langle \eta \rangle = 3.5$) - Ring vaccination consistently achieves the greatest overall reduction in female cumulative infection burden, reinforcing its gender-specific advantage in larger community networks.}\label{53}
\end{figure}

Most notably, ring vaccination maintains its clear superiority in reducing overall female cumulative cases in Fig. \ref{53}, confirming that its gender-specific advantage in containing localised female transmission chains is invariant to population scale.

\subsection{Simulation Results for the Male Cohort}
\begin{figure}[htbp]
\centering
\includegraphics[width=\linewidth]{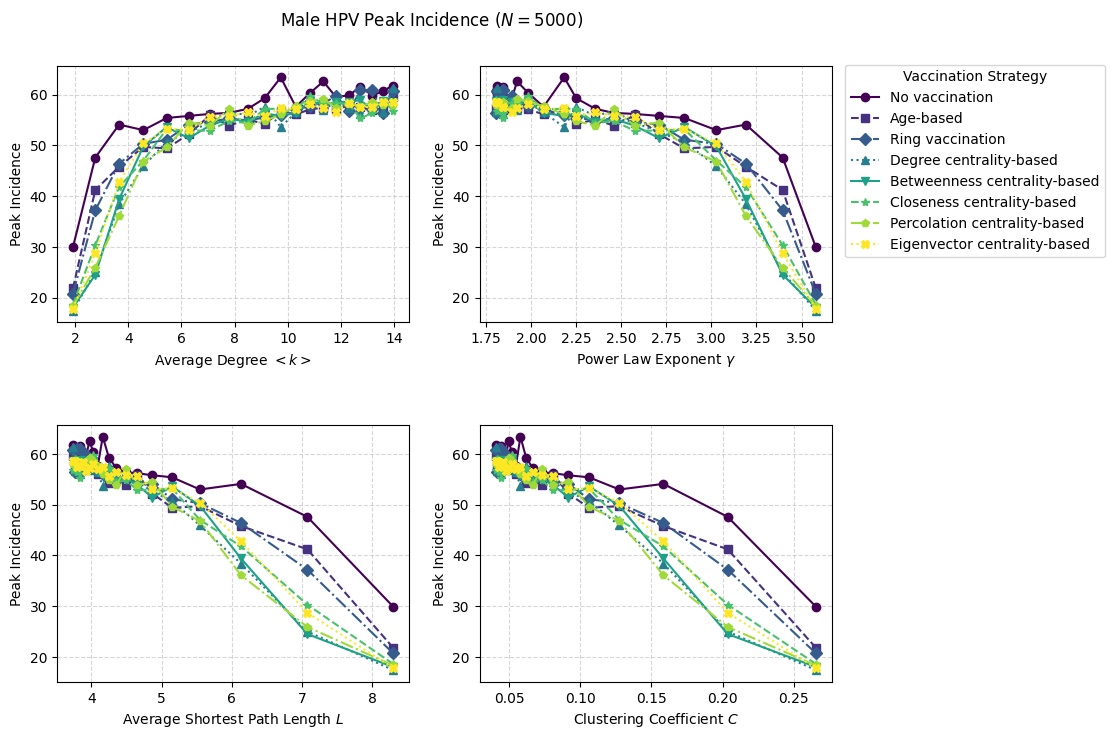}
\caption{Maximum Daily Male HPV Incidence across Network Structures and Vaccination Strategies ($N = 5000$, $\beta = 0.13$, $\langle\delta\rangle = 100$, $c =10\%$, $\langle \eta \rangle = 3.5$) - Degree-, betweenness-, and percolation centrality-based vaccination strategies effectively suppress peak male cases as population size increases.}\label{54}
\end{figure}

For the male cohort, degree-, betweenness-, and percolation centrality-based strategies remain highly effective at suppressing peak daily male cases as population size increases in Fig. \ref{54}.

\begin{figure}[htbp]
\centering
\includegraphics[width=\linewidth]{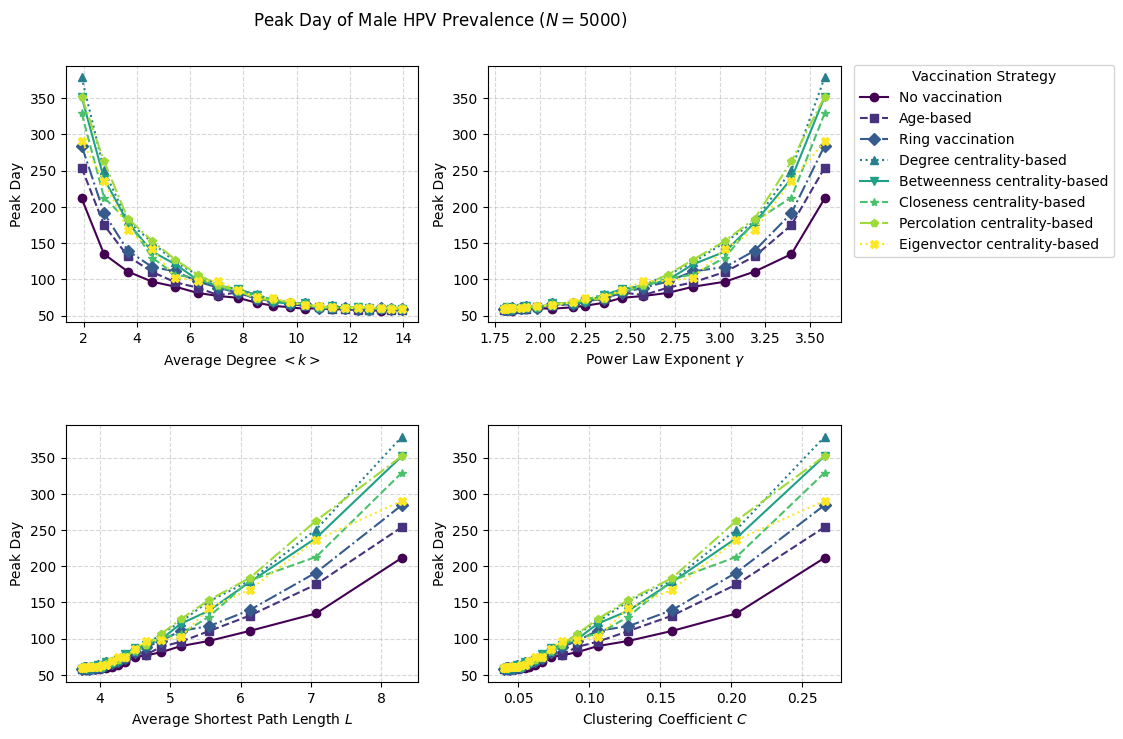}
\caption{Effect of Network Properties and Vaccination Strategies on Peak Day of Male HPV Prevalence ($N = 5000$, $\beta = 0.13$, $\langle\delta\rangle = 100$, $c =10\%$, $\langle \eta \rangle = 3.5$) - Centrality targeted vaccination strategies remain critical for postponing peak male infection prevalence, exhibiting scale-invariant temporal responses to network topology.}\label{55}
\end{figure}

Temporal trends in peak male prevalence mirror baseline observations, demonstrating that centrality-targeted schemes deliver scale-invariant delays in epidemic progression across diverse network topologies in Fig. \ref{55}. 

\begin{figure}[htbp]
\centering
\includegraphics[width=\linewidth]{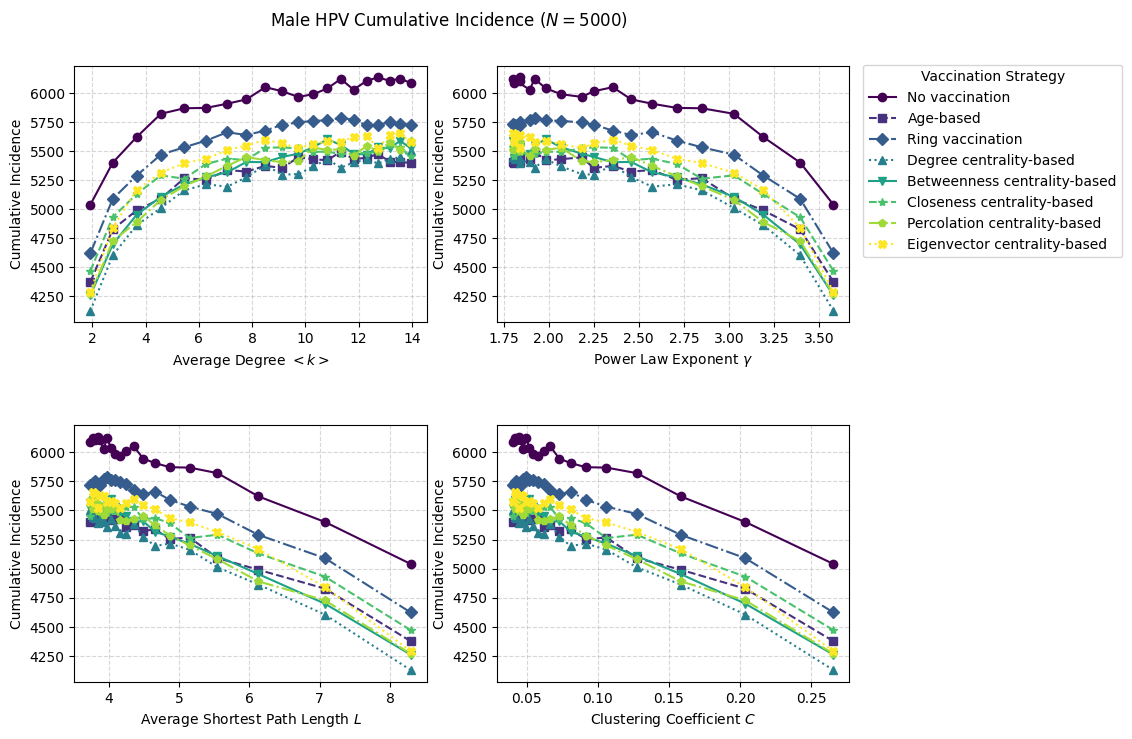}
\caption{Effect of Network Structure and Vaccination Strategy on Male HPV Cumulative Incidence ($N = 5000$, $\beta = 0.13$, $\langle\delta\rangle = 100$, $c =10\%$, $\langle \eta \rangle = 3.5$) - Degree centrality-based targeting remains the single most effective approach for mitigating cumulative disease burden in the male cohort across scaled population networks.}\label{56}
\end{figure}

Finally, degree centrality-based targeting remains the single most effective intervention for mitigating cumulative disease burden in the male cohort across scaled population networks in Fig. \ref{56}.

\section*{Acknowledgement}
We gratefully acknowledge Dr. Shailendra Sawleshwarkar for useful discussions. This research was supported by the Sydney Informatics Hub at the University of Sydney, through the use of High Performance Computing (HPC) services.

\bibliographystyle{IEEEtran}
\bibliography{main}

@article{Althaus2012TransmissionChlamydia,
  title = {Transmission of {{Chlamydia}} Trachomatis through Sexual Partnerships: A Comparison between Three Individual-Based Models and Empirical Data},
  author = {Althaus, Christian L. and Turner, Katherine M. E. and Schmid, Boris V. and Heijne, Janneke C. M. and Kretzschmar, Mirjam and Low, Nicola},
  year = 2012,
  month = jan,
  journal = {Journal of the Royal Society, Interface},
  volume = {9},
  number = {66},
  pages = {136--146},
  issn = {1742-5662},
  doi = {10.1098/rsif.2011.0131},
  language = {en-US},
  pmcid = {PMC3223622},
  pmid = {21653569}
}

@misc{AustralianDepartmentOfHealthAndAgedCare2023HumanPapillomavirus,
  title = {Human Papillomavirus ({{HPV}})},
  author = {{Australian Department of Health and Aged Care}},
  year = 2023,
  month = oct,
  journal = {The Australian Immunisation Handbook},
  urldate = {2023-10-31},
  howpublished = {https://immunisationhandbook.health.gov.au/contents/vaccine-preventable-diseases/human-papillomavirus-hpv},
  language = {en}
}

@techreport{AustralianInstituteOfHealthAndWelfare2018HumanPapillomavirus,
  title = {Human Papillomavirus in {{Australia}}},
  author = {{Australian Institute of Health and Welfare}},
  year = 2018,
  address = {Australia},
  institution = {{Australian Institute of Health and Welfare}},
  urldate = {2023-10-31},
  language = {en}
}

@article{Baldwin2003HumanPapillomavirus,
  title = {Human {{Papillomavirus Infection}} in {{Men Attending}} a {{Sexually Transmitted Disease Clinic}}},
  author = {Baldwin, Susie B and Wallace, Danelle R and Papenfuss, Mary R and Abrahamsen, Martha and Vaught, Linda C and Kornegay, Janet R and Hallum, Jennifer A and Redmond, Stacey A and Giuliano, Anna R},
  year = 2003,
  month = apr,
  journal = {The Journal of Infectious Diseases},
  volume = {187},
  number = {7},
  pages = {1064--1070},
  issn = {0022-1899},
  doi = {10.1086/368220},
  urldate = {2024-11-21}
}

@article{Barabasi1999EmergenceScaling,
  title = {Emergence of {{Scaling}} in {{Random Networks}}},
  author = {Barab{\'a}si, Albert-L{\'a}szl{\'o} and Albert, R{\'e}ka},
  year = 1999,
  month = oct,
  journal = {Science},
  volume = {286},
  number = {5439},
  pages = {509--512},
  publisher = {American Association for the Advancement of Science},
  doi = {10.1126/science.286.5439.509},
  urldate = {2023-02-01}
}

@article{Barbaro2015MeasuringHPV,
  title = {Measuring {{HPV}} Vaccination Coverage in {{Australia}}: {{Comparing}} Two Alternative Population-Based Denominators},
  author = {Barbaro, Bianca and Brotherton, Julia},
  year = 2015,
  month = jun,
  journal = {Australian and New Zealand journal of public health},
  volume = {39},
  doi = {10.1111/1753-6405.12372}
}

@article{Bell2017NetworkGrowth,
  title = {Network Growth Models: {{A}} Behavioural Basis for Attachment Proportional to Fitness},
  author = {Bell, Michael and Perera, Supun and Piraveenan, Mahendrarajah and Bliemer, Michiel and Latty, Tanya and Reid, Chris},
  year = 2017,
  month = feb,
  journal = {Scientific Reports},
  volume = {7},
  number = {1},
  pages = {42431},
  publisher = {Nature Publishing Group},
  issn = {2045-2322},
  doi = {10.1038/srep42431},
  urldate = {2023-07-02},
  copyright = {2017 The Author(s)},
  language = {en}
}

@article{Bianconi2001CompetitionMultiscaling,
  title = {Competition and Multiscaling in Evolving Networks},
  author = {Bianconi, G. and Barab{\'a}si, A.-L.},
  year = 2001,
  month = may,
  journal = {Europhysics Letters},
  volume = {54},
  number = {4},
  pages = {436},
  publisher = {IOP Publishing},
  issn = {0295-5075},
  doi = {10.1209/epl/i2001-00260-6},
  urldate = {2023-02-06},
  language = {en}
}

@article{Bruni2023GlobalRegional,
  title = {Global and Regional Estimates of Genital Human Papillomavirus Prevalence among Men: A Systematic Review and Meta-Analysis},
  author = {Bruni, Laia and Albero, Ginesa and Rowley, Jane and Alemany, Laia and Arbyn, Marc and Giuliano, Anna R. and Markowitz, Lauri E. and Broutet, Nathalie and Taylor, Melanie},
  year = 2023,
  month = sep,
  journal = {The Lancet Global Health},
  volume = {11},
  number = {9},
  pages = {e1345-e1362},
  publisher = {Elsevier},
  issn = {2214-109X},
  doi = {10.1016/S2214-109X(23)00305-4},
  urldate = {2023-12-02},
  language = {English},
  pmid = {37591583}
}

@article{Burchell2006Chapter6,
  title = {Chapter 6: {{Epidemiology}} and Transmission Dynamics of Genital {{HPV}} Infection},
  author = {Burchell, Ann N. and Winer, Rachel L. and {de Sanjos{\'e}}, Silvia and Franco, Eduardo L.},
  year = 2006,
  month = aug,
  journal = {Vaccine},
  series = {{{HPV}} Vaccines and Screening in the Prevention of Cervical Cancer},
  volume = {24},
  pages = {S52-S61},
  issn = {0264-410X},
  doi = {10.1016/j.vaccine.2006.05.031},
  urldate = {2023-03-23},
  language = {en}
}

@article{Carter2011HPVInfection,
  title = {{{HPV}} Infection and Cervical Disease: A Review: {{HPV}} and Cervical Disease},
  author = {Carter, Jonathan R. and Ding, Zongqun and Rose, Barbara R.},
  year = 2011,
  month = apr,
  journal = {Australian and New Zealand Journal of Obstetrics and Gynaecology},
  volume = {51},
  number = {2},
  pages = {103--108},
  issn = {00048666},
  doi = {10.1111/j.1479-828X.2010.01269.x},
  urldate = {2023-03-23},
  language = {en}
}

@article{Castellsague2009HPVVaccination,
  title = {{{HPV}} Vaccination against Cervical Cancer in Women above 25 Years of Age: Key Considerations and Current Perspectives},
  author = {Castellsagu{\'e}, Xavier and Schneider, Achim and Kaufmann, Andreas M. and Bosch, F. Xavier},
  year = 2009,
  month = dec,
  journal = {Gynecologic Oncology},
  series = {Cervical {{Cancer Vaccines}}: {{The Role}} of the {{Immunological Profile}} and {{Future Expectations}} for {{Vaccination II}}},
  volume = {115},
  number = {3, Supplement},
  pages = {S15-S23},
  issn = {0090-8258},
  doi = {10.1016/j.ygyno.2009.09.021},
  urldate = {2023-03-23},
  language = {en}
}

@article{Conroy-Beam2019WhyAge,
  title = {Why Is Age so Important in Human Mating? {{Evolved}} Age Preferences and Their Influences on Multiple Mating Behaviors},
  author = {{Conroy-Beam}, Daniel and Buss, David M.},
  year = 2019,
  journal = {Evolutionary Behavioral Sciences},
  volume = {13},
  number = {2},
  pages = {127--157},
  publisher = {Educational Publishing Foundation},
  address = {US},
  issn = {2330-2933},
  doi = {10.1037/ebs0000127},
  language = {en-US}
}

@article{DeBlasio2007PreferentialAttachment,
  title = {Preferential Attachment in Sexual Networks},
  author = {{de Blasio}, Birgitte Freiesleben and Svensson, {\AA}ke and Liljeros, Fredrik},
  year = 2007,
  month = jun,
  journal = {Proceedings of the National Academy of Sciences},
  volume = {104},
  number = {26},
  pages = {10762--10767},
  publisher = {Proceedings of the National Academy of Sciences},
  doi = {10.1073/pnas.0611337104},
  urldate = {2023-01-24}
}

@article{DeSanjose2018NaturalHistory,
  title = {The Natural History of Human Papillomavirus Infection},
  author = {{de Sanjos{\'e}}, Silvia and Brotons, Maria and Pav{\'o}n, Miguel Angel},
  year = 2018,
  month = feb,
  journal = {Best Practice \& Research Clinical Obstetrics \& Gynaecology},
  series = {Human {{Papilloma Virus}} in {{Gynaecology}}},
  volume = {47},
  pages = {2--13},
  issn = {1521-6934},
  doi = {10.1016/j.bpobgyn.2017.08.015},
  urldate = {2026-10-03}
}

@article{Drolet2017ImpactHuman,
  title = {The Impact of Human Papillomavirus Catch-up Vaccination in {{Australia}}: Implications for Introduction of Multiple Age Cohort Vaccination and Postvaccination Data Interpretation},
  author = {Drolet, M{\'e}lanie and Laprise, Jean-Fran{\c c}ois and Brotherton, Julia M L and Donovan, Basil and Fairley, Christopher K and Ali, Hammad and B{\'e}nard, {\'E}lodie and Martin, Dave and Brisson, Marc},
  year = 2017,
  month = dec,
  journal = {The Journal of Infectious Diseases},
  volume = {216},
  number = {10},
  pages = {1205--1209},
  issn = {0022-1899},
  doi = {10.1093/infdis/jix476},
  urldate = {2023-02-07}
}

@article{Gosak2018NetworkScience,
  title = {Network Science of Biological Systems at Different Scales: {{A}} Review},
  author = {Gosak, Marko and Markovi{\v c}, Rene and Dolen{\v s}ek, Jurij and Slak Rupnik, Marjan and Marhl, Marko and Sto{\v z}er, Andra{\v z} and Perc, Matja{\v z}},
  year = 2018,
  month = mar,
  journal = {Physics of Life Reviews},
  volume = {24},
  pages = {118--135},
  issn = {1571-0645},
  doi = {10.1016/j.plrev.2017.11.003},
  urldate = {2025-09-18}
}

@article{Helbing2015SavingHuman,
  title = {Saving {{Human Lives}}: {{What Complexity Science}} and {{Information Systems}} Can {{Contribute}}},
  author = {Helbing, Dirk and Brockmann, Dirk and Chadefaux, Thomas and Donnay, Karsten and Blanke, Ulf and {Woolley-Meza}, Olivia and Moussaid, Mehdi and Johansson, Anders and Krause, Jens and Schutte, Sebastian and Perc, Matja{\v z}},
  year = 2015,
  month = feb,
  journal = {Journal of Statistical Physics},
  volume = {158},
  number = {3},
  pages = {735--781},
  issn = {1572-9613},
  doi = {10.1007/s10955-014-1024-9},
  urldate = {2025-09-18},
  language = {en}
}

@article{Herbert2008ReducingPatient,
  title = {Reducing {{Patient Risk}} for {{Human Papillomavirus Infection}} and {{Cervical Cancer}}},
  author = {Herbert, Jennifer and Coffin, Janis},
  year = 2008,
  month = feb,
  journal = {Journal of Osteopathic Medicine},
  volume = {108},
  number = {2},
  pages = {65--70},
  publisher = {De Gruyter},
  issn = {2702-3648},
  doi = {10.7556/jaoa.2008.108.2.65},
  urldate = {2024-11-21},
  copyright = {De Gruyter expressly reserves the right to use all content for commercial text and data mining within the meaning of Section 44b of the German Copyright Act.},
  language = {en}
}

@article{Jusup2022SocialPhysics,
  title = {Social Physics},
  author = {Jusup, Marko and Holme, Petter and Kanazawa, Kiyoshi and Takayasu, Misako and Romi{\'c}, Ivan and Wang, Zhen and Ge{\v c}ek, Sun{\v c}ana and Lipi{\'c}, Tomislav and Podobnik, Boris and Wang, Lin and Luo, Wei and Klanj{\v s}{\v c}ek, Tin and Fan, Jingfang and Boccaletti, Stefano and Perc, Matja{\v z}},
  year = 2022,
  month = feb,
  journal = {Physics Reports},
  series = {Social Physics},
  volume = {948},
  pages = {1--148},
  issn = {0370-1573},
  doi = {10.1016/j.physrep.2021.10.005},
  urldate = {2025-09-18}
}

@article{Kessler2023StrategiesImprove,
  title = {Strategies to Improve Human Papillomavirus ({{HPV}}) Vaccination Rates among College Students},
  author = {Kessler, Roanna and Auwaerter, Paul},
  year = 2023,
  month = sep,
  journal = {Journal of American College Health},
  volume = {71},
  number = {7},
  pages = {2192--2199},
  publisher = {Taylor \& Francis},
  issn = {0744-8481},
  doi = {10.1080/07448481.2021.1965146},
  urldate = {2026-07-27},
  pmid = {34469255}
}

@article{Kim2008HealthEconomic,
  title = {Health and Economic Implications of {{HPV}} Vaccination in the {{United States}}},
  author = {Kim, Jane J. and Goldie, Sue J.},
  year = 2008,
  month = aug,
  journal = {The New England Journal of Medicine},
  volume = {359},
  number = {8},
  pages = {821--832},
  publisher = {Massachusetts Medical Society},
  issn = {0028-4793},
  doi = {10.1056/NEJMsa0707052},
  urldate = {2023-03-01},
  pmid = {18716299}
}

@article{Kim2009CostEffectiveness,
  title = {Cost Effectiveness Analysis of Including Boys in a Human Papillomavirus Vaccination Programme in the {{United States}}},
  author = {Kim, Jane J. and Goldie, Sue J.},
  year = 2009,
  month = oct,
  journal = {BMJ},
  volume = {339},
  pages = {b3884},
  publisher = {British Medical Journal Publishing Group},
  issn = {0959-8138, 1468-5833},
  doi = {10.1136/bmj.b3884},
  urldate = {2023-03-01},
  chapter = {Research},
  copyright = {\copyright{}  . This is an open-access article distributed under the terms of the Creative Commons Attribution Non-commercial License, which permits use, distribution, and reproduction in any medium, provided the original work is properly cited, the use is non commercial and is otherwise in compliance with the license. See: http://creativecommons.org/licenses/by-nc/2.0/  and  http://creativecommons.org/licenses/by-nc/2.0/legalcode.},
  language = {en},
  pmid = {19815582}
}

@inproceedings{Law2020PlacementMatters,
  title = {Placement Matters in Making Good Decisions Sooner: The Influence of Topology in Reaching Public Utility Thresholds},
  booktitle = {Proceedings of the 2019 {{IEEE}}/{{ACM International Conference}} on {{Advances}} in {{Social Networks Analysis}} and {{Mining}}},
  author = {Law, Sheung Yat and Kasthurirathna, Dharshana and Piraveenan, Mahendra},
  editor = {{Spezzano, Francesca}},
  year = 2020,
  month = jan,
  series = {{{ASONAM}} '19},
  pages = {787--795},
  publisher = {Association for Computing Machinery},
  address = {New York, NY, USA},
  doi = {10.1145/3341161.3343674},
  urldate = {2023-07-02},
  isbn = {978-1-4503-6868-1}
}

@article{Lowy2008HumanPapillomavirus,
  title = {Human Papillomavirus Infection and the Primary and Secondary Prevention of Cervical Cancer},
  author = {Lowy, Douglas R. and Solomon, Diane and Hildesheim, Allan and Schiller, John T. and Schiffman, Mark},
  year = 2008,
  journal = {Cancer},
  volume = {113},
  number = {S7},
  pages = {1980--1993},
  issn = {1097-0142},
  doi = {10.1002/cncr.23704},
  urldate = {2023-03-23},
  language = {en}
}

@article{Mossong2008SocialContacts,
  title = {Social {{Contacts}} and {{Mixing Patterns Relevant}} to the {{Spread}} of {{Infectious Diseases}}},
  author = {Mossong, Jo{\"e}l and Hens, Niel and Jit, Mark and Beutels, Philippe and Auranen, Kari and Mikolajczyk, Rafael and Massari, Marco and Salmaso, Stefania and Tomba, Gianpaolo Scalia and Wallinga, Jacco and Heijne, Janneke and {Sadkowska-Todys}, Malgorzata and Rosinska, Magdalena and Edmunds, W. John},
  year = 2008,
  month = mar,
  journal = {PLOS Medicine},
  volume = {5},
  number = {3},
  pages = {e74},
  publisher = {Public Library of Science},
  issn = {1549-1676},
  doi = {10.1371/journal.pmed.0050074},
  urldate = {2026-10-03},
  language = {en}
}

@article{Munoz-Quiles2021EliminationInfections,
  title = {On the Elimination of Infections Related to Oncogenic Human Papillomavirus: An Approach Using a Computational Network Model},
  author = {{Mu{\~n}oz-Quiles}, Cintia and {D{\'i}ez-Domingo}, Javier and Acedo, Luis and {S{\'a}nchez-Alonso}, V{\'i}ctor and Villanueva, Rafael J.},
  year = 2021,
  month = may,
  journal = {Viruses},
  volume = {13},
  number = {5},
  pages = {906},
  issn = {1999-4915},
  doi = {10.3390/v13050906},
  language = {en-US},
  pmcid = {PMC8153310},
  pmid = {34068358}
}

@article{Munoz-Quiles2024QuantifyingDelay,
  title = {Quantifying the Delay in Eliminating Vaccine-Targeted Human Papillomavirus after a Drop in the Coverage Using a Lifetime Sexual Partners Network},
  author = {{Mu{\~n}oz-Quiles}, Cintia and {Orrico-S{\'a}nchez}, Alejandro and {S{\'a}nchez-Alonso}, V{\'i}ctor and {Andreu-Vilarroig}, Carlos and Villanueva, Rafael-Jacinto},
  year = 2024,
  month = nov,
  journal = {Chaos, Solitons \& Fractals},
  volume = {188},
  pages = {115547},
  issn = {0960-0779},
  doi = {10.1016/j.chaos.2024.115547},
  urldate = {2024-10-08}
}

@article{Nepomuceno2020ComputationalChaos,
  title = {Computational Chaos in Complex Networks},
  author = {Nepomuceno, Erivelton G and Perc, Matja{\v z}},
  year = 2020,
  month = feb,
  journal = {Journal of Complex Networks},
  volume = {8},
  number = {1},
  pages = {cnz015},
  issn = {2051-1329},
  doi = {10.1093/comnet/cnz015},
  urldate = {2025-09-18}
}

@article{Olsen2010HumanPapillomavirus,
  title = {Human Papillomavirus Transmission and Cost-Effectiveness of Introducing Quadrivalent {{HPV}} Vaccination in {{Denmark}}},
  author = {Olsen, Jens and Jepsen, Martin Rudbeck},
  year = 2010,
  month = apr,
  journal = {International Journal of Technology Assessment in Health Care},
  volume = {26},
  number = {2},
  pages = {183--191},
  issn = {1471-6348},
  doi = {10.1017/S0266462310000085},
  language = {en-US},
  pmid = {20392322}
}

@article{Perc2008StochasticResonance,
  title = {Stochastic Resonance on Weakly Paced Scale-Free Networks},
  author = {Perc, Matja{\v z}},
  year = 2008,
  journal = {Physical Review E},
  volume = {78},
  number = {3},
  doi = {10.1103/PhysRevE.78.036105},
  language = {en-US}
}

@inproceedings{Piraveenan2013EffectVaccination,
  title = {Effect of {{Vaccination Strategies}} on the {{Herd Immunity}} of {{Growing Networks}}},
  booktitle = {2013 {{International Conference}} on {{Social Computing}}},
  author = {Piraveenan, Mahendra and Uddin, Shahadat and Thedchanamoorthy, Gnana},
  year = 2013,
  month = sep,
  pages = {288--294},
  doi = {10.1109/SocialCom.2013.47},
  urldate = {2024-06-24}
}

@article{Piraveenan2013PercolationCentrality,
  title = {Percolation {{Centrality}}: {{Quantifying Graph-Theoretic Impact}} of {{Nodes}} during {{Percolation}} in {{Networks}}},
  author = {Piraveenan, Mahendra and Prokopenko, Mikhail and Hossain, Liaquat},
  year = 2013,
  month = jan,
  journal = {PLOS ONE},
  volume = {8},
  number = {1},
  pages = {e53095},
  publisher = {Public Library of Science},
  issn = {1932-6203},
  doi = {10.1371/journal.pone.0053095},
  urldate = {2024-04-16},
  language = {en}
}

@article{Schwarz1985StructureTranscription,
  title = {Structure and Transcription of Human Papillomavirus Sequences in Cervical Carcinoma Cells},
  author = {Schwarz, Elisabeth and Freese, Ulrich Karl and Gissmann, Lutz and Mayer, Wolfgang and Roggenbuck, Birgit and Stremlau, Armin and zur Hausen, Harald},
  year = 1985,
  month = mar,
  journal = {Nature},
  volume = {314},
  number = {6006},
  pages = {111--114},
  publisher = {Nature Publishing Group},
  issn = {1476-4687},
  doi = {10.1038/314111a0},
  urldate = {2023-03-23},
  copyright = {1985 Nature Publishing Group},
  language = {en}
}

@article{Scianna2025ComparingEffectiveness,
  title = {Comparing the Effectiveness of Ring and Block-Vaccination Strategies on Networks},
  author = {Scianna, Matteo and Gallotti, Riccardo and Tizzoni, Michele and Artime, Oriol and {Alvarez-Zuzek}, Lucila G.},
  year = 2025,
  month = aug,
  journal = {PLOS Computational Biology},
  volume = {21},
  number = {8},
  pages = {e1013274},
  publisher = {Public Library of Science},
  issn = {1553-7358},
  doi = {10.1371/journal.pcbi.1013274},
  urldate = {2026-08-12},
  language = {en}
}

@article{Steffens2026ImmunisationProgram,
  title = {Immunisation Program Managers' Experiences of Implementing the Change from a Two Dose to a Single Dose Course of {{HPV}} Vaccination in {{Australia}}'s School-Based Program},
  author = {Steffens, Maryke and Bolsewicz, Katarzyna and Prokopovich, Kathleen and Beard, Frank and Brotherton, Julia M. L.},
  year = 2026,
  month = mar,
  journal = {Vaccine},
  volume = {76},
  pages = {128300},
  issn = {0264-410X},
  doi = {10.1016/j.vaccine.2026.128300},
  urldate = {2026-07-27}
}

@article{Sung2021GlobalCancer,
  title = {Global Cancer Statistics 2020: {{GLOBOCAN}} Estimates of Incidence and Mortality Worldwide for 36 Cancers in 185 Countries},
  author = {Sung, Hyuna and Ferlay, Jacques and Siegel, Rebecca L. and Laversanne, Mathieu and Soerjomataram, Isabelle and Jemal, Ahmedin and Bray, Freddie},
  year = 2021,
  journal = {CA: a cancer journal for clinicians},
  volume = {71},
  number = {3},
  pages = {209--249},
  issn = {1542-4863},
  doi = {10.3322/caac.21660},
  urldate = {2023-03-23},
  language = {en}
}

@article{Tabrizi2014HPVGenotype,
  title = {{{HPV}} Genotype Prevalence in {{Australian}} Women Undergoing Routine Cervical Screening by Cytology Status Prior to Implementation of an {{HPV}} Vaccination Program},
  author = {Tabrizi, Sepehr N. and Brotherton, Julia M. L. and Stevens, Matthew P. and Condon, John R. and McIntyre, Peter and Smith, David and Garland, Suzanne M.},
  year = 2014,
  month = jul,
  journal = {Journal of Clinical Virology},
  volume = {60},
  number = {3},
  pages = {250--256},
  issn = {1386-6532},
  doi = {10.1016/j.jcv.2014.04.013},
  urldate = {2023-12-02}
}

@article{Thedchanamoorthy2014InfluenceVaccination,
  title = {Influence of Vaccination Strategies and Topology on the Herd Immunity of Complex Networks},
  author = {Thedchanamoorthy, Gnana and Piraveenan, Mahendra and Uddin, Shahadat and Senanayake, Upul},
  year = 2014,
  month = aug,
  journal = {Social Network Analysis and Mining},
  volume = {4},
  number = {1},
  pages = {213},
  issn = {1869-5469},
  doi = {10.1007/s13278-014-0213-5},
  urldate = {2026-07-31},
  language = {en}
}

@misc{TheUniversityOfSydney2022SnapshotStudent,
  title = {A Snapshot of Student Diversity at {{Sydney}} -- {{Teaching}}@{{Sydney}}},
  author = {{The University of Sydney}},
  year = 2022,
  month = feb,
  urldate = {2024-07-27},
  language = {en-US}
}

@article{Vassallo2020RingVaccination,
  title = {Ring Vaccination Strategy in Networks: {{A}} Mixed Percolation Approach},
  author = {Vassallo, Lautaro},
  year = 2020,
  journal = {Physical Review E},
  volume = {101},
  number = {5},
  doi = {10.1103/PhysRevE.101.052309}
}

@incollection{Villa1997HumanPapillomaviruses,
  title = {Human Papillomaviruses and Cervical Cancer},
  booktitle = {Advances in Cancer Research},
  author = {Villa, Luisa Lina},
  editor = {Woude, George F. Vande and Klein, George},
  year = 1997,
  month = jan,
  volume = {71},
  pages = {321--341},
  publisher = {Academic press},
  doi = {10.1016/S0065-230X(08)60102-5},
  urldate = {2023-03-23},
  language = {en}
}

@article{Villanueva2022MathematicalModel,
  title = {A Mathematical Model for Human Papillomavirus Vaccination Strategies in a Random Network},
  author = {Villanueva, Rafael-Jacinto and {S{\'a}nchez-Alonso}, V{\'i}ctor and Acedo, Luis},
  year = 2022,
  journal = {Mathematical Methods in the Applied Sciences},
  volume = {45},
  number = {6},
  pages = {3284--3294},
  issn = {1099-1476},
  doi = {10.1002/mma.7205},
  urldate = {2023-02-28},
  language = {en}
}

@article{Wang2016StatisticalPhysics,
  title = {Statistical Physics of Vaccination},
  author = {Wang, Zhen and Bauch, Chris T. and Bhattacharyya, Samit and {d'Onofrio}, Alberto and Manfredi, Piero and Perc, Matjaz and Perra, Nicola and Salath{\'e}, Marcel and Zhao, Dawei},
  year = 2016,
  month = dec,
  journal = {Physics Reports},
  volume = {664},
  pages = {1--113},
  issn = {03701573},
  doi = {10.1016/j.physrep.2016.10.006},
  urldate = {2023-05-06}
}

@article{Wang2017VaccinationEpidemics,
  title = {Vaccination and Epidemics in Networked Populations---{{An}} Introduction},
  author = {Wang, Zhen and Moreno, Yamir and Boccaletti, Stefano and Perc, Matja{\v z}},
  year = 2017,
  month = oct,
  journal = {Chaos, Solitons \& Fractals},
  volume = {103},
  pages = {177--183},
  issn = {0960-0779},
  doi = {10.1016/j.chaos.2017.06.004},
  urldate = {2025-09-18}
}

@inproceedings{Wang2024SeCoNetHeterosexual,
  title = {{{SeCoNet}}: A Heterosexual Contact Network Growth Model for Human Papillomavirus Disease Simulation},
  booktitle = {Proceedings of the 2023 {{IEEE}}/{{ACM International Conference}} on {{Advances}} in {{Social Networks Analysis}} and {{Mining}}},
  author = {Wang, Weiyi and Piraveenan, Mahendra},
  year = 2024,
  month = mar,
  series = {{{ASONAM}} '23},
  pages = {98--102},
  publisher = {Association for Computing Machinery},
  address = {New York, NY, USA},
  doi = {10.1145/3625007.3627478},
  urldate = {2024-04-01},
  isbn = {979-8-4007-0409-3}
}

@misc{Wang2026SeCoNet,
  title = {{{SeCoNet}}},
  author = {Wang, Weiyi},
  year = 2026,
  month = jun,
  urldate = {2026-07-27}
}

@article{Watts1998CollectiveDynamics,
  title = {Collective Dynamics of `Small-World' Networks},
  author = {Watts, Duncan J. and Strogatz, Steven H.},
  year = 1998,
  month = jun,
  journal = {Nature},
  volume = {393},
  number = {6684},
  pages = {440--442},
  publisher = {Nature Publishing Group},
  issn = {1476-4687},
  doi = {10.1038/30918},
  urldate = {2023-05-13},
  copyright = {1998 Macmillan Magazines Ltd.},
  language = {en}
}

@article{Weeden2020SmallWorldNetwork,
  title = {The {{Small-World Network}} of {{College Classes}}: {{Implications}} for {{Epidemic Spread}} on a {{University Campus}}},
  author = {Weeden, Kim A. and Cornwell, Benjamin},
  year = 2020,
  month = may,
  journal = {Sociological Science},
  volume = {7},
  pages = {222--241},
  issn = {2330-6696},
  doi = {10.15195/v7.a9},
  urldate = {2026-10-03},
  language = {en-US}
}

@misc{WorldHealthOrganization2022CervicalCancer,
  title = {Cervical Cancer},
  author = {{World Health Organization}},
  year = 2022,
  month = feb,
  urldate = {2023-03-23},
  howpublished = {https://www.who.int/news-room/fact-sheets/detail/cervical-cancer},
  language = {en}
}

@article{WorldHealthOrganization2022HumanPapillomavirus,
  title = {Human Papillomavirus Vaccines: {{WHO}} Position Paper (2022 Update)},
  author = {{World Health Organization}},
  year = 2022,
  month = dec,
  journal = {Weekly Epidemiological Record},
  volume = {97},
  number = {50},
  pages = {645--672},
  publisher = {World Health Organization},
  urldate = {2023-05-22},
  language = {en}
}
\end{document}